\documentclass[11pt]{article}
\usepackage{verbatim,amsmath,amssymb}
\usepackage{epsfig,float,color}
\usepackage[font=small,labelfont=bf]{caption}
\usepackage{subcaption}
\usepackage{geometry}
\usepackage{natbib}
\usepackage{footnote}   
\usepackage{array}
\usepackage{ragged2e}
\usepackage[T1]{fontenc}

\usepackage{graphicx}
\usepackage [english]{babel}
\usepackage [autostyle, english = american]{csquotes}
\usepackage{hyperref,url}

\usepackage{titlesec}
\titleformat{\subsection}
  {\normalfont\normalsize\bfseries}{\thesubsection}{1em}{}

\titleformat{\subsubsection}
  {\normalfont\normalsize\bfseries}{\thesubsubsection}{1em}{}

\usepackage{hyperref}
\usepackage[dvipsnames]{xcolor}
\usepackage{xpatch}
\titleclass{\subsubsubsection}{straight}[\subsection]

\newcounter{subsubsubsection}[subsubsection]
\renewcommand\thesubsubsubsection{\thesubsubsection.\arabic{subsubsubsection}}

\titleformat{\subsubsubsection}
  {\normalfont\normalsize\bfseries}{\thesubsubsubsection}{1em}{}
\titlespacing*{\subsubsubsection}
{0pt}{3.25ex plus 1ex minus .2ex}{1.5ex plus .2ex}

\titleformat{\section}
  {\normalfont\Large\bfseries}{\thesection}{1em}{}

\titleformat{\subsection}
  {\normalfont\large\bfseries}{\thesubsection}{1em}{}

\titleformat{\subsubsection}
  {\normalfont\normalsize\bfseries}{\thesubsubsection}{1em}{}

\titleformat{\subsubsubsection}
  {\normalfont\normalsize\bfseries}{\thesubsubsubsection}{1em}{}

\makeatletter
\renewcommand\paragraph{\@startsection{paragraph}{5}{\z@}%
  {3.25ex \@plus1ex \@minus.2ex}%
  {-1em}%
  {\normalfont\normalsize\bfseries}}
\renewcommand\subparagraph{\@startsection{subparagraph}{6}{\parindent}%
  {3.25ex \@plus1ex \@minus .2ex}%
  {-1em}%
  {\normalfont\normalsize\bfseries}}
\def\toclevel@subsubsubsection{4}
\def\toclevel@paragraph{5}
\def\toclevel@subparagraph{6}
\def\l@subsubsubsection{\@dottedtocline{4}{7em}{4em}}
\def\l@paragraph{\@dottedtocline{5}{10em}{5em}}
\def\l@subparagraph{\@dottedtocline{6}{14em}{6em}}
\makeatother

\DeclareUnicodeCharacter{0301}{\'{e}}
\usepackage[labelfont=bf]{caption}
\usepackage{array}
\usepackage{makecell}
\usepackage{xcolor}
\usepackage[normalem]{ulem}
\usepackage{setspace}

\definecolor{darkblue}{rgb}{0,0,1}
\definecolor{col1}{rgb}{1,0,1}
\definecolor{col2}{rgb}{0,0.5,0}
\definecolor{col3}{rgb}{0.5,0,1}
\definecolor{col4}{rgb}{0.1,.75,0}

\hypersetup{pdftex=true, colorlinks=true, breaklinks=true, linkcolor=darkblue, menucolor=darkblue, pagecolor=darkblue, citecolor=darkblue, urlcolor=darkblue}

\newcommand{\bitm}{\begin{itemize}}
\newcommand{\eitm}{\end{itemize}}
\newcommand{\bnumr}{\begin{enumerate}}
\newcommand{\enumr}{\end{enumerate}}

\newcommand {\eqb}[1]{\begin{equation}\begin{array}{#1}}
\newcommand {\eqe}{\end{array}\end{equation}}

\newcommand {\esb}[1]{\begin{equation*}\begin{array}{#1}}
\newcommand {\ese}{\end{array}\end{equation*}}
\newcommand {\ds}{\displaystyle}

\newcommand {\II}{{I\kern-.3em I}}
\newcommand {\III}{{I\kern-.3em I\kern-.3em I}}

\newcommand {\mra}{\mathrm{a}}
\newcommand {\mrb}{\mathrm{b}}
\newcommand {\mrc}{\mathrm{c}}
\newcommand {\mrd}{\mathrm{d}}

\newcommand {\mrk}{\mathrm{k}}

\newcommand {\mrm}{\mathrm{m}}
\newcommand {\mrn}{\mathrm{n}}

\newcommand {\mrp}{\mathrm{p}}

\newcommand {\mrs}{\mathrm{s}}
\newcommand {\mrt}{\mathrm{t}}

\newcommand {\mrz}{\mathrm{z}}

\newcommand {\mg}{\mathbf{g}}
\newcommand {\mh}{\mathbf{h}}

\newcommand {\ba}{\boldsymbol{a}}

\newcommand {\be}{\boldsymbol{e}}

\newcommand {\bg}{\boldsymbol{g}}
\newcommand {\bh}{\boldsymbol{h}}

\newcommand {\bl}{\boldsymbol{l}}

\newcommand {\bn}{\boldsymbol{n}}

\newcommand {\br}{\boldsymbol{r}}

\newcommand {\bt}{\boldsymbol{t}}
\newcommand {\bu}{\boldsymbol{u}}

\newcommand {\bx}{\boldsymbol{x}}

\newcommand {\mI}{\mathbf{I}}

\newcommand {\bA}{\boldsymbol{A}}
\newcommand {\bB}{\boldsymbol{B}}

\newcommand {\bF}{\boldsymbol{F}}
\newcommand {\bG}{\boldsymbol{G}}
\newcommand {\bH}{\boldsymbol{H}}

\newcommand {\bK}{\boldsymbol{K}}

\newcommand {\bR}{\boldsymbol{R}}

\newcommand {\IR}{{\rm\kern.24em
   \vrule width.02em height1.53ex depth-.05ex
   \kern-.3em R}}
\newcommand {\ic}{{\rm\kern.20em
   \vrule width.02em height1.0ex depth-.05ex
   \kern-.22em c}}
\newcommand {\ia}{{\rm\kern.20em
   \vrule width.02em height1.05ex depth-.0ex
   \kern-.25em a}}
\newcommand {\IC}{{\rm\kern.24em
   \vrule width.02em height1.4ex depth-.05ex
   \kern-.26em C}}
\newcommand {\ID}{{\rm\kern.34em
   \vrule width.02em height1.5ex depth-.05ex
   \kern-.36em D}}
\newcommand {\IS}{{\rm\kern.24em
   \vrule width.02em height1.6ex depth.05ex
   \kern-.26em S}}
\newcommand {\IT}{{\rm\kern.50em
   \vrule width.02em height1.55ex depth-.05ex
   \kern-.52em T}}

\newcommand {\IE}{{\rm\kern.24em
   \vrule width.02em height1.55ex depth-.05ex
   \kern-.33em E}}
\newcommand {\IEa}{{\rm\kern.24em
   \vrule width.02em height1.55ex depth-.05ex
   \kern-.33em E}^{1}_{ijkl}}
\newcommand {\IEb}{{\rm\kern.24em
   \vrule width.02em height1.55ex depth-.05ex
   \kern-.33em E}^{2}_{ijkl}}

\newcommand {\Ass}[2]{\kern 0.9ex \vrule width0.45em height0.2ex depth0ex \kern -2.1ex \bigwedge_{#1}^{#2}}
\newcommand {\ASS}[2]{\kern 1.45ex \vrule width0.5em height0.2ex depth0ex \kern -2.65ex \bigwedge_{#1}^{#2}}

\definecolor{col1}{rgb}{1,0,1}
\definecolor{col2}{rgb}{0,.7,0}

\usepackage{etoolbox}



\graphicspath{ {images/} }

\usepackage{hyperref}
\usepackage{fancyhdr}

\fancypagestyle{firstpage}{
  \fancyhf{}
  \fancyhead[L]{\small Published in: \textit{Advanced Materials Interfaces} (2026), e70553}
  \fancyhead[R]{\small DOI: \href{https://doi.org/10.1002/admi.70553}{10.1002/admi.70553}}
  \fancyfoot[C]{\thepage}
}

\begin{document}

\begin{center}
\Large{\bf{A survey of interlayer interaction models for graphene and other 2D materials}}
\end{center}
\renewcommand{\thefootnote}{\fnsymbol{footnote}}

\begin{center}
\large{Gourav Yadav$^{\mra}$, Shakti S.~Gupta$^{\mra}$, and Roger A.~Sauer$^{\mrb,\mrc,\mrd}$\footnote[1]{corresponding author, email: roger.sauer@rub.de}
}\\
\vspace{4mm}

\small{\textit{
$^\mra$Department of Mechanical Engineering, Indian Institute of Technology Kanpur, UP 208016, India \\[1.1mm]
$^\mrb$Institute for Structural Mechanics, Ruhr University Bochum, 44801 Bochum, Germany \\[1.1mm]
$^\mrc$Department of Structural Mechanics, Gda\'{n}sk University of Technology, 80-233 Gda\'{n}sk, Poland \\[1.1mm]
$^\mrd$Department of Mechanical Engineering, Indian Institute of Technology Guwahati, Assam 781039, India
\\
}}

\end{center}

\vspace{-1mm}

\renewcommand{\thefootnote}{\arabic{footnote}}

\vspace{-5mm}

\vspace{2mm}

\vspace{-1mm}


\rule{\linewidth}{.15mm}
{\bf Abstract}:
This work presents a survey of mechanical models describing van der Waals interactions between 2D materials, encompassing both continuous elastomer-like materials and discrete (crystalline) 2D materials such as graphene. These interactions give rise to a range of physical phenomena, including contact instabilities, Moiré patterns, surface reconstructions, and superlubricity. The underlying contact forces follow from the variation of an interfacial interaction potential. The presentation first discusses normal contact models, and then tangential contact models. Both atomistic and continuum approaches are considered. In addition, the influence of external loading and changes in length scale on the ground state configuration and frictional contact behavior are analyzed. A particular emphasis is placed on discussing strategies that reduce computational cost in multiscale modeling.


\vspace{-5mm}
\rule{\linewidth}{.15mm}

\tableofcontents

\vspace{2mm}
\section{Introduction} \label{sec_1}
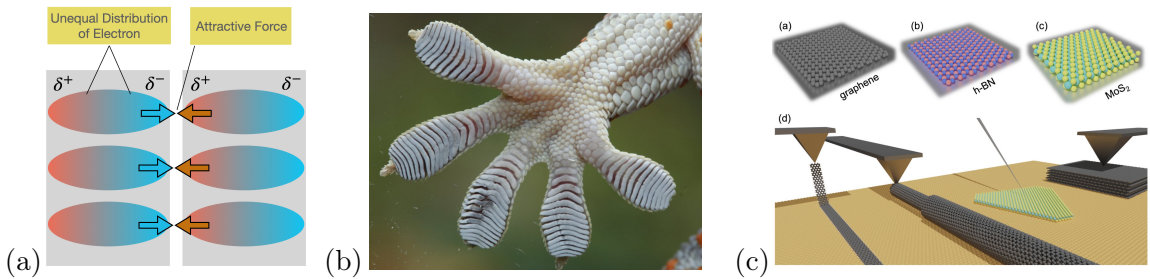
\begin{figure}[H]
\begin{center} \unitlength1cm
\begin{picture}(0,4.0)
\put(-7,0){\includegraphics[height=35mm]{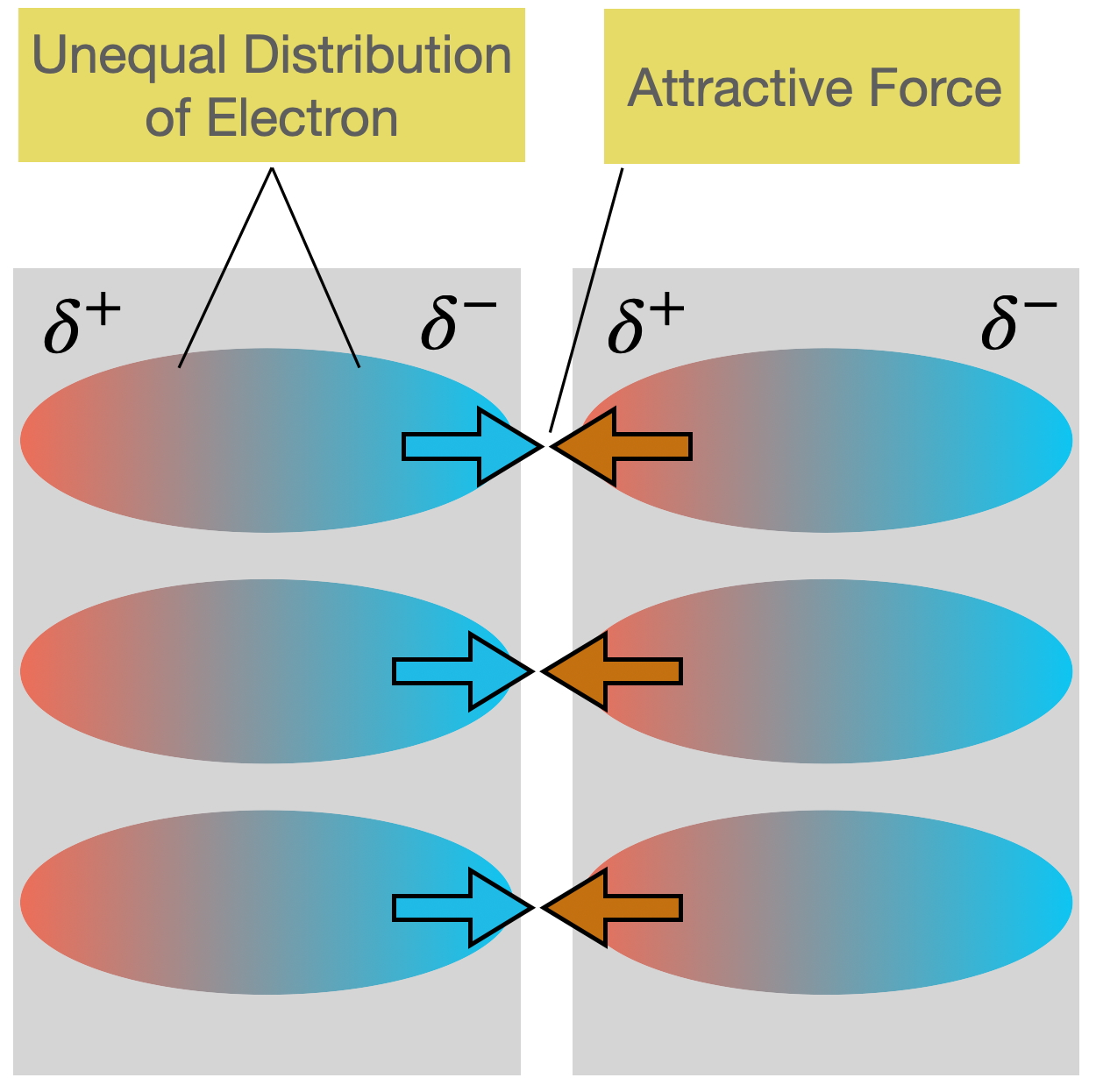}}
\put(-2.7,0){\includegraphics[height=34mm]{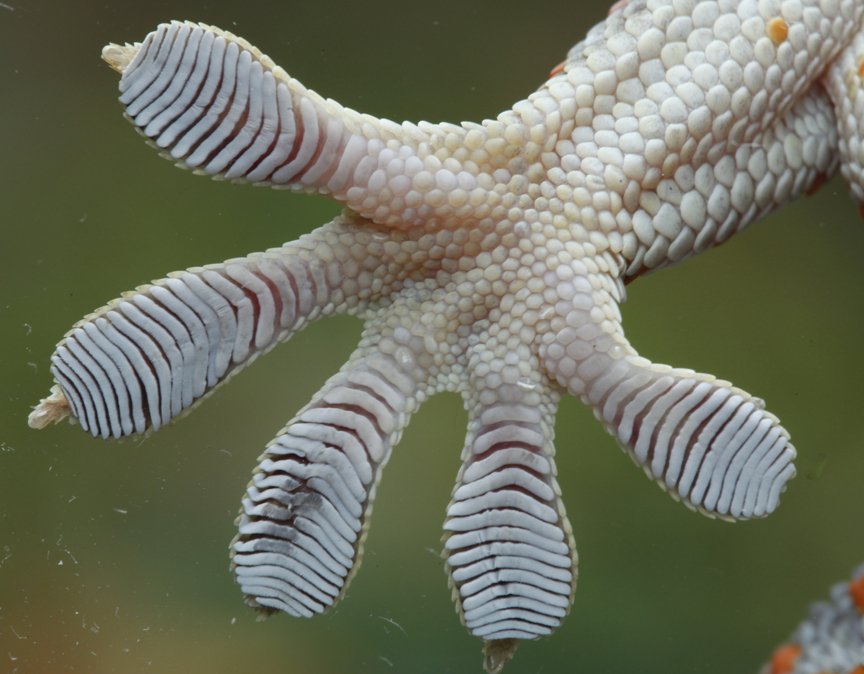}}
\put(2.6,0){\includegraphics[height=34mm]{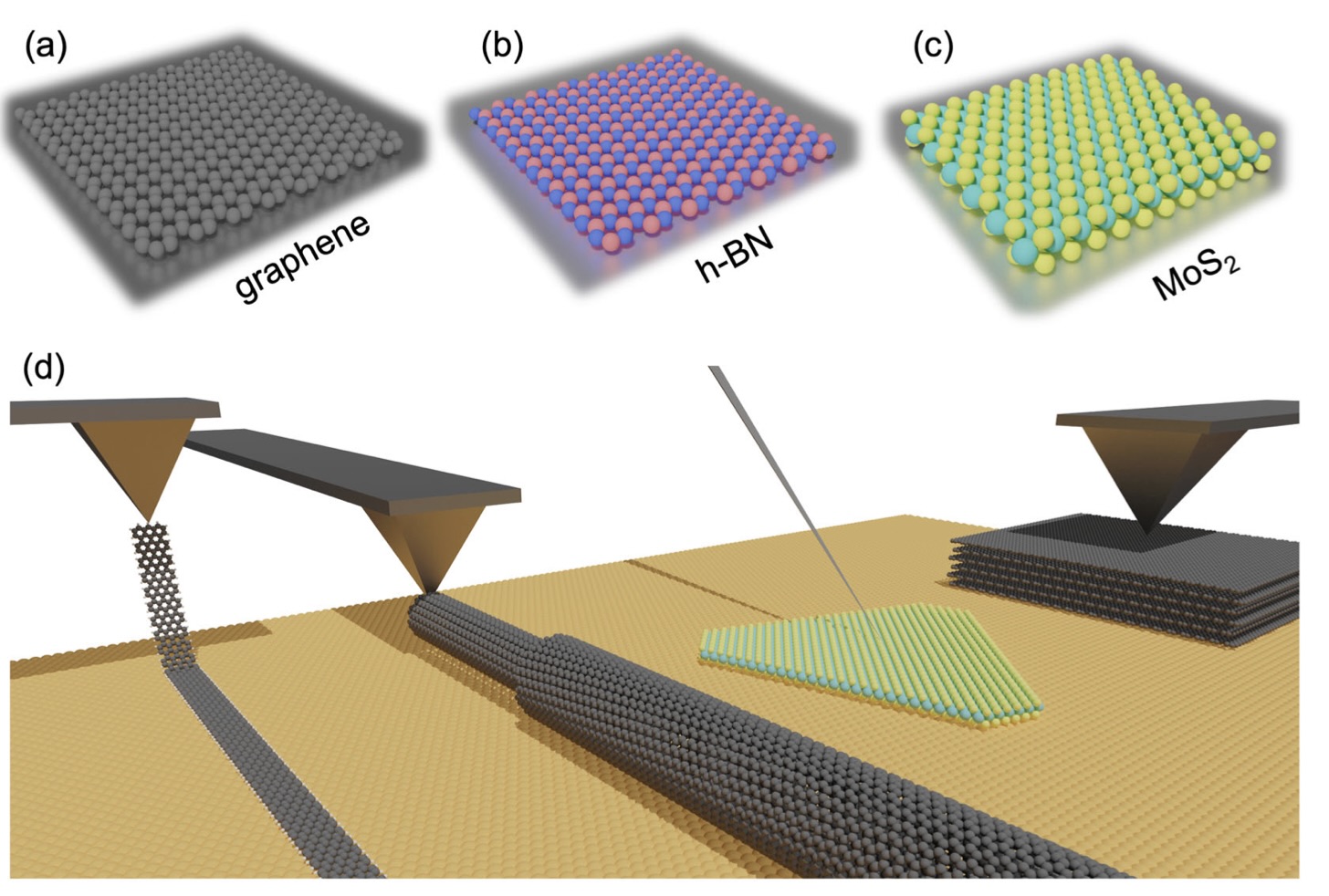}}
\put(-7.5,0){(a)}
\put(-3.3,0){(b)}
\put(2.1,0){(c)}
\end{picture}
\caption{(a) Van der Waals forces between temporary charges (b) Adhesive structure of the gecko's foot (c) Some common nanomaterial structures whose adhesion has been investigated using atomic force microscopy (AFM); reprinted from \citet{wang2024colloquium} with permission from American Physical Society.} 
\label{Intro-pictures}
\end{center}
\end{figure}

Due to the electron cloud's low probability of being uniformly distributed around the nucleus of an atom or a molecule at any given time, it is very likely to be polarized, forming a temporary dipole (see Fig.~\ref{Intro-pictures}a). This dipole induces polarization in surrounding nonpolar or polar molecules, which then propagate like an electromagnetic wave 
\citep{parsegian1973van, israelachvili2011intermolecular,   rodriguez2011casimir,luo2014van}. This phenomenon of polarization results in a force called the London dispersion force \citep{london1937general}. Other similar types of force/interaction that involve at least one permanent dipole are the Keesom force \citep{keesom1915second} and Debye force \citep{roberts1938induced}. All three interactions, London, Kessom, and Debye, are called van der Waals (vdW) interactions, named after the Dutch physicist Johannes Diderik van der Waals. He initially proposed these interactions in 1873 while formulating a hypothesis to explain the properties of real gases. Since these interactions do not involve the displacement, exchange, or pooling of electrons, they are weaker and of longer range than chemical bonds. For most ranges, vdW forces are attractive, resulting in a sticking tendency between bodies known as \textit{adhesion} (see Figs.~\ref{Intro-pictures}b and \ref{Intro-pictures}c). This term is somewhat broad, as there can be various sources of adhesion, as indicated by \citet{israelachvili2011intermolecular} and \citet{sauer2016survey}. In the frictional context of non-polar molecules like crystalline materials, generally, vdW forces refer to London dispersion forces; however, in the context of adhesion, they can refer to any vdW interaction. A comprehensive discussion on the molecular origins of adhesion can be found in \citet{dzyaloshinskii1961general, gerberich2006physics} and \citet{ israelachvili2011intermolecular}.

Although the individual vdW interactions are very weak, adhesive stress from vdW interaction can reach up to 10 GPa in intimately contacting surfaces \citep{popov2010contact}. Hence, these forces are often strong enough to deform bodies elastically or plastically when they come into contact with each other. However, in general, contact surface intimacy is hindered due to the presence of roughness. The vdW forces can be found to govern many applications in nanotechnology \citep{rance2010van, nerngchamnong2013role, li2019probing}, structural biology \citep{Israelachvili_1973, leckband2001intermolecular}, and bio-adhesive systems of small animals \citep{autumn2000adhesive}. Being a molecular force, the vdW force increases in importance as the length scale of the system decreases. This trend is evident across a range of engineering applications, such as micro-electro-mechanical systems (MEMS) \citep{dechev2004microassembly, meitl2006transfer, purtov2015switchable, cecil2016review}, debonding and delamination of thin films \citep{hendrickx2005spectral, roy2007cross}, flow and aggregation of adhesive particles \citep{kendall2007effect, liu2010applicability, li2011adhesive}, adhesive bonding technologies \citep{banea2009adhesively, he2011review}, and functionally graded materials \citep{hess2016simple}.
In parallel, vdW forces also govern a variety of interfacial mechanical phenomena, including loading/unloading hysteresis in rough contacts \citep{carbone2015loading}, superlubric sliding in incommensurate interfaces \citep{vanossi2013colloquium}, and adhesion-controlled friction under tangential loading \citep{popov2017friction}.  Generally, in adhesive contact interfaces, sliding can also occur, for example, during the peeling of adhesive tapes \citep{mergel2021contact}, when lubrication is present \citep{berman2014graphene}, and in case of bending of multilayer van der Waals structures \citep{pan2019bending}. This tendency modifies vdW interactions, giving rise to tangential tractions in addition to normal tractions, which together form the basis of a vdW contact model. The tangential component of the vdW contact produces tractions that manifest as adhesive friction. 

\begin{figure}[H]
\begin{center} \unitlength1cm
\begin{picture}(0,5.5)
\put(-6.2,0){\includegraphics[height=50mm]{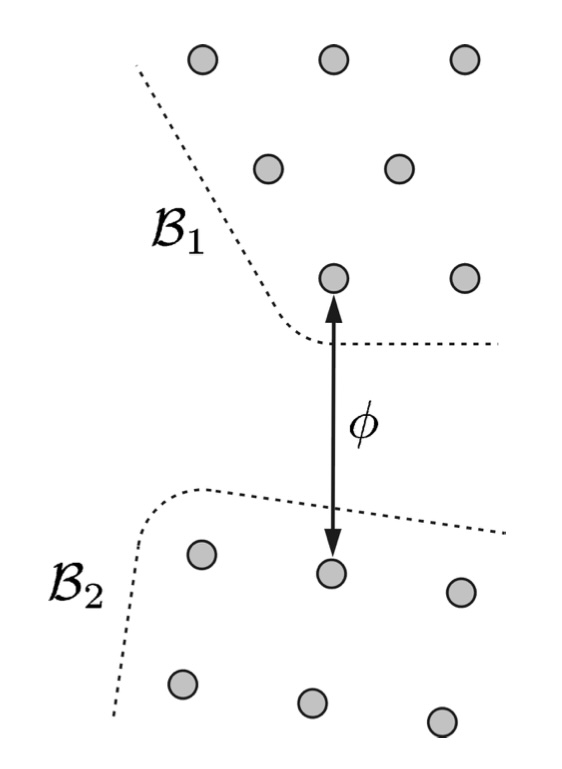}}
\put(0,0){\includegraphics[height=55mm]{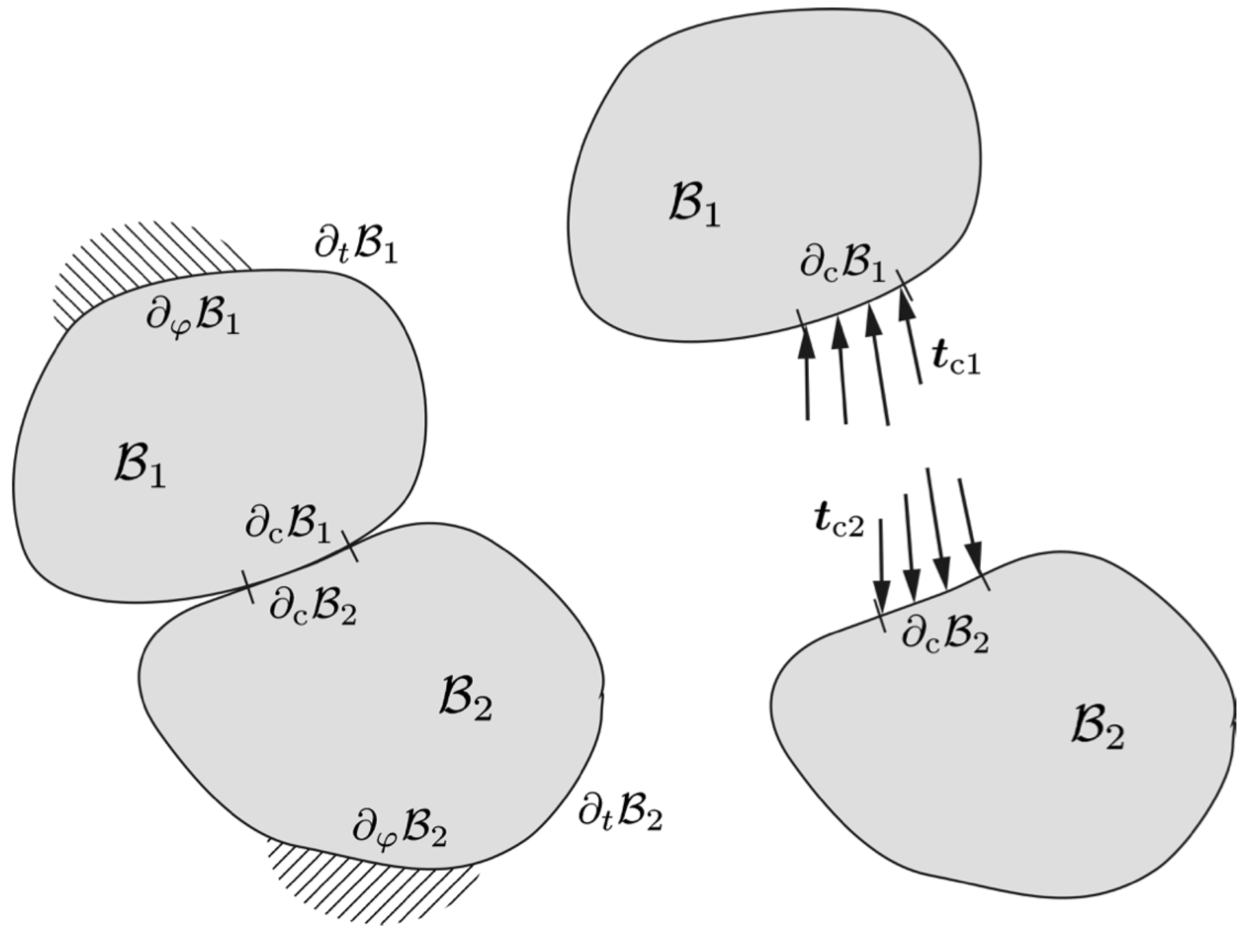}}
\put(-6.8,0){(a)}
\put(0.0,0){(b)}
\end{picture}
\caption{Description of (a) nano and (b) macro level contact. Adopted from \citet{sauer2006, sauer2011challenges}.}
    \label{Nanosclae versus macroscale and contact description}
\end{center}
\end{figure}

At the nanometer scale and below, contact is resolved into the interactions of individual atomic particles, as illustrated in Fig.~\ref{Nanosclae versus macroscale and contact description}a. In contrast, at the macroscopic level, adhesive contact is typically described using continuum quantities such as contact tractions and deformations see Fig.~\ref{Nanosclae versus macroscale and contact description}b.
The interactions are typically modeled using pairwise interaction potentials. Some of the widely used potentials include  the Mie potential \citep{mie1903kinetischen}, the Morse potential \citep{morse1929diatomic}, the Lennard-Jones (LJ) potential \citep{lennard1931cohesion}, the Buckingham potential \citep{buckingham1938classical}, and the Tang-Tönnies potential \citep{tang1984improved}. Among these, the LJ potential is the most commonly employed in modeling vdW interactions, primarily due to its relatively simple analytical form and its ability to capture both short-range repulsion and long-range attraction. The LJ potential can be expressed as
\begin{equation}
\ds \phi(r)=\epsilon\left(\frac{r_0}{r}\right)^{\!\!12}- 2\epsilon\left(\frac{r_0}{r}\right)^{\!\!6}~,
\label{LJ}
\end{equation}
where $r$ denotes the distance between interacting particles, and $\epsilon$ and $r_0$ are model parameters that characterize the strength and equilibrium separation of the interaction, respectively. 
This paper reviews and discusses contact models of two-dimensional (2D) materials, with a focus on graphene and hexagonal boron nitride (h-BN). These materials are chosen both due to the extensive literature available and because their structures are widely used as representative systems for developing models that provide general insights into the behavior of 2D materials. When stacked, these lamellar solid materials display easy sliding between neighboring atomic layers and are used as solid lubricants, \citep{GRATTAN1967453, bowden2001friction, doi:10.1126/science.aab0930}. These materials have revolutionized the field of tribology, as the friction coefficient between their layers is found to be very low ($<$0.01) even under large normal pressures \citep{tomanik2023review}. Because of the strong in-plane covalent bonds and weak vdW interactions between the layers, it is possible
to achieve tunable properties, one of which is structural superlubricity. It is achieved through
different stacking configurations, relative twisting, and strain engineering \citep{Dienwiebel2004a, Feng2013, mandelli2017sliding, Cao2018, morovati2022interlayer}. One cause of the very low friction is the formation of long-range patterns, also called Moir{\'e} patterns or Moir{\'e} supercells, resulting from rotational misalignment and lattice mismatch between the stacked layers \citep{ru2020interlayer}. Achieving superlubricity across multiple lengths and load scales was difficult in the past, but recently it has been observed well into macro and mesoscales in layered 2D materials \citep{berman2015macroscale}.

Friction between ideally smooth surfaces arises from atomic-scale stick-slip behavior -- an inherently
unstable process. This dynamic phenomenon primarily governs frictional energy dissipation
through mechanisms such as phononic excitation \citep{park2006electronic, qi2008electronic}. Additionally, energy is also lost via the scattering
of excited electrons, known as Coulomb drag between closely spaced layers \citep{liu2023edge}. 2D materials, with their unique atomic structures, exhibit distinctive electronic transport
properties and phonon-electron coupling, both of which can infuence their frictional behavior  \citep{park2006electronic, qi2008electronic, filleter2009friction,prasad2017phononic}. Notably, the strength of electron-phonon interactions in graphene has been found to depend on the
number of layers and local interactions with substrates \citep{filleter2009friction, castellanos2013periodic}. Further, multiple studies have demonstrated that the frictional response of 2D materials is strongly influenced by their size, shape, and sliding direction \citep{verhoeven2004model, Wang2017size, mandelli2017sliding, ouyang2018nanoserpents, Xue2022, song2023atomic, gao2025frictional, yadav2025investigating}.

There are several existing reviews in the field of mechanics of adhesive interfaces. The review \citet{autumn2002evidence} focuses on gecko adhesion. \citet{nosonovsky2007multiscale} address friction mechanisms in nano and bio-tribology. \citet{sauer2016survey} provides a survey of computational methods for adhesive contact focusing on general continuum mechanical models for the attractive adhesion of solids. The review by \citet{ciavarella2019role} examines the advantages and drawbacks of different techniques used to analyze contact problems that involve adhesive tractions. Their focus is on how the surface roughness affects contact. \citet{dai2020mechanics} conducted a comprehensive analysis of recent experimental and theoretical investigations concerning the mechanical properties of interfaces between 2D materials. Their analysis encompassed both normal contacts (adhesion) and tangential interactions (shear/friction) occurring at the interface. Multiple reviews \citep{hu2013energy, vanossi2013colloquium, krylov2014physics, guo2014friction, penkov2014tribology, zhai2017carbon, yang2017frictional, zhang2019tribology, luo2020superlubricitive, liu2020graphene, song2020structural, shekhar2021overcoming, guo2021recent, luo2021origin, zhang2022friction, yang2023review, wang2024colloquium} have covered the advances in friction research for low-dimensional nanomaterials, like graphene, h-BN, and transition-metal dichalcogenide (TMD) such as Tungsten diselenide (WSe$_2$) and Molybdenum disulfide (MoS$_2$), and related nanostructures such as carbon nanotubes (CNTs), and their derivatives. 

This review focuses on combined adhesion and friction phenomena and is motivated by bringing new understanding to the topic. Its primary aim is to provide a comprehensive overview of existing contact modeling techniques, with a specific focus on continuum-based methods. A secondary aim is to discuss models that can potentially be used in structural analysis of layered vdW materials. A special focus is given to works whose contributions are significant either in establishing foundational models or in developing efficient methodologies capable of capturing vdW adhesive friction. 

The remainder of this paper is organized as follows: Section~\ref{sec_2}, presents normal contact, its mechanism, and models for continuous and discrete interfaces. Models that reduce the order of integration of the vdW energy are covered, and continuum-based contact models are discussed. Further, the origin of the analytical form of the adhesion energy for various types of crystalline structures is discussed. Section~\ref{sec_3} discusses tangential contact models for nanoscale sticking and sliding. Friction laws and the origin of friction are discussed. Models that cover the frictional modeling in the quasi-static sliding regime are discussed for continuous and discrete interfaces. Section \ref{sec_4} draws conclusions.

\section{Normal contact models } \label{sec_2}

Having established the significance of vdW contact in the introduction, this section now turns to a detailed discussion of its mechanical and computational modeling for normal contact of continuous and discrete interfaces.

\begin{figure}[H]
\begin{center} \unitlength1cm
\begin{picture}(0,5.0)
\put(-5,0){\includegraphics[height=50mm]{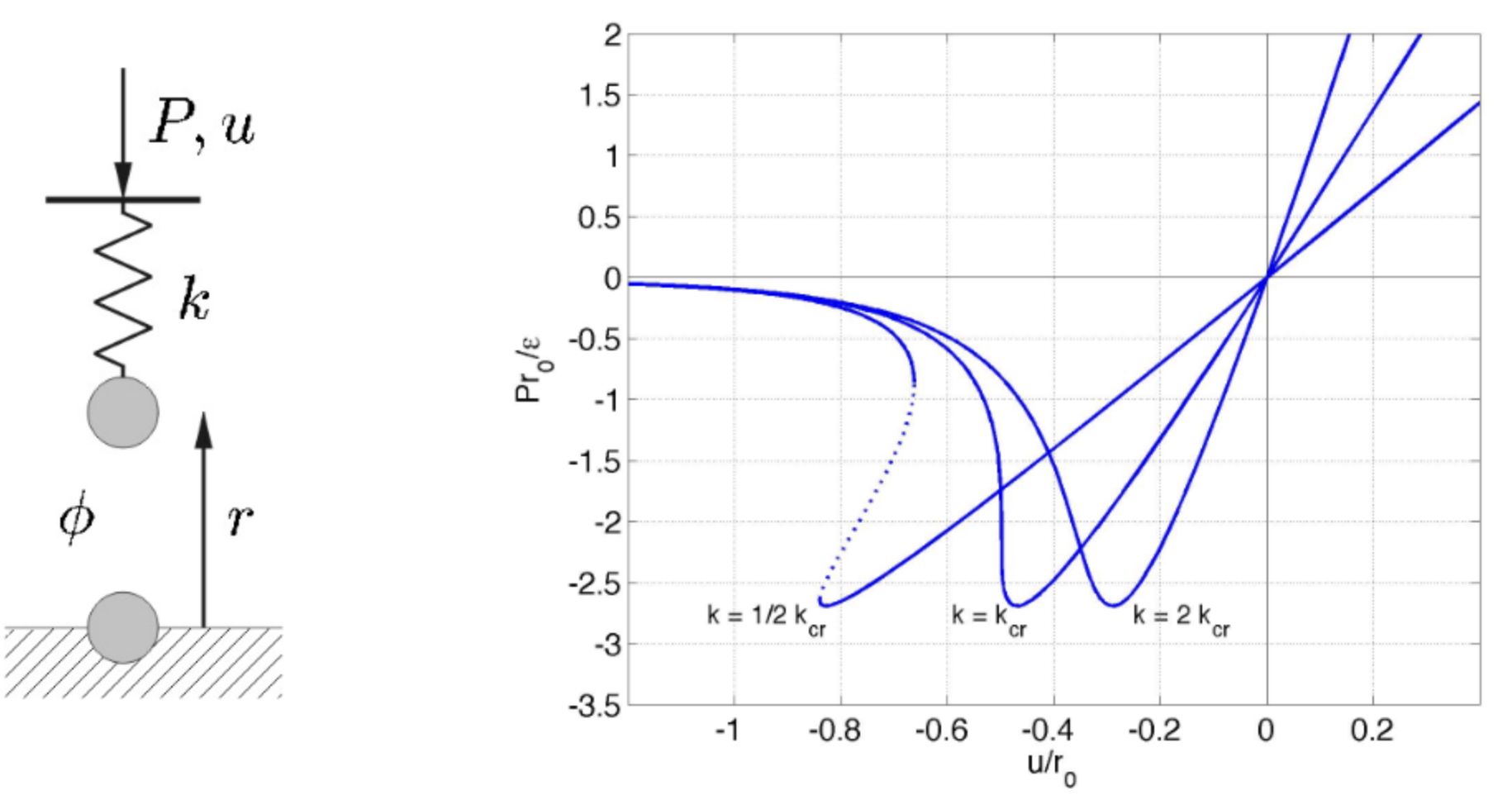}}
\put(-5.5,0){(a)}
\put(-1.5,0){(b)}
\end{picture}
\caption{(a) 1D adhesive contact model, (b) load-displacement curve. The dotted curve represents unstable solutions. Adopted from \citet{sauer2006}.}
    \label{normal contact instabilty}
\end{center}
\end{figure}

The most well-known analytical theories for adhesive normal contact are the models proposed by Johnson, Kendall, and Roberts (JKR) \citep{johnson1971surface}, and Derjaguin, Muller, and Toporov (DMT) \citep{derjaguin1975effect}. These classical models were later unified through the work of \citet{maugis1992adhesion}, who introduced an intermediate solution bridging the JKR and DMT regimes. Notably, these early adhesive contact theories assumed that indentation and detachment processes are reversible and non-dissipative during normal contact. However, both the JKR and DMT models exhibit sudden \textit{jump-in} and \textit{jump-out} at the onset of contact and during detachment, respectively. These abrupt transitions, known as \textit{adhesion instabilities}, result in energy dissipation and give rise to contact hysteresis \citep{israelachvili1995irreversibility}. The mechanics of such vdW driven adhesive contacts can be effectively captured using a one-dimensional (1D) model \citep{sauer2006, sauer2011challenges}. It considers two particles, as illustrated in Fig.~\ref{normal contact instabilty}a, interacting via the LJ potential $\phi$ given in Eq.~\eqref{LJ}. In this setup, the lower particle is fixed, while the upper particle is connected to a spring and displaced downward by an imposed displacement $u$, which requires an external force $P$. The total potential energy of the system is
\begin{equation}
 \ds   \Pi (r) = \frac{1}{2}k\big(u+(r-r_0)\big)^2 + \phi(r)~.
    \label{1-D bond-debond}
\end{equation}
For a prescribed displacement $u =\bar{u}$, the equilibrium condition $(\partial \Pi/\partial r)_{u=\bar{u}}=0$ together with the stability requirement $(\partial^2 \Pi/\partial r^2)_{u=\bar{u}}\geq 0$ lead to the following condition

\begin{equation}
    \ds k + \underbrace{13\cdot12 \frac{\epsilon}{r_0^2} \Big(\frac{r_0}{r}\Big)^{14} - 12\cdot7 \frac{\epsilon}{r_0^2} \Big(\frac{r_0}{r}\Big)^{8}} \geq 0~.
    \label{adhesion stabiltiy condition}
\end{equation}
The limiting value of spring stiffness $k$ satisfying Eq.~\eqref{adhesion stabiltiy condition} is obtained from the minimum of the underbrace term, which occurs at $r_0/r = (4/13)^{1/6}$. The resulting critical stiffness then is given as
\begin{equation}
 \ds   k_{\text{cr}}=36\left(\frac{4
}{13}\right)^\frac{4}{3}\frac{\epsilon}{r_0^2}~.
\label{critical stiffness adhesion}
\end{equation}
This mean that for $k > k_{\text{cr}}$, the system is stable. However, if $k < k_{\text{cr}}$, the system becomes unstable, as illustrated in Fig.~\ref{normal contact instabilty}b. This analysis leads to the conclusion that during strong adhesion between soft bodies like elastomers, instabilities may arise because the adhesive forces can become strong enough to overcome the internal elastic resistance of the material.

While 1D models can capture basic aspects of normal contact, they do not account for varying interface conditions.  These are addressed in the following -- distinguishing between \textit{continuous interfaces}, where interactions are homogenized (i.e., integrated) into smooth formulations, and \textit{discrete interfaces}, where the discrete nature of atomic lattice structure is explicitly retained. This distinction does not refer to the physical nature of the interface itself, but rather to the level of description employed in the modeling.

\subsection{Continuous interfaces }\label{sec_2_1}
 Engineering problems involving vdW interfaces often span over several orders of magnitude -- from nanometers to millimeters and larger -- making atomistic simulations computationally expensive. As a result, multiscale modeling approaches have become increasingly important for capturing the essential physics, while maintaining computational efficiency. 
\citet{tadmor1996quasicontinuum} developed an approach that combines atomistic and continuum descriptions -- the quasicontinuum (QC) method. Unlike strict domain-decomposition techniques, QC allows particles in the computational domain to transition smoothly between atoms governed by interatomic potentials and finite element nodes obeying the corresponding Cauchy-Born rule \citep{cousins1978inner, zanzotto1996cauchy}, as discussed in the QC context by \citet{ericksen2008cauchy}.  \citet{tadmor1999mixed} extended the QC method to complex crystals, while \citet{arroyo2002atomistic, arroyo2003finite, arroyo2004finite} generalized it to curved cyrstalline monolayer sheets. The QC method and related multiscale approaches are comprehensively reviewed in \citet{miller2002quasicontinuum, tadmor2011modeling} and \citet{kochmann2016quasicontinuum}. 
 
A promising strategy within the QC framework involves coarse-graining the material behavior at the atomic scale and transforming it into an effective continuum contact formulation. This approach relies on homogenization of the medium and involves the volume integration of pairwise atomic interactions, as originally proposed by \citet{hamaker1937london}. However, the classical formulation of \citet{hamaker1937london} is limited to rigid bodies and does not account for deformation during contact. To overcome this limitation, \citet{sauer2006} developed the coarse-grained contact (CGC) model, which incorporates material deformation during contact within a nonlinear continuum framework. The formulation of the contact energy, $\Pi_{\mathrm{c}}$, obtained following the coarse-graining procedure described in \citet{sauer2006} and \citet{sauer2007contact}, is discussed here. Also, discussed are some of the methods of reducing the region of influence, in order to reduce the order of numerical integration.

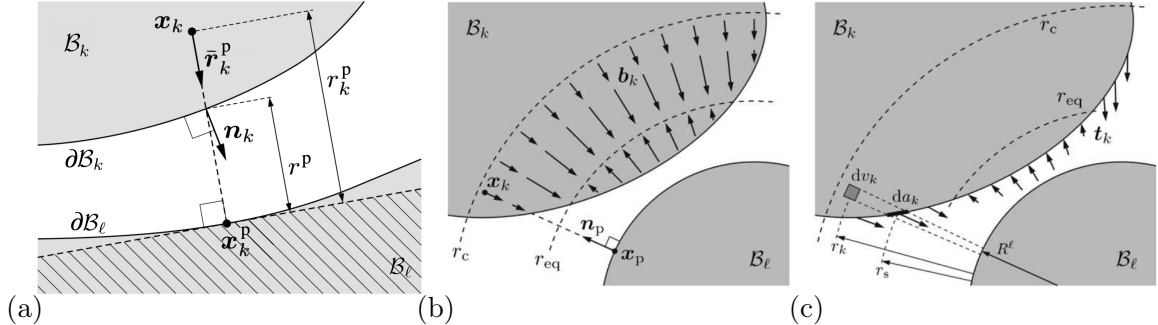
\begin{figure}[H]
\begin{center} \unitlength1cm
\begin{picture}(0,4.5)
\put(-7.4,0.3){\includegraphics[height=39mm]{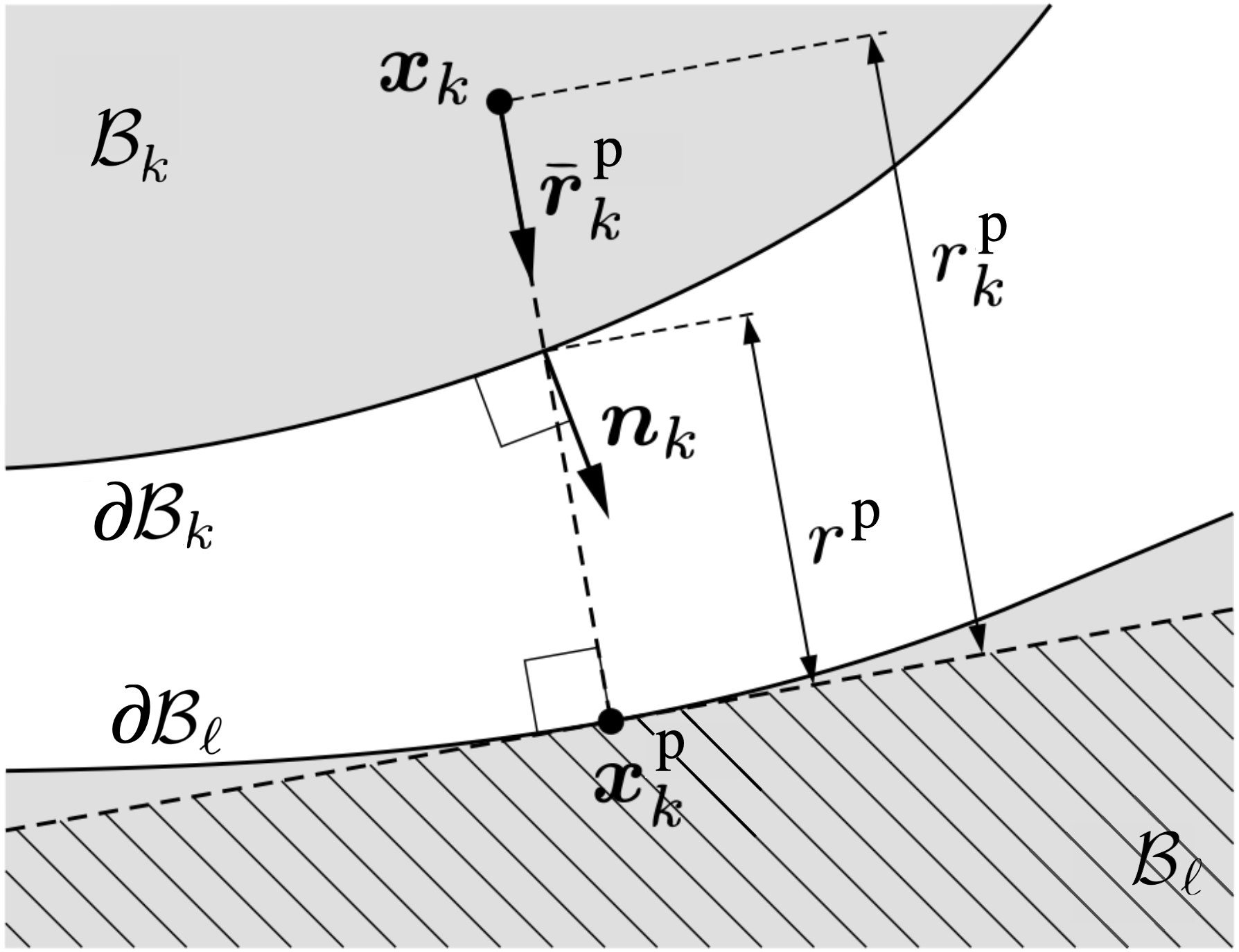}}
\put(-2,0.4){\includegraphics[height=38mm]{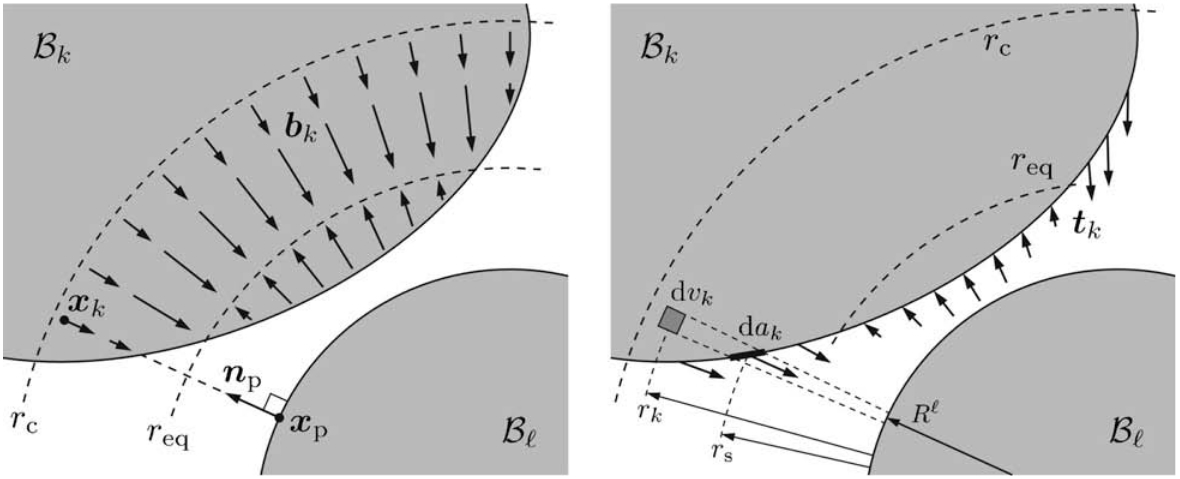}}
\put(-7.8,0){(a)}
\put(-2.4,0){(b)}
\put(2.5,0){(c)}
\end{picture}
\caption{Interaction forces of the Coarse Grained Contact model \citep{sauer2006}. (a) Closest point projection and approximation of $\mathcal{B}_\ell$ by a flat half-space, (b) body force formulation, and (c)  surface force formulation. (b) and (c) reprinted from \citet{sauer2009formulation} with permission from Wiley. } 
\label{sauer sir's work}
\end{center}
\end{figure}

The CGC model replaces the discrete summation of the pairwise interactions with the continuous double volume integral over the two interacting bodies, $\mathcal{B}_1$ and $\mathcal{B}_2$,
\begin{equation}
\ds \Pi_{\text{c}}= \int_{\mathcal{B}_1}\int_{\mathcal{B}_2}\beta_1 \,\beta_2 \,\phi(r)\,\text{d}v_2\,\text{d}v_1~.
\label{double integral contact energy}
\end{equation}
Here $\beta_1$ and $\beta_2$ are the particle number densities of bodies $\mathcal{B}_1$ and $\mathcal{B}_2$, respectively, in the current configuration. $\Pi_{\text{c}}$ has been evaluated analytically for rigid spheres \citep{hamaker1937london, bradley1932lxxix}, shells \citep{tadmor2001london}, and long slender bodies \citep{grill2023analytical, borkovic2024analytical}. However, for general contact problems, continuum mechanical formulations require numerical integration, for example within the finite element method (FEM), which can become computationally expensive. Therefore, approximate methods, which result in the partial analytical intergation of $\phi$ are particularly attractive, as they reduce computational cost while maintaining the accuracy of the formulation. Although \citet{argento1997surface} already evaluated integral \eqref{double integral contact energy} for small deformations, the CGC model also applies to large deformations. In the CGC approach, the reduction in computational cost is achieved by employing partial analytical integration and thus lowering the dimensionality of the remaining numerical integration domain. For instance, the double volume integration can be reduced to a single surface intergration, by mapping the body forces onto the surface and approximating the neighboring body by flat half-space, such that four of the six integrals are integrated analytically. Implementation of the CGC model into the FEM is achieved by formulating the governing weak form of the system. Following \citet{sauer2006} and \citet{sauer2009formulation}, the variation of the contact energy $\Pi_{\text{c}}$ for the two contacting bodies ($\mathcal{B}_k$, $k$ = 1, 2) can be written as $ \delta \Pi_{\text{c}} :=  \delta \Pi_{\text{c},1} +  \delta \Pi_{\text{c},2} $, where 
\begin{equation}
\ds  \delta \Pi_{\text{c},k} =  -\int_{\mathcal{B}_k} \delta\boldsymbol{\varphi}_k \cdot \beta_k \, \boldsymbol{b}_k \,\text{d}v_k~.\hspace{1cm} \text{(With no summation on $k$.)}
\label{variation of double integral contact energy _ 2}
\end{equation}
Here, $\boldsymbol{b}_k$ is a body force given by

\begin{equation}
\ds    \boldsymbol{b}_k = \ds \frac{A_\mathrm{H}}{2\pi r_0^4 J_{\ell} \beta_{0k}}\left[\frac{1}{5}\left(\frac{r_0}{r_k^\mrp}\right)^{\!\!10} - \left(\frac{r_0}{r_k^\mrp}\right)^{\!\!4} \right]\boldsymbol{n}_\mrp~,
    \label{full force force vector}
\end{equation}
where $A_\mathrm{H} = 2 \pi^2\beta_{01}\beta_{02}\epsilon r_0^6$ is the Hamaker constant\footnote{It is noted that the Hamaker constant $A_\mathrm{H}$, while expressed here in terms of atomistic parameters ($\beta_{01}$, $\beta_{02}$, $\epsilon$, $r_0$) for consistency with the underlying interaction potential, is a fundamentally material property that can be independently determined from experiments \cite{israelachvili2011intermolecular} or computed using continuum approaches such as Lifshitz theory \cite{lifshitz1992theory}.}, $\beta_{01}$ and $\beta_{02}$ are the particle number density in the reference configuration, $J_\ell$ is the local volume change of $\mathcal{B}_\ell$ and $r_k^\mrp$ denotes the distance between the position $\boldsymbol{x}_k$ and the surface $\partial\mathcal{B}_\ell$; see Fig.~\ref{sauer sir's work}a.

 A surface traction is obtained from projecting the body force $\boldsymbol{b}_k$ onto the body's surface $\partial\mathcal{B}_k$ (see Fig.~\ref{sauer sir's work}b and \ref{sauer sir's work}c). This projection restricts the interaction region and introduces an additional approximation, enabling the reduction of the volume integration over $\mathcal{B}_k$ to a surface integration over $\partial\mathcal{B}_k$. This is achieved by expressing the volume element as $\text{d}v_k = \Bar{\boldsymbol{r}}_k^\mrp \cdot \boldsymbol{n}_k \, \text{d}r \, \text{d}a_k$ in the weak form of the governing equation, where $\Bar{\boldsymbol{r}}_k^\mrp$ is a unit vector pointing toward the projection point on the opposing surface, and $\boldsymbol{n}_k$ is the unit outward normal to $\partial\mathcal{B}_k$, as illustrated in Fig.~\ref{sauer sir's work}a. 

\begin{figure}[H]
    \centering
    \includegraphics[scale=0.26]{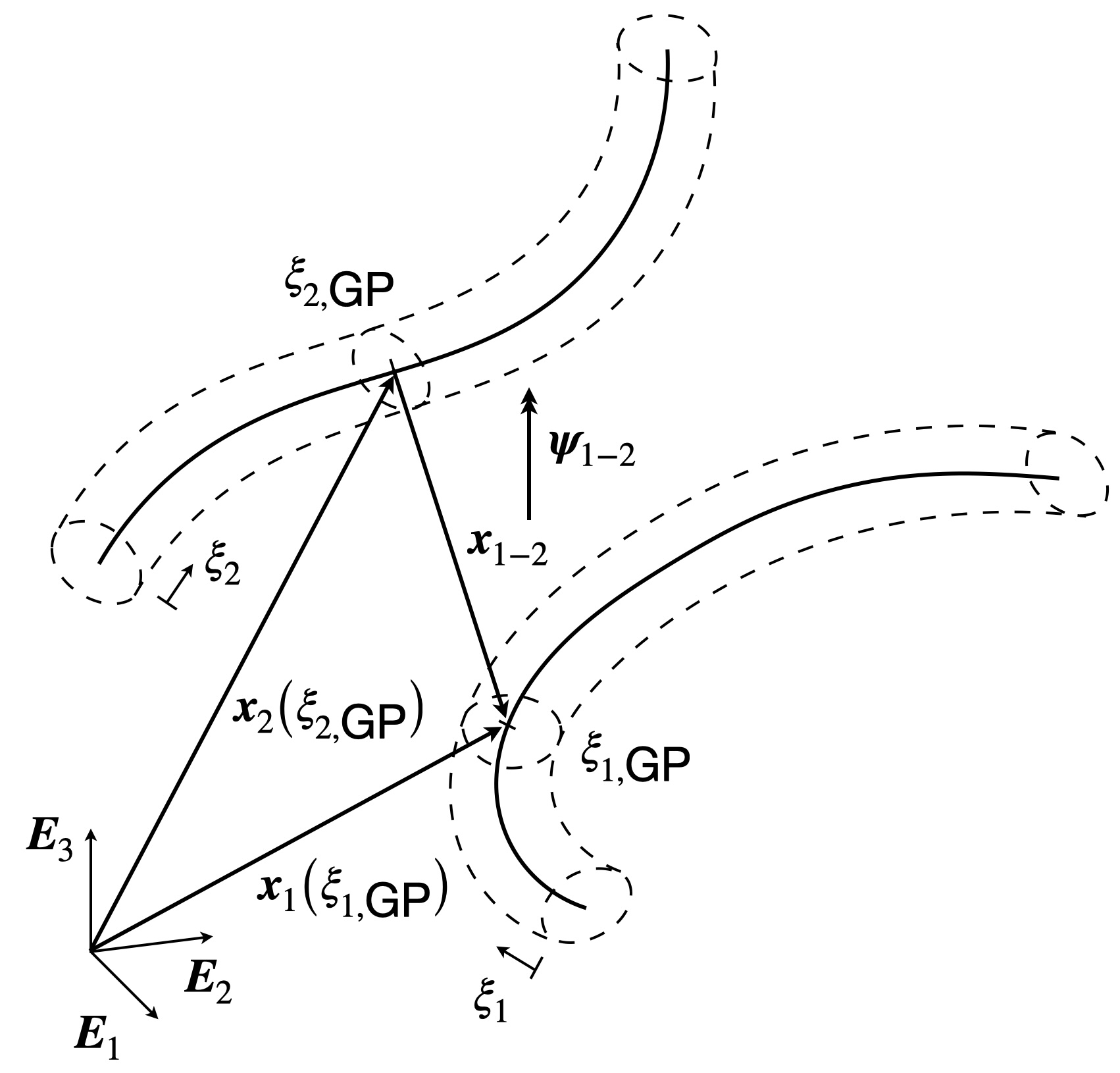}
    \caption{ Illustration of the SSIP approach of \citet{grill2020computational}. Shown are two cross-sections at Gaussian quadrature points (GP) $\xi_{1,\text{GP}}$ and $\xi_{2,\text{GP}}$ of beams 1 and 2, characterized by their separation $\boldsymbol{x}_{1-2}$ and relative rotation $\boldsymbol{\psi}_{1-2}$. The geometric quantities of a representative pair $(\xi_{1,\text{GP}}, \xi_{2,\text{GP}})$ are depicted. Figure adapted from \citet{grill2020computational} with permission from Wiley.} 
\label{SSIP}
\end{figure}

The resulting contact traction then follows as \citep{sauer2009formulation}
 \begin{equation}
\ds \boldsymbol{t}_k= \ds \pi \beta_k\beta_{\ell} \epsilon r_0^3 \left[ \frac{1}{45}\left(\frac{r_0}{r_\mrs}\right)^{\!\!9} - \frac{1}{3}\left(\frac{r_0}{r_\mrs}\right)^{\!\!3} \right]\boldsymbol{n}_\mrp~,
\label{contact traction_2}
\end{equation}
where $r_\mrs$ is the normal distance between the surfaces.
Using this traction, Eq.~\eqref{variation of double integral contact energy _ 2} can be rewritten as 
\begin{equation}
  \ds   \delta \Pi_{\text{c},k} = - \int_{\partial\mathcal{B}_k} \delta \boldsymbol{\varphi}_k \cdot  \boldsymbol{t}_k \, \cos{\alpha_k}\, \text{d}a_k~, 
  \label{final weak form}
\end{equation}
with $\cos{\alpha_k}$ = $-\bn_\mrp\cdot\bn_k$, see \citet{sauer2009formulation}. This approach results in a more efficient surface force formulation, that can directly be utilised in case of surface potentials \citep{sauer2013computational}. Similar to the approaches of \citet{sauer2006} and \citet{sauer2007contact}, several other studies have proposed methods to reduce the level of integration in Eq.~\eqref{double integral contact energy} by introducing suitable assumptions. For instance, \citet{fan2016three} developed an adhesive contact formulation that replaces the conventional double volume integration with a more efficient double-layer surface integral to compute the adhesive contact force vector. Furthermore, \citet{grill2020computational} proposed a 3D beam-beam interaction model that substantially simplifies the overall integration effort. Their Section-to-Section Interaction Potential (SSIP) model, illustrated by Fig.~\ref{SSIP}, reduces the original 6D integral in Eq.~\eqref{double integral contact energy} to a computationally more efficient 2D integration of the form
\begin{equation}
  \begin{split}
        \Pi_{\mathrm{c}} &=\int\int_{l_1,l_2}\int\int_{A_1,A_2}\rho_1(\bx_1)\,\rho_2(\bx_2)\,\phi(r)\,\mathrm{d}A_2 \,\mathrm{d}A_1\,\mathrm{d}s_2\,\mathrm{d}s_1, \,\,\,\,\, \text{with} \,\,\,\,\,r =\|\bx_1 -\bx_2\|\\
          &= \int\int_{l_1,l_2}\tilde{\pi}(\boldsymbol{x}_{1-2}, \boldsymbol{\psi}_{1-2})\,\mathrm{d}s_2\,\mathrm{d}s_1.\\
    \end{split}
    \label{SSIP equation}
\end{equation}

The resulting SSIP $\tilde{\pi}(\boldsymbol{x}_{1-2}, \boldsymbol{\psi}_{1-2})$ is intended to represent the interaction energy per unit length between two beam cross-sections (modeled as disks) at arbitrary separation and relative orientation. However, the role of the relative orientation of the interacting cross-sections is not accounted for in their section–section law. Resulting SSIP yields accurate predictions for long-range interactions, such as electrostatics, but becomes inaccurate for short-range van der Waals interactions. This limitation for short-range interactions has been addressed by the Section-to-Beam Interaction Potential (SBIP) approach introduced by \citet{grill2024asymptotically}. More recently, improved SSIP formulations that resolve short-range deficiencies have been developed by \citet{borkovic2024novel, borkovic2026efficient} for planar beam deformations. In addition, \citet{borkovic2024analytical} derived a general and exact interaction law for spatial beam--half-space contact, valid for both short-range and longe-range interactions.
While models like the CGC and SSIP provide an efficient and accurate representation of normal interactions for continuous interfaces, they inherently rely on spatial averaging of the underlying atomic interactions. As a consequence, they are unable to capture the discrete variation of interfacial energies at atomic-scale, which becomes particularly important in crystalline vdW interfaces. In such systems, relative tangential displacements between lattices can lead to significant variations in adhesion, friction, and superlubric behavior. Therefore, to accurately describe these phenomena, it is necessary to adopt models that explicitly account for atomic-scale periodicity and registry effects, which is discussed in the following subsection.

\subsection{Discrete interfaces }\label{sec_2_2}
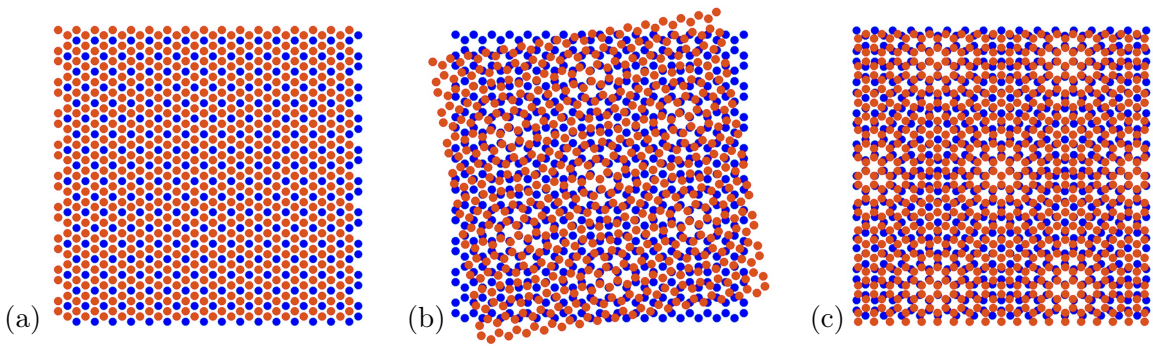
\begin{figure}[H]
\begin{center} \unitlength1cm
\begin{picture}(0,4.5)
\put(-7.2,0){\includegraphics[height=40mm]{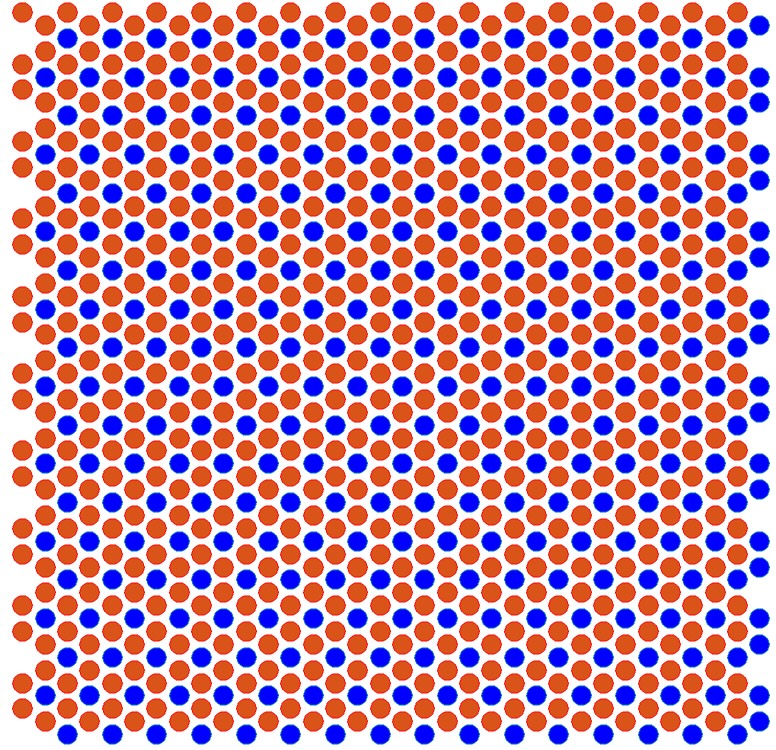}}
\put(-2.3,-0.25){\includegraphics[height=45mm]{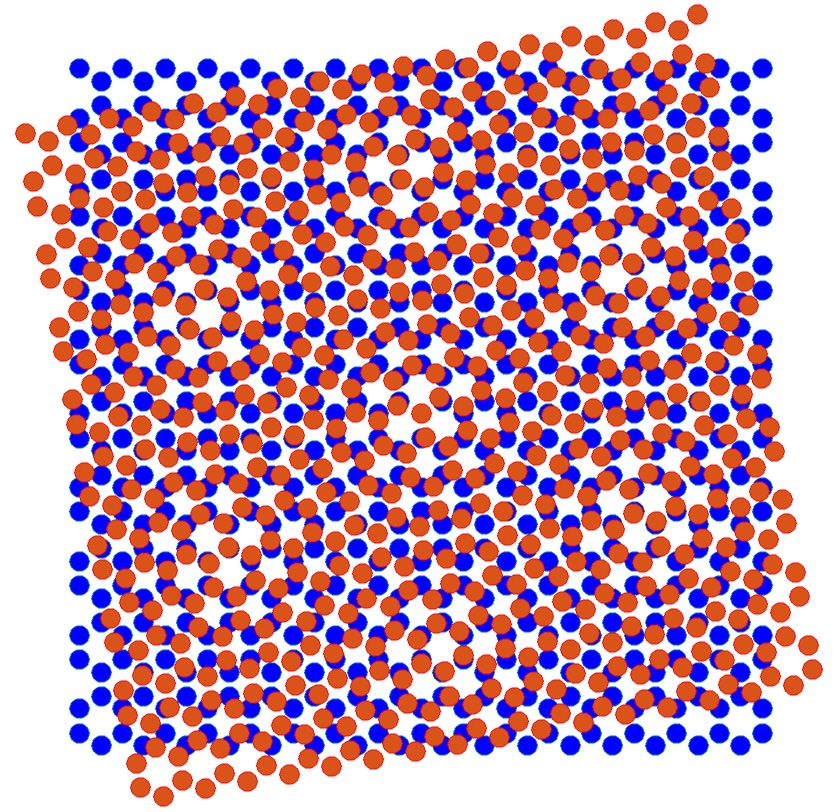}}
\put(3.3,0){\includegraphics[height=40mm]{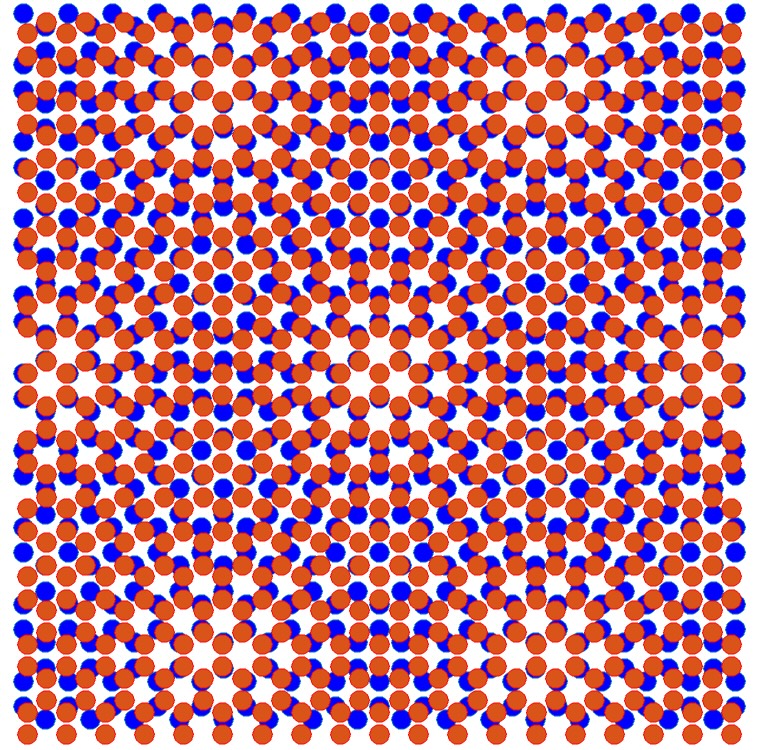}}
\put(-7.8,0){(a)}
\put(-2.5,0){(b)}
\put(2.8,0){(c)}
\end{picture}
\caption{Schematics of different 2D material interfaces. Although not shown, the upper (red) layer is defined by primitive lattice vectors $\bA_1$ and $\bA_2$, and the lower (blue) layer by $\boldsymbol{B}_1$ and $\bB_2$. (a) Homointerface ($\bA_i = \bB_i$): locally and globally commensurate. (b,c) Heterointerface ($\bA_i \neq \bB_i$): locally incommensurate yet globally commensurate, leading to Moiré superlattice (MSL) formation.} 
\label{Commensurate and incommensurate}
\end{center}
\end{figure}

Owing to the inherently periodic but discrete atomic arrangement in crystalline solids, their interfaces become discrete interfaces. For crystalline interfaces, evaluation of the contact energy $\Pi_{\text{c}}$ using Eq.~\eqref{double integral contact energy} inherently smooths out the tangential energy variations associated with the discrete atomic structure of the surface. In the following, the term \enquote{homointerface} is used to describe the stacking of layers that are perfectly aligned (Fig.~\ref{Commensurate and incommensurate}a), while \enquote{heterointerface} refers to the stacking of layers exhibiting angular misalignment or lattice mismatch (Fig.~\ref{Commensurate and incommensurate}b and \ref{Commensurate and incommensurate}c). This usage is consistent with what is commonly referred to as a \enquote{homojunction} and \enquote{heterojunction} in the literature \citep{Leven2013_GhBN_heterojunction, song2018robust}, but emphasizes more the mechanical interface nature rather than electronic junction behavior. In what follows, an overview of the formulation of the contact energy for crystalline interfaces is presented.

\subsubsection{ Homointerfaces}\label{sec_2_2_1}
We begin by determining the interaction energy of a single atom, denoted as $\Psi_{\text{atom}}$, with a crystalline interface, considered here to be graphene, though the formulation is applicable to any crystalline surface. For a solid-gas (or crystal-atom) adsorption system, where the $i^{\text{th}}$ gas-phase atom interacts with the $j^{\text{th}}$ atom of the solid, and $r_{ij}$ is the distance between them, the total interaction energy for the $i^{\text{th}}$ gas atom is expressed as the sum of pairwise interactions
\begin{equation}
\ds \Psi_{\text{atom}}(\boldsymbol{r}_i)= \ds \sum_j \phi_{\text{gs}}(r_{ij})~,
\label{pair wise crystalline}
\end{equation}
where $\boldsymbol{r}_i$ is the position of $i^{th}$ gas atom relative to some reference point in the solid (absorbent), and $\phi_{\text{gs}}(r_{ij})$ is the pair-wise interaction between the gas atom and the absorbent atom. This type of interaction force is generally modeled by the 12-6 LJ expression given in Eq.~\eqref{LJ}.

Many researchers have emphasized the importance of deriving an analytical expression for $\Psi_{\text{atom}}$ to avoid numerical approximations and to construct a reliable adsorption field. Crystalline materials are characterized by periodic atomic arrangements defined by symmetry operations. Consequently, analytical descriptions of such structures and expression for scalar field variations of physical properties, such as the adhesion energy, requires the use of reciprocal space. \citet{hove1953evaluation} first proposed an expression for $\Psi_{\text{atom}}$ as a function of the normal distance $g_\mrn$ and the tangential surface vector $\boldsymbol{\tau}$ in the lattice plane utilizing the Fourier series expansion for surface lattices with square symmetry.

The necessity of such an analytical formulation was further highlighted by \citet{novaco1972adsorption}, who studied the discrete and configuration-dependent nature of $\Psi_{\text{atom}}$ in the context of adsorption energy states of isolated helium atoms on graphite basal planes. This approach was further advanced when \citet{steele1973physical} through comparison studies demonstrated that $\Psi_{\text{atom}}$ can be accurately represented by a truncated Fourier series with only a few terms. The approach followed in the works of \citet{steele1973physical} and \citet{CARLOS1980339} is elaborated here for flat and curved homointerfaces. 

\subsubsubsection{Flat homointerfaces}\label{Flat homointerfaces}

\begin{figure}[H]
\begin{center} \unitlength1cm
\begin{picture}(0,5)
\put(-5.5,-0.3){\includegraphics[height=50mm]{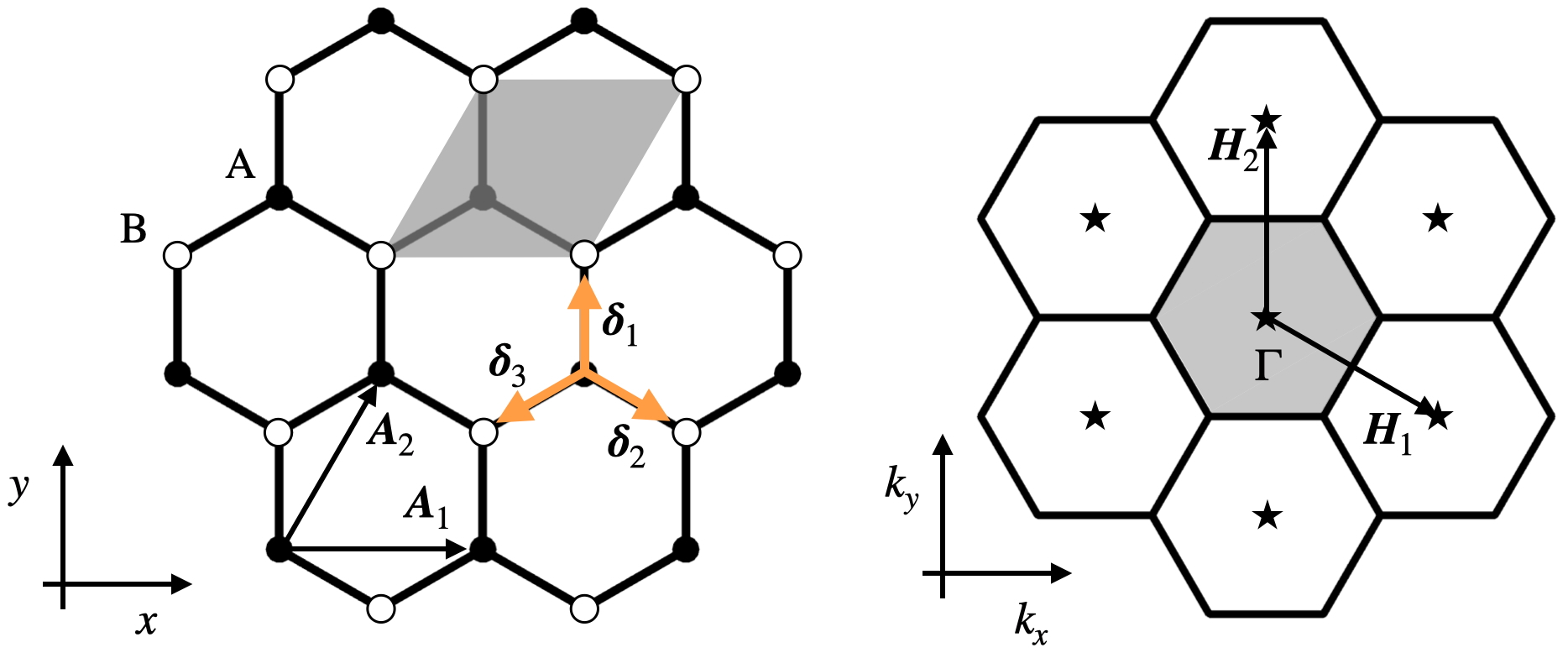}}
\put(-6,-0.5){(a)}
\put(0.75,-0.5){(b)}
\end{picture}
\vspace{5mm}
\caption{Sketches of Bravais and reciprocal lattice of undeformed graphene. (a) Bravais lattice vectors $\bA_1$ and $\bA_2$ along with its unit cell (shaded). The relative positions of the sublattice-B atoms with respect to the sublattice-A atoms are given by the vectors $\boldsymbol{\delta}_i$ ($i = 1, 2, 3$). (b) First harmonics reciprocal space wave vectors {$\bH_1$, $\bH_2$}, where the star denotes the corresponding reciprocal lattice points, and the shaded region represents the Brillouin zone with its center $\Gamma$.} 
\label{Graphene lattice}
\end{center}
\end{figure}

Consider a two-dimensional lattice such as graphene. A translation $\boldsymbol{l}$ along the crystal surface is defined as
\begin{equation}
\ds \bl= \ds l_1\bA_1+l_2\bA_2~,
    \label{general lattice vector}
\end{equation}
where $l_i$ $\in$ $\mathbb{Z}$ and $\bA_i$ are the primitive lattice vectors as shown in Fig.~\ref{Graphene lattice}a. The corresponding primitive reciprocal lattice vectors $\bH_j$ can be seen in Fig.~\ref{Graphene lattice}b. From translation symmetry then follows
\begin{equation}
   \ds   \Psi_{\text{atom}}(\boldsymbol{\tau})= \Psi_{\text{atom}}(\boldsymbol{\tau}+ \boldsymbol{l})\,.
     \label{crystal energy variation with translation}
 \end{equation}

The position vector $\boldsymbol{r}$ is decomposed into a normal gap component $g_\mrn$ and into a tangential gap component $\boldsymbol{\tau}$. The atom/surface potential, originally expressed as a pairwise summation in Eq.~\eqref{pair wise crystalline}, can be expressed in terms of its truncated Fourier components (harmonics) due to the substrate's two-dimensional periodicity \citep{steele1973physical, Carlos1979, CARLOS1980339}. For the graphite structure, \citet{steele1973physical, CHOW1976225} and \citet{jung2014ab} reported that the contribution from the Fourier terms other than the first harmonics is very small compared to the rest of the terms and hence can be neglected. The potential can thus be written as 
\begin{equation}
\ds \Psi_{\text{atom}}(\boldsymbol{r})=\Psi_0(g_\mrn) + \sum_{j=1}^3 \Psi_{\mrs}(g_\mrn) \,e^{i\bH_j \cdot \boldsymbol{\tau}}~,
\label{energy as fourier sum}
\end{equation}
where $\Psi_\mrs$ is the Fourier transform of $\phi_{\text{gs}}$. From Eqs.~\eqref{crystal energy variation with translation} and \eqref{energy as fourier sum}, the relation between the primitive Bravais and reciprocal lattice vectors is given by $\bA_i \cdot \bH_j = 2\pi \delta_{ij}$. Here, $\delta_{ij}$ denotes the Kronecker-delta function, with $\delta_{ij}$ =1 for $i=j$ and 0 otherwise. Eq.~\eqref{energy as fourier sum} is further expanded along with the form of $\Psi_0$ as

\begin{equation}
\begin{aligned}
\Psi_{\text{atom}}(\boldsymbol{r})
&= \frac{\pi \epsilon_{\text{gs}}}{A_s}\,
q\!\left(\frac{1}{5}\frac{r_{\text{gs}}^{12}}{g_\mathrm{n}^{10}}
        - \frac{r_{\text{gs}}^6}{g_\mathrm{n}^{4}}\right) \\[6pt]
&\;\;+\sum_{j=1}^3\sum_{k=1}^q 
   \exp(i\,\boldsymbol{H}_j\!\cdot\! \boldsymbol{\tau}_k)\,
   \frac{\pi \epsilon_{\text{gs}}}{A_s}  
   \left[
      \frac{r_{\text{gs}}^{12}}{60}
      \left(\frac{H}{2g_\mathrm{n}}\right)^{\!5} 
      K_5(H g_\mathrm{n})
      - 2 r_{\text{gs}}^{6}
      \left(\frac{H}{2g_\mathrm{n}}\right)^{\!2} 
      K_2(H g_\mathrm{n})
   \right].
\end{aligned}
\label{expanded crystal fourier sum energy}
\end{equation}

where $A_s$, $q$, $H$, and $\boldsymbol{\tau}_k$ are the area, total number of atoms, magnitude of the reciprocal lattice vector, and the tangential surface vector of atoms of the unit lattice cell, respectively. $K_2$ and $K_5$ are the modified Bessel function of the second kind, while $\epsilon_{\text{gs}}$ and $r_{\text{gs}}$ are the LJ potential model parameters that characterize the interaction strength and equilibrium separation, respectively, between the gas atom and the surface. From now onwards subscript \enquote{atom} is skipped in $\Psi_{\text{atom}}$.

Utilizing the derivation provided in Appendix \ref{Appendix A}, and decomposing the planar translational vector $\boldsymbol{\tau}$ into components $g_\mra$ and $g_\mrz$ along the armchair and zigzag directions, denoted by $\be_\mra$ and $\be_\mrz$, respectively, Eq.~\eqref{energy as fourier sum} can be further simplified into

\begin{equation}
  \ds  \Psi(g_\mra, g_\mrz,g_\mrn)= \Psi_0(g_\mrn)+\Psi_1(g_\mrn)\Psi_\mrt(g_\mra, g_\mrz)~,
    \label{energy magnitude and variation form}
\end{equation}
with
\begin{equation}
\ds \Psi_\mrt(g_\mrz, g_\mra)=  4\cos{\left(\frac{H}{2}g_\mra\right)}\cos{\left(\frac{\sqrt{3} H}{2}g_\mrz\right)} + 2\cos{\left( H g_\mra\right)}~.
\label{energy variation}
\end{equation}
Here, $H=4\pi/(3\mathrm{a}_{\text{cc}})$ is the magnitude of the reciprocal lattice vector for graphene, where $\mathrm{a}_{\text{cc}}$ denotes the carbon-carbon bond length in graphene. The energy $\Psi$ can represent both an energy per atom or per unit area, depending on the convention used for the stacking energy. In Eq.~\eqref{energy magnitude and variation form}, $\Psi_0$ denotes the average energy, given by the form appearing in the first term on the right-hand side of Eq.~\eqref{expanded crystal fourier sum energy}. This form of $\Psi_0$ can also be found in several other studies (e.g., \citet{girifalco2000carbon, sauer2006, sauer2008atomistically, lu2007cohesive, zhang2017energy}). The function $\Psi_\mrt$ is the modulation function defined through the structure factor of the triangular lattice, and $\Psi_1$ represents the amplitude of the energy modulations, i.e., the registry-dependent variation of the interlayer interaction that governs frictional resistance. A convenient way to determine $\Psi_1$ is via numerical curve fitting \citep{Xue2022, mokhalingam2024continuum}. The expression from \citet{mokhalingam2024continuum} is 
\begin{equation}
\ds \Psi_1(g_\mrn)=p_{02}\,g_{02}\,\exp\bigg(\!\!-\ds\frac{g_\mrn}{g_{02}}\bigg)~,
\label{psi_1 form}
\end{equation}
where $p_{02}$, and $g_{02}$ are constants that are calibrated from molecular dynamics (MD) simulations. Some authors have conducted experiments and simulations to obtain the numerical value of $\Psi_1$ at equilibrium gap \citep{Dienwiebel2004a, verhoeven2004model, Lebedeva2010, Lebedeva2011}. By taking the gradient of the interaction energy with respect to the in-plane displacement, the traction components along $\be_\mra$ and $\be_\mrz$ directions are obtained as
\begin{subequations}
\begin{align}
\ds t_\mra(g_\mra, g_\mrz) = \,&-\frac{\partial \Psi}{\partial g_\mra}= 2 \Psi_1 H \left( \sin{\bigg(\frac{H}{2}g_\mra\bigg)}\cos{\bigg(\frac{\sqrt{3}H}{2}g_\mrz\bigg)} + \sin{\big(H g_\mra\big)} \right )~, \\
\ds t_\mrz(g_\mrz, g_\mra) = \,& -\frac{\partial\Psi}{\partial g_\mrz} = 2 \sqrt{3} \Psi_1 H \left(\cos{\bigg(\frac{H}{2}g_\mra\bigg)}\sin{\bigg(\frac{\sqrt{3}H}{2}g_\mrz\bigg)} \right )~. 
\end{align}
\label{armchair and zigzag tractions}
\end{subequations}

\subsubsubsection{Curved homointerfaces}\label{curved homointerfaces}
An earlier theoretical study by \citet{damnjanovic1999full} formulated a general potential for describing physical properties of carbon nanotubes (CNT) using the Fourier series representation 
\begin{equation}
  \ds  \Psi_{\text{CNT}}(\boldsymbol{r}) = \ds \sum_{K,M=-\infty}^{\infty}\,\, \alpha_K^M(\rho)\,e^{i\,n'M\,\phi}e^{i\,(2\pi/a_t)\,K\,z}~,
    \label{eq20}
\end{equation}
where $\rho$, $\phi$, and $z$ are the radial, circumferential, and axial components of the position vector $\boldsymbol{r}$, respectively. Further, for CNT with chirality $(n,m)$, $a_t$ is defined as 
\begin{equation}
    \ds a_t := \frac{3\sqrt{(n^2+m^2+nm)}} {nR} \mra_{\text{cc}}~, 
\end{equation}
where $R = 3$ if $\ds (n-m)/3n'$ is an integer and $R = 1$ otherwise. Here, $n'$ is the greatest common divisor of $(n,m)$. The condition on the sum is such that $Mr_t+K$ should be a multiple of $q/n'$. Here, $q$ is defined as 
\begin{equation}
    \ds q:= 2\frac{n^2+nm+m^2}{n'R}\,,  
\end{equation}
and $r_t$ is defined as 
\begin{equation}
\ds r_t:=\frac{q}{n'}\text{Fr} \left[\frac{n'}{qR}\left(3-2\frac{n-m}{n'}\right)  + \frac{n'}{n}\left(\frac{n-m}{n'}\right)^{\varphi(n/n')-1} \right].
\end{equation}
Here, $\text{Fr}[\cdot]$ is the fractional part of the rational number $[\cdot]$ and $\varphi({\cdot})$ is the Euler-Totient function, giving numbers that are prime to $({\cdot})$. Furthermore, the function $\alpha_K^M(\rho)$ represents the Fourier series coefficient; it quantifies the amplitude of the periodic energy modulation associated with the underlying lattice symmetry.  For more details about the physical geometry of CNTs, refer to \citet{dresselhaus1995physics}.

The studies of \citet{yin2011shape, wu2012interaction, wang2015curvature, wang2019curvature, wang2020van} and \citet{wang2021surface} examined the influence of curvature on the interactions between bodies, but assumed homointerface. Using this assumption, \citet{mokhalingam2024continuum} derived an expression for the interaction energy between curved surfaces based on shell theory. In systems involving curved surfaces, such as coaxial double-walled carbon nanotubes (DWCNTs), the interacting surfaces generally possess unequal areas. As a result, the total interaction energy, expressed as $\Pi_{\text{c}} = \int_{S} \Psi \mathrm{d}A$, is not equivalent when evaluated over each surface independently. This necessitates a reformulation of the interaction potential on a common reference surface. Accordingly, Eq.~\eqref{energy magnitude and variation form} is integrated over the reference surface, yielding
\begin{equation}
\ds \Pi_{\text{c}} = \int_{S_0} \Psi \, \text{d}A_0~.
\label{curved surface energy}
\end{equation}
Following classical shell theory, e.g. see \citet{basar1996finite} and \citet{arciniega2007tensor}, the relation between the reference curved area element $\mrd A_0$ and aligned curved area element $\mrd A$ located at a distance $\xi_0$ is given by 
\begin{equation}
\ds dA_0 = S(\xi_0)\,\text{d}A\,, \,\,\,\,\,\,\,\,\,\,\,\,\,\,\,\,\,\,\,\,\,\,\,\,\, S(\xi_0):=1-2H_0\,\xi_0 +\kappa_0\,\xi_0^2\,,
\label{curved surface area}
\end{equation}
here $H_0$ and $\kappa_0$ are the mean and Gaussian curvature of d$A$, respectively. Considering an imaginary mid surface $\Bar{S}$ of the bilayer at initial distance $G_\mathrm{n}/2$ from either graphene layer as the reference surface, Eq.~\eqref{curved surface energy} gives
\begin{equation}
\ds \Pi_{\text{c}} =\int_{S} \Psi_{\text{c}} \, \text{d}A~,
\label{curved surface energy 2}
\end{equation}
where,\\
\begin{equation}
\ds \Psi_{\text{c}} = \Bar{S}\,\Psi~, \,\,\,\,\,\,\,\,\,\text{and}\hspace{1cm} \Bar{S}:=S\left(\frac{G_\mathrm{n}}{2}\right)=1-H_0\,G_\mathrm{n} + \kappa_0\,\frac{G_\mathrm{n}^2}{4}~.
\label{curved energy parameters}
\end{equation}

Although the amplitude of higher-order harmonics naturally decays exponentially with harmonics order \citep{popov2009electromechanical}, it is important to note that the formulation of \citet{damnjanovic1999full} is general and accurate than \eqref{energy in reciprocal lattice vectors expanded} and \eqref{moire energy expression}. Their approach does not impose any assumptions on the contact state (commensurate or incommensurate) nor on the number of harmonics to be retained, in contrast to the representation used in Eq.~\eqref{energy magnitude and variation form}.


\subsubsection{Heterointerfaces}\label{heterojunctions}
\begin{figure}[H]
\begin{center} \unitlength1cm
\begin{picture}(0,4.5)
\put(-6.4,0){\includegraphics[height=40mm]{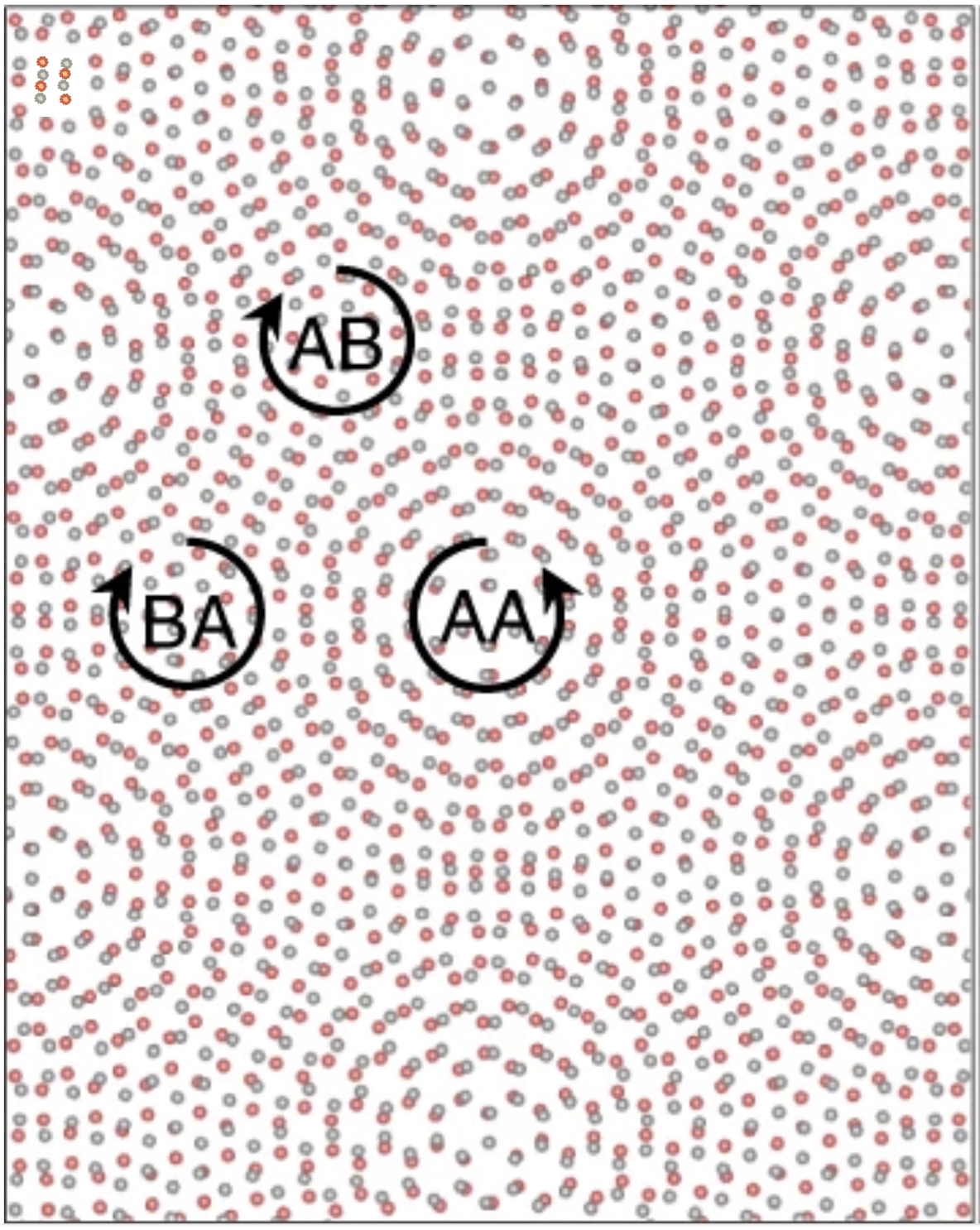}}
\put(-1.7,0)
{\includegraphics[height=40mm]{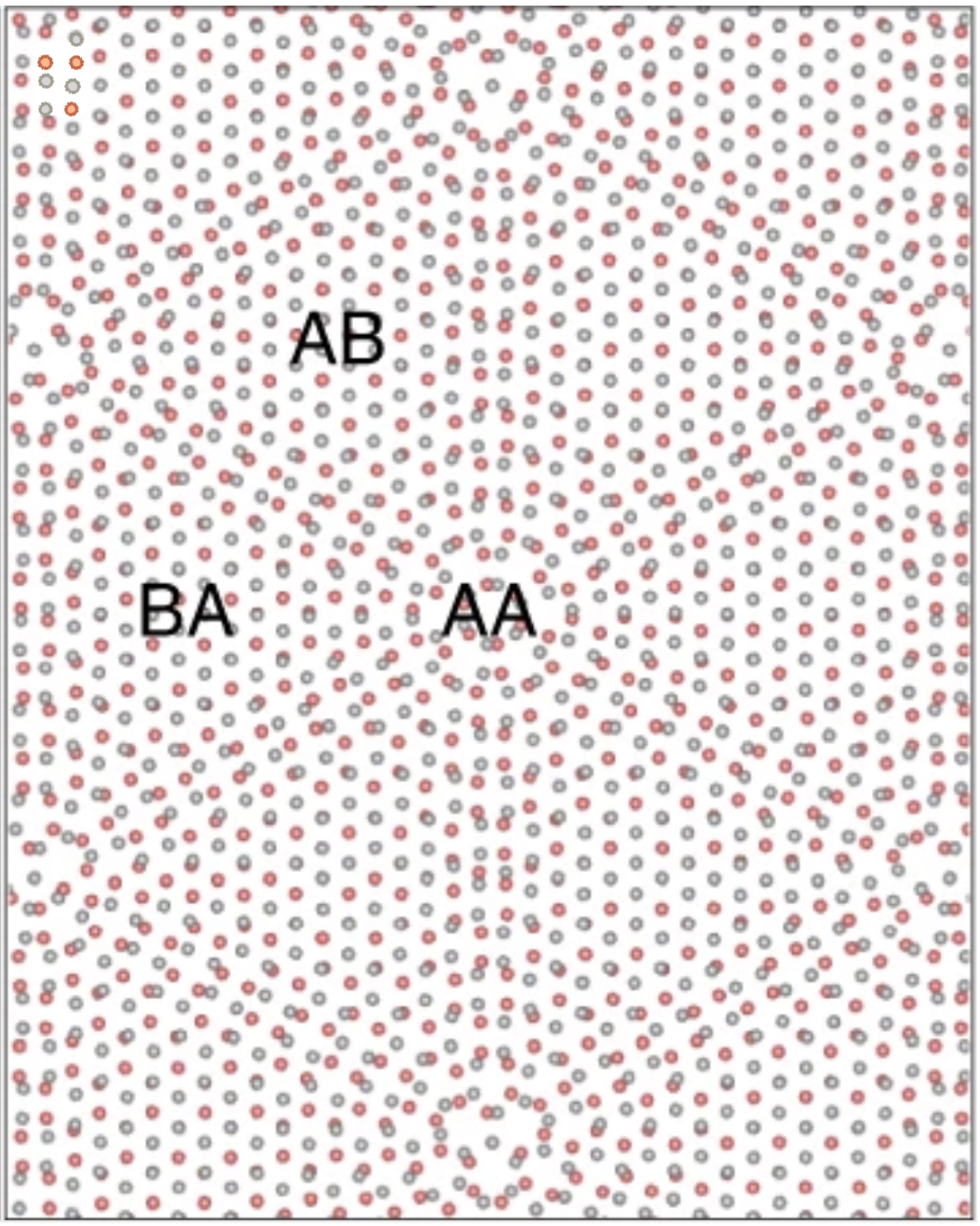}}
\put(2.8,0){\includegraphics[height=40mm]{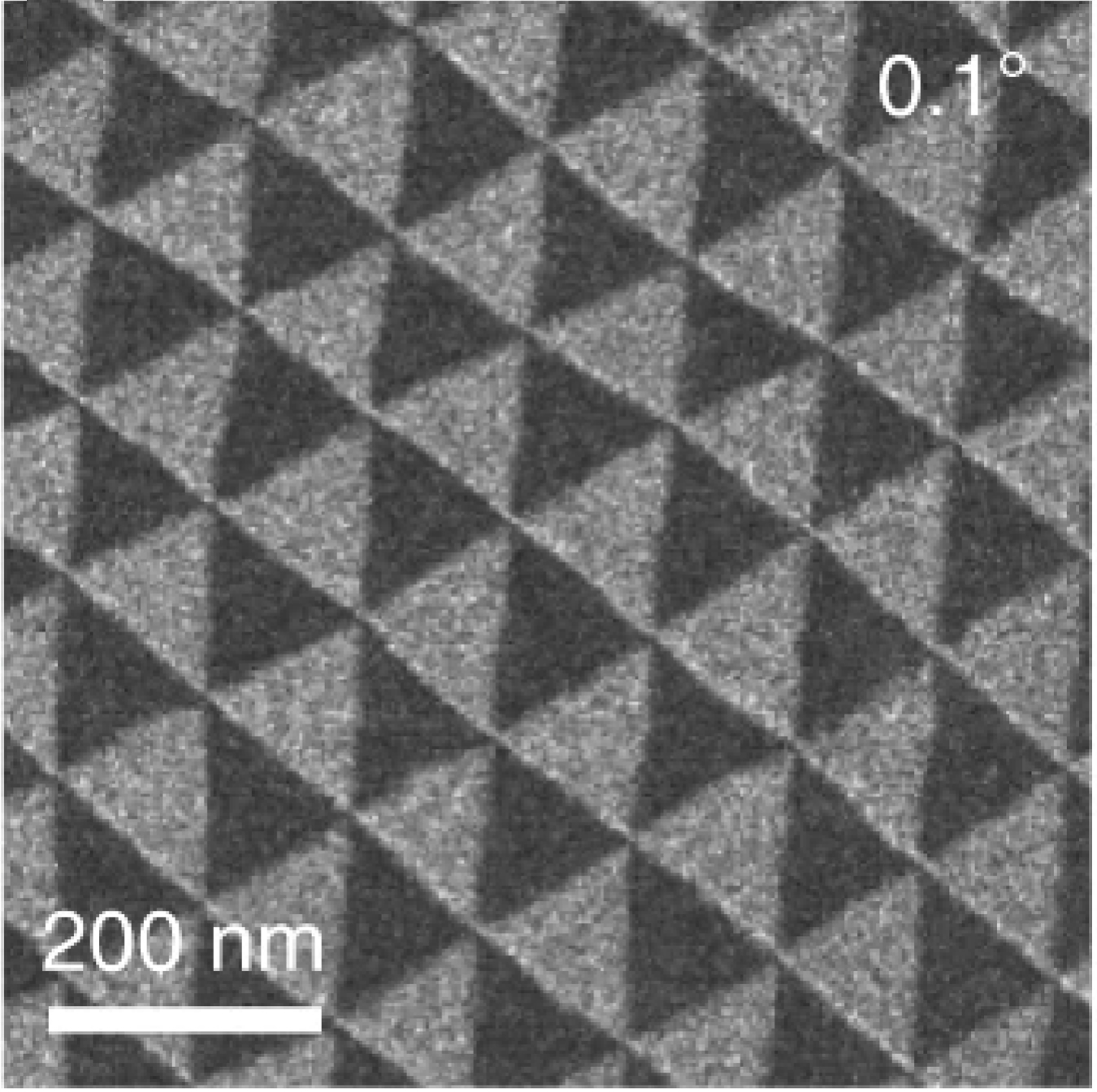}}
\put(-7,0){(a)}
\put(-2.4,0){(b)}
\put(2.1,0){(c)}
\end{picture}
\caption{Atomic-scale reconstruction in tBLG. Schematic of tBLG (a) before atomic reconstruction and (b) after atomic reconstruction. (c) Dark-field transmission electron microscopy (TEM)  picture of tBLG with a twist angle of 0.1$^\circ$. The arrows in (a) illustrate the periodic rotational modulation of the lattice. Reprinted from \citet{yoo2019atomic} with permission from Nature Portfolio.} 
\label{Surface reconstruction experimental}
\end{center}
\end{figure}
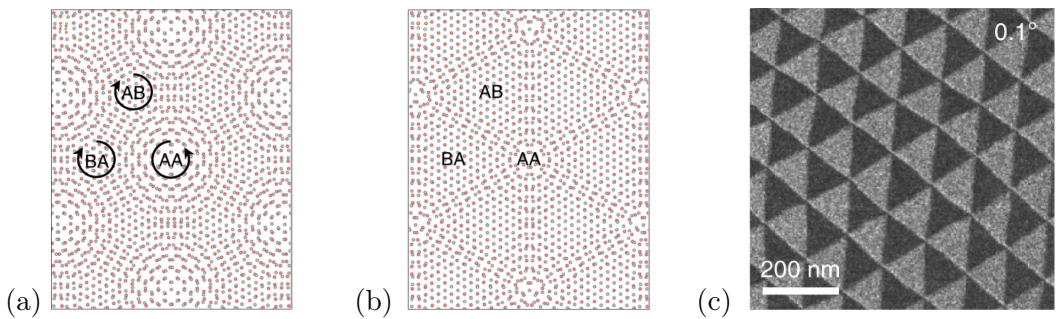

Bilayer systems such as graphene/graphene (Gr/Gr) and graphene/hexagonal boron nitride (Gr/h-BN) have large potential applications in electronic nanodevices \citep{novoselov2005two, giovannetti2007substrate, dean2010boron, britnell2012field}. These systems are often engineered to tailor their electronic characteristics, for example, to produce metamaterial-like behavior with periodic variations in the band structure \citep{zhou2015van}. In such devices, the relative rotation and translation between the two layers have a substantial influence on their properties and performance \citep{rong1993}. These bilayer structures are also refered as van der Waals heterostructures (vdWHs). The resulting interference between their atomic lattices produces long-range periodic patterns known as MSLs; see Fig.~\ref{Commensurate and incommensurate}. These patterns represent a continuous spatial variation in the stacking configuration. However, due to the stacking-dependent adhesion energy and the elasticity of the layers, structural reconstruction occurs to minimize the system's total energy. This leads to the localization of commensurate stacking, such as AA and AB/BA stacking, near Moiré centers, while accumulated strain localizes in narrow incommensurate domain walls (the MSL boundaries) \citep{butz2014dislocations, kim2017evidence, yoo2019atomic, ni2019soliton}. One such example for twisted bilayer graphene (tBLG) is shown in Fig.~\ref{Surface reconstruction experimental}. Figure~\ref{Surface reconstruction experimental}a illustrates the structure before and after relaxation, while Fig.~\ref{Surface reconstruction experimental}b presents its TEM images. The formation of triangular domains upon relaxation is observed. Moiré patterns and domain reconstruction are observed in DWCNTs as well \citep{zhao2022interlayer}.  In these domains, the graphene layers undergo local stretching or compression to accommodate energy variations \citep{lin2013ac, woods2014commensurate, uchida2014atomic,neek2014graphene, lee2016nature, arora2020superconductivity}, with typical strain magnitudes on the order of 0.1$\%$ \citep{van2015relaxation, liang2020effect}. These surface reconstructions can be well described by the classical Frenkel-Kontorova (FK) model \citep{frenkel1939theory}, see Section \ref{FK model}.
 
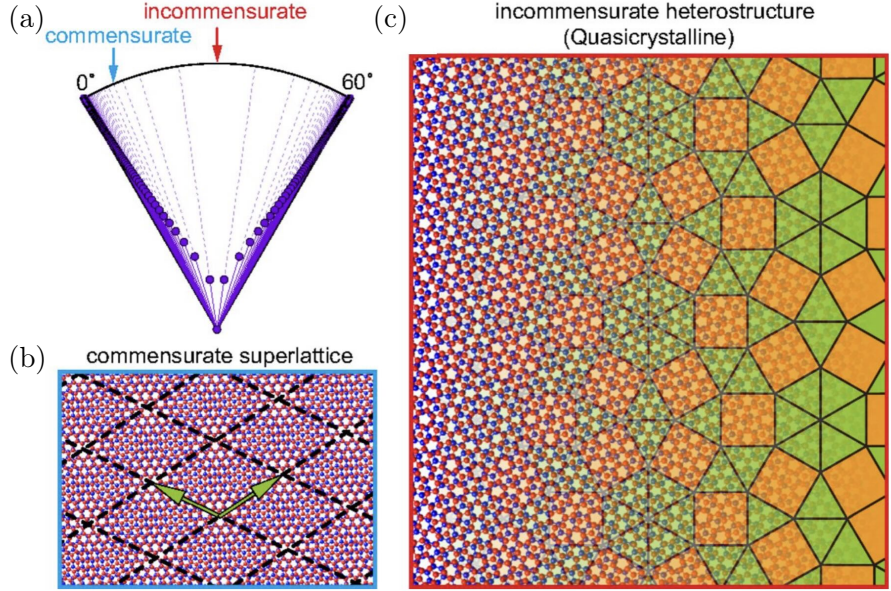
\begin{figure}[H]
\begin{center} \unitlength1cm
\begin{picture}(0,8)
\put(-5.5,0){\includegraphics[height=80mm]{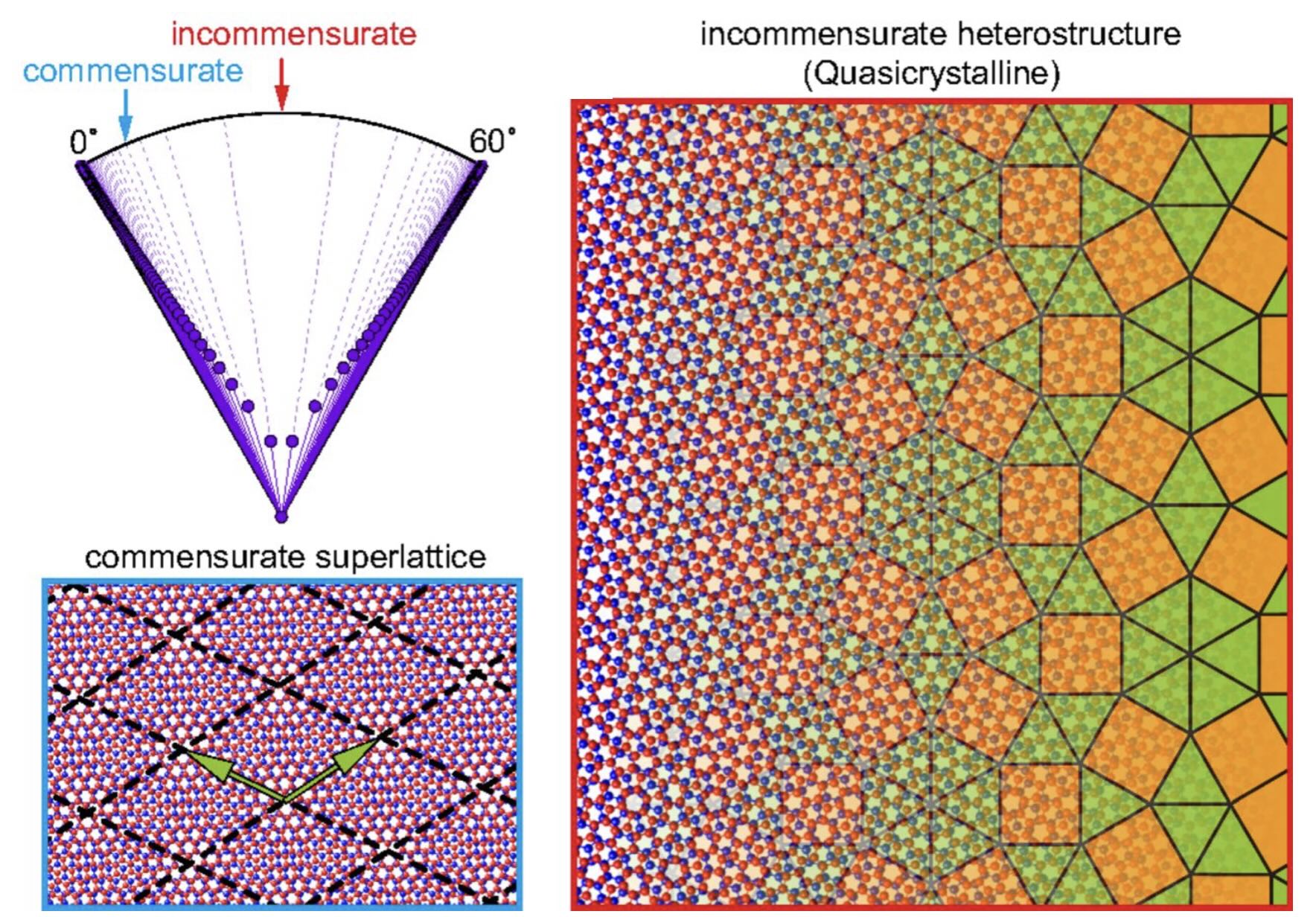}}
\put(-5.8,7.5){(a)}
\put(-5.8,3){(b)}
\put(-1,7.5){(c)}
\end{picture}
\caption{(a) The distribution of all possible twisting angles $\theta^*$. The Moir{\'e} wavelength (distance between two nearest Moir{\'e} centers, shown by green arrows in (b)) for corresponding twisting angles is quantified by the length of the solid line, which is displayed using a logarithmic scale. (b) A tBLG with a twisting angle of 7.34$^{\circ} $. The green arrows correspond to the Moir{\'e} lattice vectors. (c) The incommensurate 30$^{\circ} $ tBLG corresponds to a quasiperiodic structure  (marked by the red arrow in a). Reprinted from \citet{yao2018quasicrystalline} with permission from National Academy of Sciences.}
\label{Incommensurate angles}
\end{center}
\end{figure}

The resulting phase-separated structure, characterized by incommensurate domain walls, plays a key role in modifying the electronic band structure, as shown in the works of \citet{dos2007graphene, giovannetti2007substrate, trambly2010localization, shallcross2010electronic,  zeng2022formation} and \citet{luo2022strain}, and significantly influences the electronic, optical, magnetic, and tribological properties of vdWHs.  To analyze the mechanical and adhesive behavior of such reconstructed configurations, accurate modeling of the atomic registry and stacking variations is essential. Fig.~\ref{Incommensurate angles}a distinguishes between commensurate and incommensurate twist angles; Fig.~\ref{Incommensurate angles}b shows a commensurate structure with well-defined MSLs and their wavelengths indicated by the green arrows; and Fig.~\ref{Incommensurate angles}c presents the incommensurate structure. Identifying special twist angles or lattice mismatches that lead to commensurate stacking -- resulting in periodic MSLs -- is crucial for enabling feasible computational modeling. For tBLG, methods to determine such commensurate angles are proposed in \citet{shallcross2010electronic, zhang2017energy}. Similarly, for vdWHs formed by dissimilar materials, commensurability is achieved only when the lattice constant ratio is a rational number, allowing for a shared superlattice vector. These geometric constraints and angle selection criteria are central to the study of atomic-scale contact and adhesion in layered crystalline systems \citep{shallcross2010electronic, zhang2017energy, yao2018quasicrystalline}.  In \citet{shallcross2010electronic}, it is reported that the commensurate stacking angle is given by 
\begin{equation}
 \ds   \theta^* = \arccos{\frac{3q^2-p^2}{3q^2+p^2}}~,
 \label{incommensurate angle formula} 
\end{equation}
where $p$ and $q$ are any two integers satisfying $0<p\geq q$.

The interaction energy expression $\Psi^{\text{M}}$ for vdWHs is provided in Appendix \ref{Appendix B}. A closely related way of describing the adhesive energy in vdWH systems is through the generalized stacking-fault energy (GSFE), denoted by $V_{\text{GSFE}}$. The GSFE is defined as the energy difference (per unit area) between the ground-state bilayer configuration and a uniformly disregistered configuration, where \enquote{disregistry} denotes the relative in-plane displacement between the two layers \citep{vitek1968intrinsic, zhou2015van}. Similar to various interaction energies (e.g.~Eqs.~\eqref{energy magnitude and variation form} and \eqref{moire energy expression}) presented above, the GSFE is typically expressed using a truncated Fourier expansion of the interlayer energy, which naturally incorporates the lattice symmetry. GSFE landscapes are commonly calibrated from DFT calculations by fitting the energy around selected high-symmetry stacking points \citep{zhou2015van}, similar to the procedure used for homointerfaces (see Eqs.~\eqref{AA} and \eqref{AB and BA}). This approach has been widely used in the literature \citep{zhou2015van, lebedev2016interlayer, lebedeva2016dislocations, dai2016structure, carr2018relaxation, lebedeva2019commensurate, zhu2020modeling, leconte2022relaxation, cazeaux2022relaxation}. 
However, other studies \citep{espanol2017discrete, espanol2018discrete, espanol2023discrete} have adopted a discrete atomistic-to-continuum approach to derive the stacking-fault energy. These works start from atomistic models using LJ or Kolmogorov-Crespi (KC) potentials and then approximate the resulting lattice sums with integrals to obtain a continuum interlayer energy of the Ginzburg-Landau type.

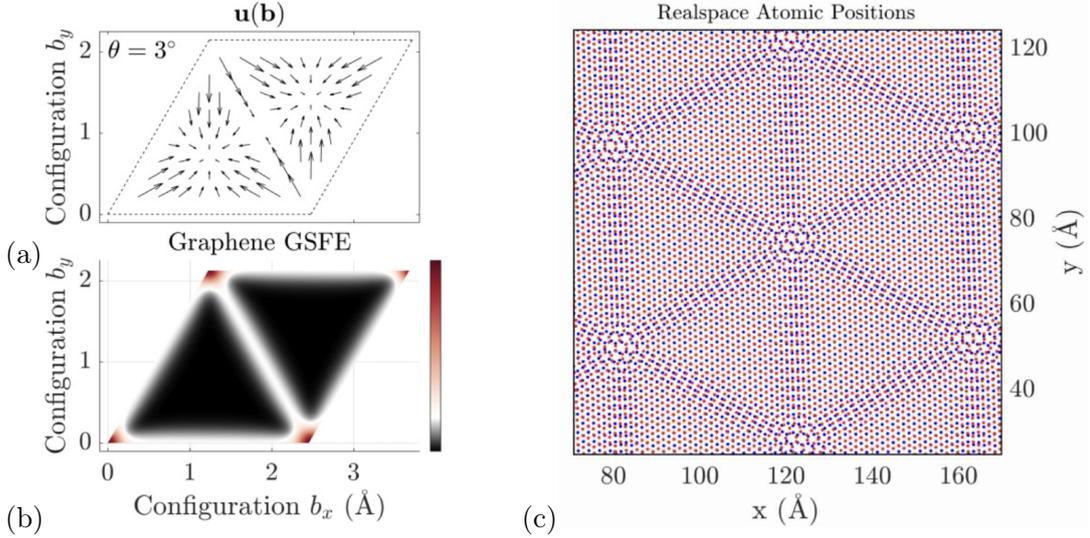
\begin{figure}[H]
\begin{center} \unitlength1cm
\begin{picture}(0,7)
\put(-6.5,0){\includegraphics[height=70mm]{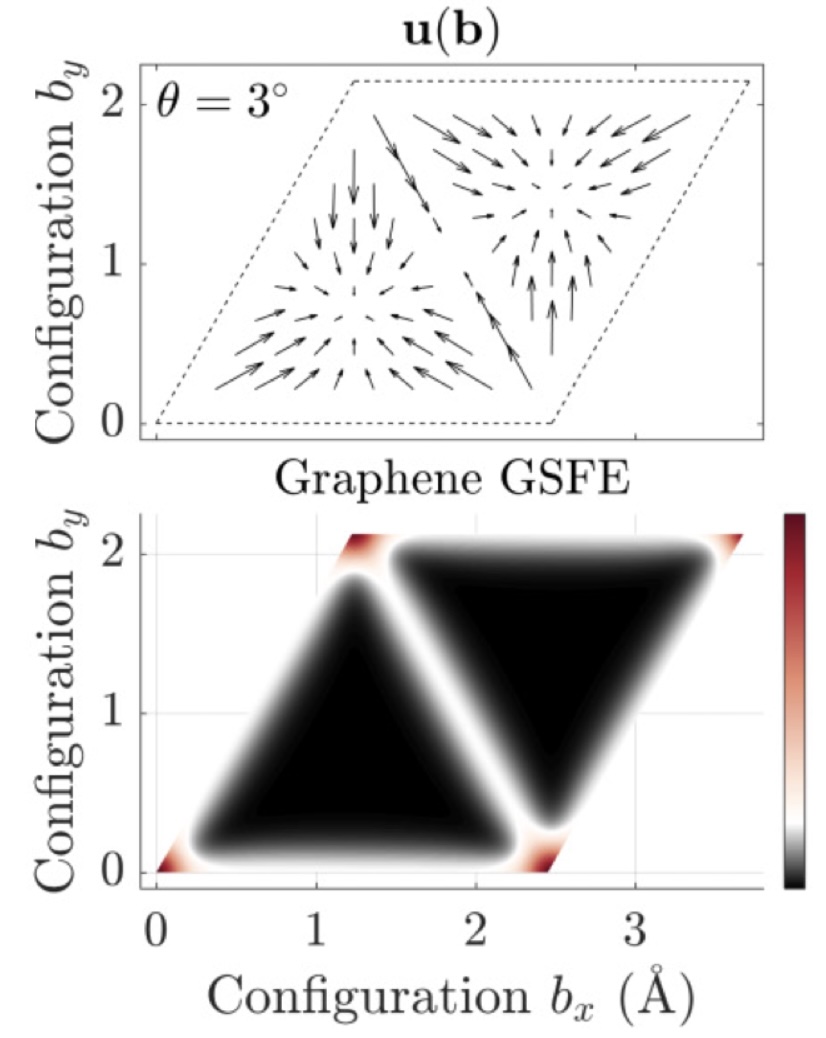}}
\put(0.5,0){\includegraphics[height=70mm]{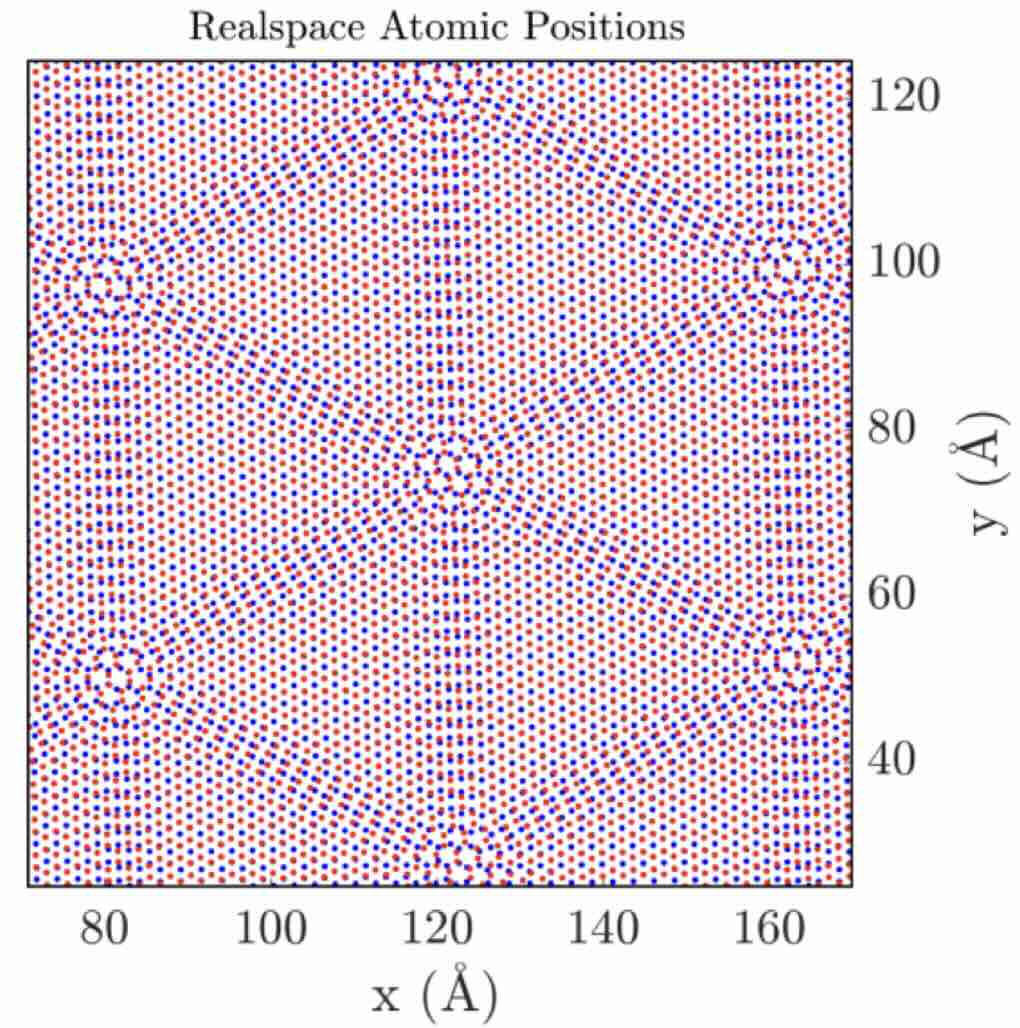}}
\put(-6.8,0){(b)}
\put(-6.8,3.5){(a)}
\put(0,0){(c)}
\end{picture}
\caption{Bilayer graphene relaxation under twisting: (a) displacement field $\bu$, (b) stacking fault energy $V_{\text{GSFE}}$ in one Moir{\'e} unit cell of graphene, and (c) relaxed configuration. Reprinted from \citet{carr2018relaxation} with permission from American Physical Society.}
\label{relax_GSFE}
\end{center}
\end{figure}

 Using the GSFE framework, \citet{carr2018relaxation} predicted the relaxed configurations shown in Fig.~\ref{relax_GSFE}c, which exhibit strong agreement with experimental observations -- for example, those reported by \citet{yoo2019atomic} (see Fig.~\ref{Surface reconstruction experimental}). Upon relaxation the regions with lowest energy stackings expand and highest-energy stackings shrink. The associated displacement field is shown in Fig.~\ref{relax_GSFE}a, and the resulting $V_{\text{GSFE}}$ is presented in Fig.~\ref{relax_GSFE}b. Several MD-based investigations \citep{neek2014graphene, leven2016interlayer, yang2020molecular, Dey2023} have demonstrated the reconstructed geometries of stacked 2D structures as well. 

\section{Tangential contact models}\label{sec_3}
Section \ref{sec_2} presented the contact energy and the corresponding interfacial tractions for normal contact of both continuous and discrete interfaces. This is now followed by tangential contact for both interface types.

\subsection{Continuous interfaces }\label{sec_3_1}
Following experimental observation, macroscopic sliding friction is usually described by the classical \textit{Amontons-Coulomb law}, 
\begin{equation}
\ds F_{\text{t}} = \mu F_{\text{n}}, \hspace{1cm} F_{\text{n}} > 0,
\label{AC law}
\end{equation}
where $\mu$ is the coefficient of friction, relating the tangential (sliding) friction force $F_{\text{t}}$ to the applied normal force $F_{\text{n}}$. This law defines the frictional behavior for many engineering application.  However, over the past century, experimental studies \citep{derjaguin1934molekulartheorie, bowden1939area, schallamach1952load, gao2004frictional, persson2008origin, cohen2011incidence, ruths2011surface, jagota2011adhesion, sahli2018evolution, mergel2018continuum} have indicated that in many cases the tangential force is proportional to the area of contact, thereby violating Amontons’ second law that states their independence. Furthermore, studies such as those of \citet{persson2008origin, cohen2011incidence} and \citet{sahli2018evolution} have shown that friction forces may remain non-zero even under vanishing or tensile normal loads. To address these observations, modified formulations of Amontons’ law have been proposed, incorporating effects such as adhesion and contact area evolution \citep{gao2004frictional, ruths2011surface, jagota2011adhesion, mergel2021contact}. This led to the so-called $\textit{extended Amontons’ law}$
\begin{equation}
\ds F_{\text{t}} = \mu F_{\text{n}} +\tau_0 A_{\text{real}}~,
\label{extended amontons law}
\end{equation}
where $\tau_0$ is the interfacial shear strength and $A_{\text{real}}$ is the total area of contacting microasperities \citep{carpick1997scratching, degrandi2012sliding, sahli2018evolution}. While this model captures a broader range of phenomena, it still falls short in describing cases involving evolving contact areas as appear in bio-adhesive systems and soft materials under sliding \citep{sahli2018evolution}. Experimental and numerical studies have reported tangential forces even under negative (i.e., tensile) normal loads in such systems \citep{autumn2006frictional, drechsler2006biomechanics, zhao2008adhesion, eason2015stress}. \citet{mergel2018continuum} experimentally verified the relation $F_{\text{t}} \propto A_{\text{real}}$.

\begin{figure}[H]
\centering
\includegraphics[scale=0.2]{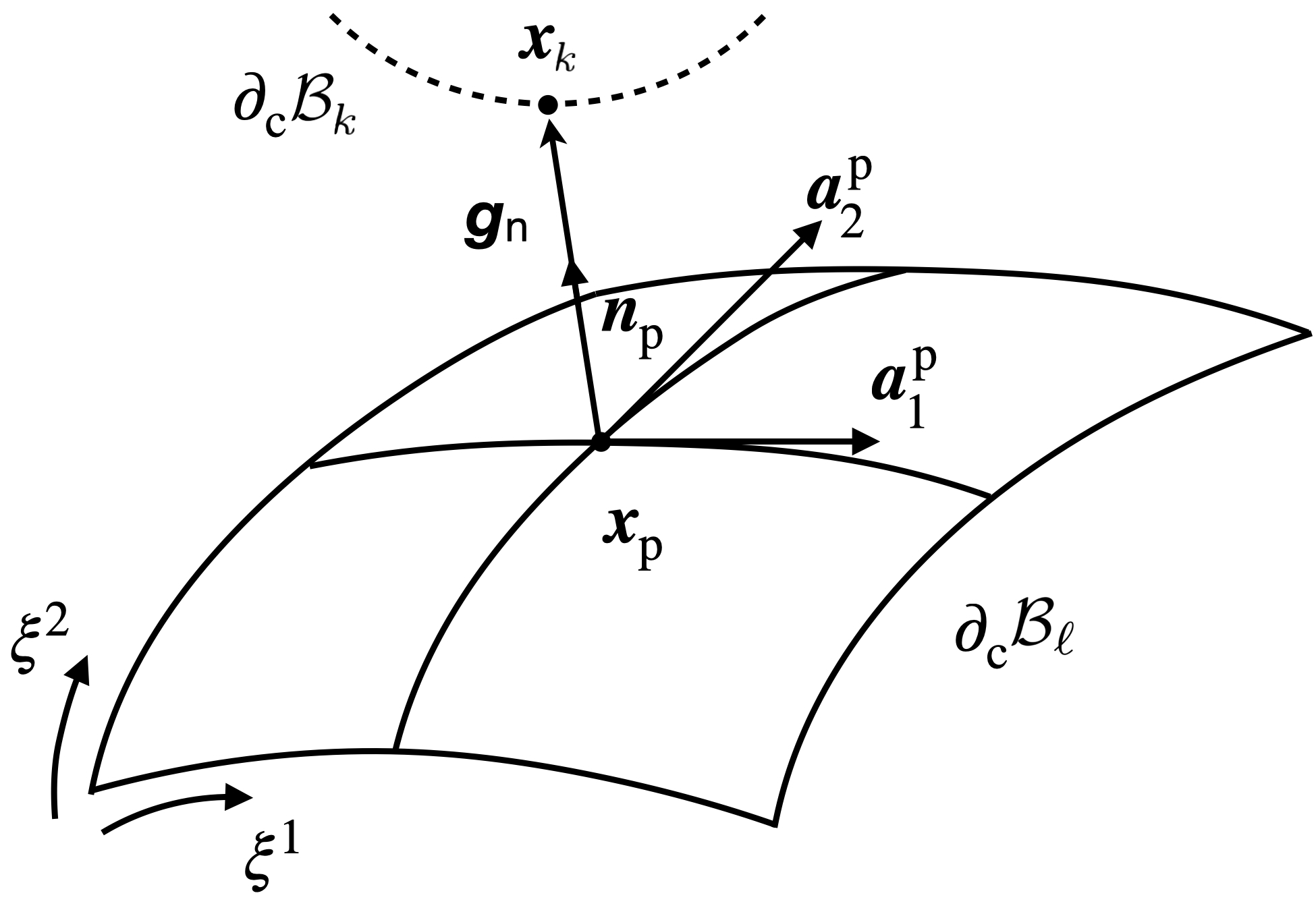} 
\caption{Contact gap vector $\boldsymbol{g}_\text{n}$ between point $\boldsymbol{x}_k$ and contact surface $\partial_\text{c} \mathcal{B}_{\ell}$. Figure adapted from \citet{mergel2021contact}.}
\label{contact kinematics}
\end{figure}

Motivated by these observations and focusing on nanoscale tribological systems, \citet{mergel2018continuum} proposed two continuum friction laws that are briefly presented here. They require a covariant description of the contacting surfaces (see Fig.~\ref{contact kinematics}). To model tangential stick-slip behavior, the tangential gap vector $\bg_{\text{t}}$ (which is equal to in-plane translational component $\boldsymbol{\tau}$ of position vector $\boldsymbol{r}$ in Eq.~\eqref{energy as fourier sum}), is decomposed into a reversible (elastic) part $\Delta \bg_{\text{e}}$ and an irreversible (inelastic) part $\bg_{\text{s}}$ as
\begin{equation}
\ds \bg_{\text{t}} = \Delta \bg_{\text{e}} + \bg_{\text{s}}~.
\end{equation}

This decomposition is central to the algorithmic treatment of friction. Following standard contact formulations, the contact traction $\bt_{{\text{c}},k}$ is decomposed into normal $\bt_{{\text{n}},k}$ and tangential $\bt_{{\text{t}},k}$ components, where the subscript $k$ refers to body $k = 1, 2$ as before. The elastic tangential gap $\Delta \bg_{\text{e}}$ is associated with the tangential stiffness $\epsilon_t \geq 0$, obtained from the interaction potential. This behavior is illustrated in Figs.~\ref{Traction from DI and EA}b and \ref{Traction from DI and EA}d. The description of nanoscale contact thus becomes similar to the penalty formulation in classical computational contact mechanics \citep{sauer2013computational}. Further, given the possibility of stick-slip, the upper bound on the tangential traction is defined as
\begin{equation}
\ds \|\bt_{{\text{t}},k}\|= \text{min} \big(\epsilon_t \|\Delta \bg_{\text{e}}\|, \mu     t_{{\text{n}},k}\big)\,.
\end{equation}

The normal traction at a point $\boldsymbol{x}_k \in \partial_\text{c} \mathcal{B}_k$ due to body $\mathcal{B}_\ell$ is still given by Eq.~\eqref{contact traction_2}. The tangential traction $\bt_\mrt(g_\mrn,\,\bg_\mrt)$ which becomes a function of the normal gap $g_\mrn$ and tangential gap vector $\bg_\mrt$, is defined by the non-negative function $t_{\text{slide}}(g_\mrn)$ in the form
\begin{equation}
\ds \|\bt_\mrt(g_\mrn,\bg_\mrt)\|\,\,\,\bigg\{ \begin{matrix}  \,\,\,\leq t_{\text{slide}} (g_\mrn) \hspace{1cm} \text{during sticking}\,,
\\=t_{\text{slide}} (g_\mrn) \hspace{1cm} \text{during sliding}\,.
 \end{matrix} 
 \label{tangentaial traction}
 \end{equation}
 In their work, \citet{mergel2018continuum} proposed two approaches to determine $t_{\text{slide}}$: Distance-independent (DI) friction and Extended Amontons (EA) law in local form.

\subsubsection{Distance-independent model}\label{sec_3_1_1}

This model considers that the sliding threshold remains constant inside the current contact region, regardless of the distance $g_\mrn$. After defining a suitable cutoff distance $g_{\text{cut}}$, the sliding threshold or resistance is defined as
\begin{equation}
\ds t_{\text{slide}}(g_\mrn)=\bigg\{ \begin{matrix} \tau_{\text{DI}},  \hspace{0.7cm} g_\mrn\leq g_{\text{cut}}~,
\\0,  \hspace{1cm} g_\mrn \ge g_{\text{cut}}~.
 \end{matrix} 
 \label{DI traction}
 \end{equation}
The discontinuity at $g_\mrn$ = $g_{\text{cut}}$ can be regularized in the computational framework by
 \begin{equation}
     \ds t_{\text{slide}}(g_\mrn)= \frac{\tau_{\text{DI}}}{1+e^{k_{\text{DI}}(g_n-g_{\text{cut}})}}~,
     \label{DI traction reqularised}
 \end{equation}
 where $k_{\text{DI}}$ > 0 is a large regularization parameter.
 Fig.~\ref{Traction from DI and EA}a shows the variation of $t_{\text{slide}}$. Since the model does not depend on $g_\mrn$ (except at the cutoff value $g_{\text{cut}}$), it yields $F_\mrt = \tau_0 A_{\text{real}}$. This model assumes a constant sliding resistance, similar to the works of \citet{deng2012adhesion} for realtive sliding of bilayer graphene and \citet{lengiewicz2020finite} for non-adhesive frictional sliding in rubber.
 
\subsubsection{Extended Amontons model }\label{sec_3_1_2}
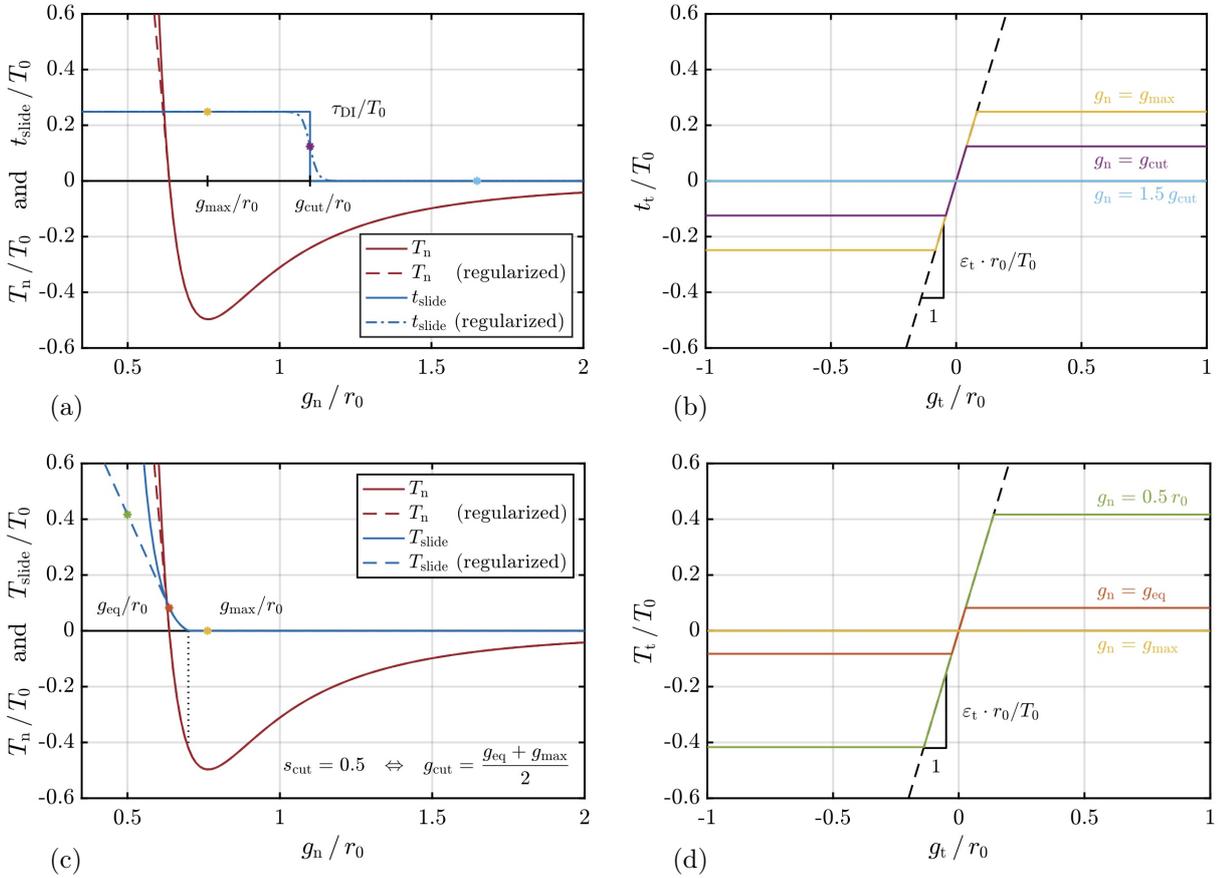
\begin{figure}[H]
\begin{center} \unitlength1cm
\begin{picture}(0,12)
\put(-8.5,6){\includegraphics[height=55mm]{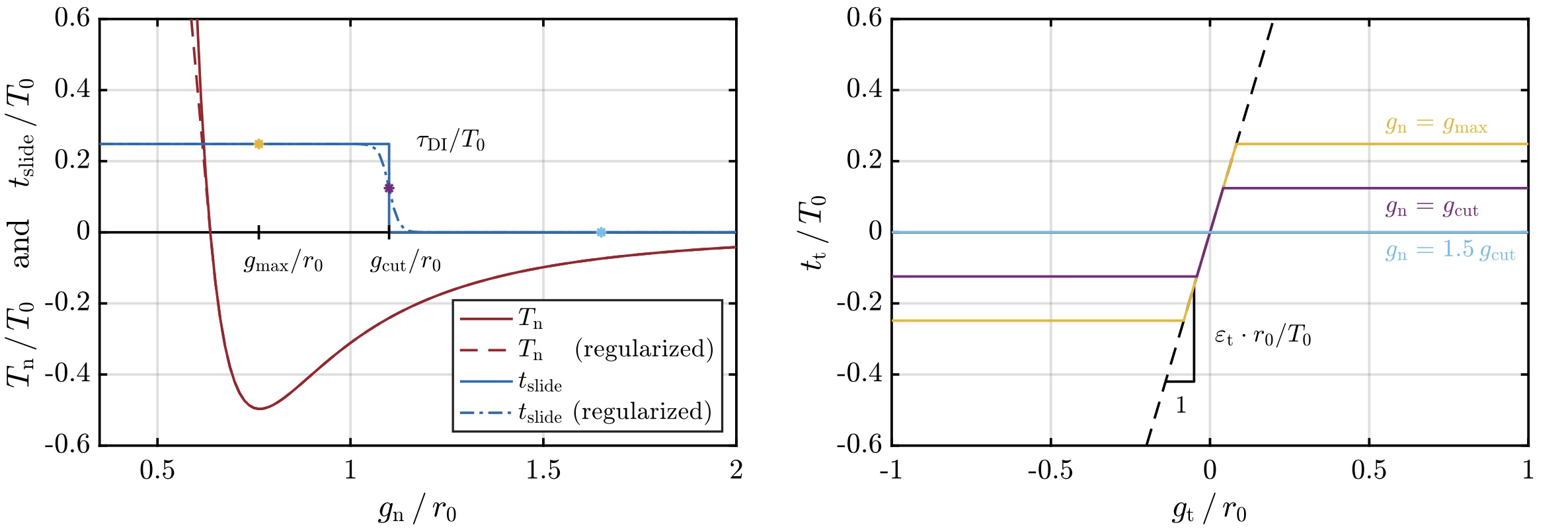}}
\put(-8.7,0){\includegraphics[height=56mm]{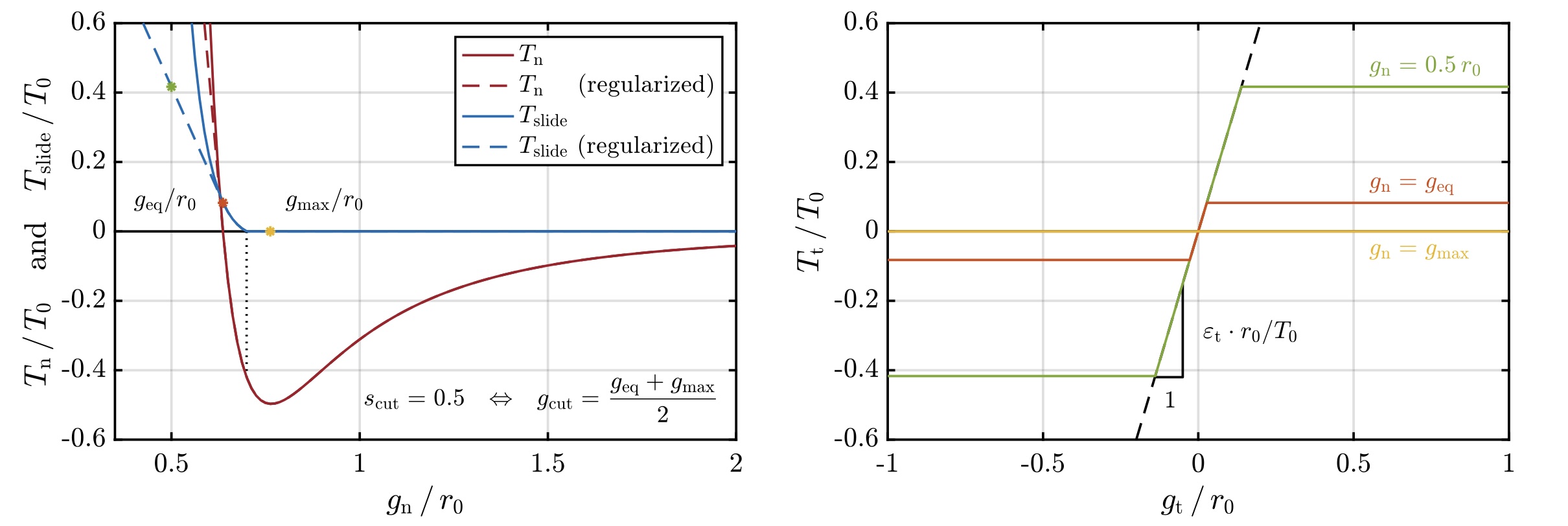}}
\put(-7.9,6){\fontsize{10pt}{20pt}\selectfont (a)}
\put(0.3,6){\fontsize{10pt}{20pt}\selectfont (b)} 
\put(-7.9,0){\fontsize{10pt}{20pt}\selectfont (c)}
\put(0.3,0){\fontsize{10pt}{20pt}\selectfont (d)}
\end{picture}
\caption{Frictional adhesion models DI (a,b) and EA (c,d). (a,c) Normal traction $T_\mrn$ and sliding threshold $t_{\text{slide}}$ and (b,d) corresponding tangential tractions. Here, $T_0$ = $\ds A_\text{H}/(2\pi r_0^3)$ and \enquote{regularized} corresponds to a regularization of the tractions at $g_{\text{cut}}$ and $g_\mrn \leq 0$. Reprinted from \citet{mergel2021contact} with permission from Elsevier.} 
\label{Traction from DI and EA}
\end{center}
\end{figure}

This model provides a continuum formulation of Eq.~\eqref{extended amontons law}. In contrast to the DI model, this model depends directly on the local normal contact tractions $T_\mrn(g_\mrn)$ of the adhesion model, making it dependent on the normal gap/load. To prevent the system from being ill-posed, a cutoff distance $g_{\text{cut}}$ is selected between the equilibrium distance $g_{\text{eq}}$ of $T_\mrn$ and the location $g_{\text{max}}$ of $-T_{\text{max}}$. The cutoff is defined as
\begin{equation}
g_{\text{cut}}=s_{\text{cut}}\,g_{\text{max}}+(1-s_{\text{cut}})g_{\text{eq}},\hspace{2cm}s_{\text{cut}} \, \in \, [0,1]~. \label{cutoff gap for EA traction}
\end{equation}
This is done to ensure that the sliding resistance $T_{\text{slide}}(g_\mrn)$ does not become negative for any normal gap $g_\mrn$. Accordingly, it is defined as
\begin{equation}
\ds T_{\text{slide}}(g_\mrn)=\Bigg\{ \begin{matrix} \frac{\mu_{\text{EA}}}{J_{\text{cl}}}[T_\mrn(g_\mrn)-T_\mrn(g_{\text{cut}})],  \hspace{1cm} g_\mrn \le g_{\text{cut}}\,,
\\0,  \hspace{4.5cm} g_\mrn \geq g_{\text{cut}}\,.
 \end{matrix} 
 \label{EA traction}
 \end{equation}
 By appropriately selecting $g_{\text{cut}} \leq g_{\text{max}}$, the EA model provides a physically consistent frictional response under both compressive and tensile normal tractions. Fig.~\ref{Traction from DI and EA}b shows the variation of $T_{\text{slide}}$. Both DI and EA models are capable of capturing frictional behavior even under zero or tensile normal tractions -- an important distinction from conventional friction models. The resulting tangential tractions can be incorporated in a nonlinear FEM code as discussed in \citet{mergel2021contact}.

Following the discussion of continuous interfaces, we now turn to discrete interfaces and present the corresponding models and modeling approaches in the context of tangential contact.

\subsection{Discrete interfaces}\label{sec_3_2}

For periodic lattice structures such as graphene, the interaction potential $\Psi$ varies spatially; see Eq.~\eqref{energy magnitude and variation form}. This spatial variation gives rise to tangential tractions $\boldsymbol{t}_{\text{t}}$, as shown in Eq.~\eqref{armchair and zigzag tractions}, typically exhibiting the same periodicity as the adhesive energy. As discussed in the previous section, the system attains its minimum energy configuration or ground state by structural reconstruction. This ground-state configuration inherently resists perturbations, so that even in a quasi-static regime a finite external force, commonly referred to as static friction, is required to drive the system from one equilibrium state to another. Key characteristics of this frictional force arises from local restructuring of the lattice into energetically favorable states, induced by the periodic variation of the interaction energy in crystals.

This section reviews various theoretical and computational models that describe static and kinetic friction in sliding contact systems. These include classical atomistic models such as the Prandtl-Tomlinson (PT), Frenkel-Kontorova, and Frenkel-Kontorova-Tomlinson (FKT) models, which are widely used to study qualitative aspects of friction in systems with a low number of degrees of freedom. Additionally, discrete methods including DFT and MD simulations and continuum-based numerical approaches such as the FEM are discussed. 

\subsubsection{Analytical models}\label{sec_3_2_1}
\begin{figure}[H]
\begin{center} \unitlength1cm
\begin{picture}(0,9)
\put(-5.2,0){\includegraphics[height=90mm]{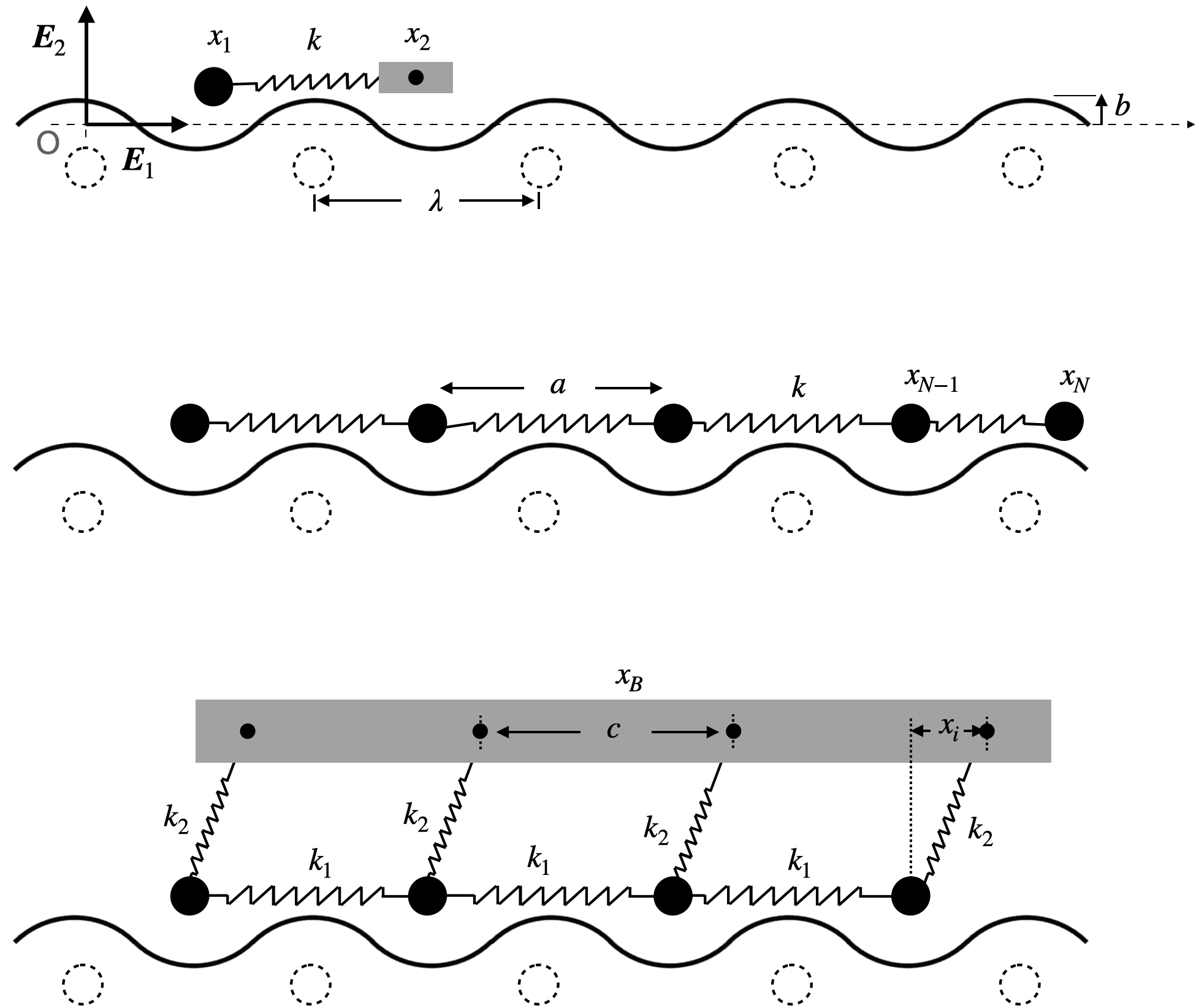}}
\put(-6,7.5){(a)}
\put(-6,4){(b)}
\put(-6,0){(c)}
\end{picture}
\caption{Illustration of the 1D (a) PT, (b) FK, and (c) FKT models. Here $k$, $k_1$ and $k_2$ denote spring constants,  $\lambda$ is the wavelength of the potential, $a$ and $c$ are the lattice constants, and $x_i$, $i$ = 1, 2, ...,\,$N$ mark the position of point masses (filled circles). Dashed circles represents substrate atoms. $x_B$ is the position of the loading device (shown in grey). All three system are usually driven at constant velocity $v$ i.e, $x_2 = vt$, $x_N=vt$, and $x_B=vt$.} 
\label{Theoretical model of friction}
\end{center}
\end{figure}
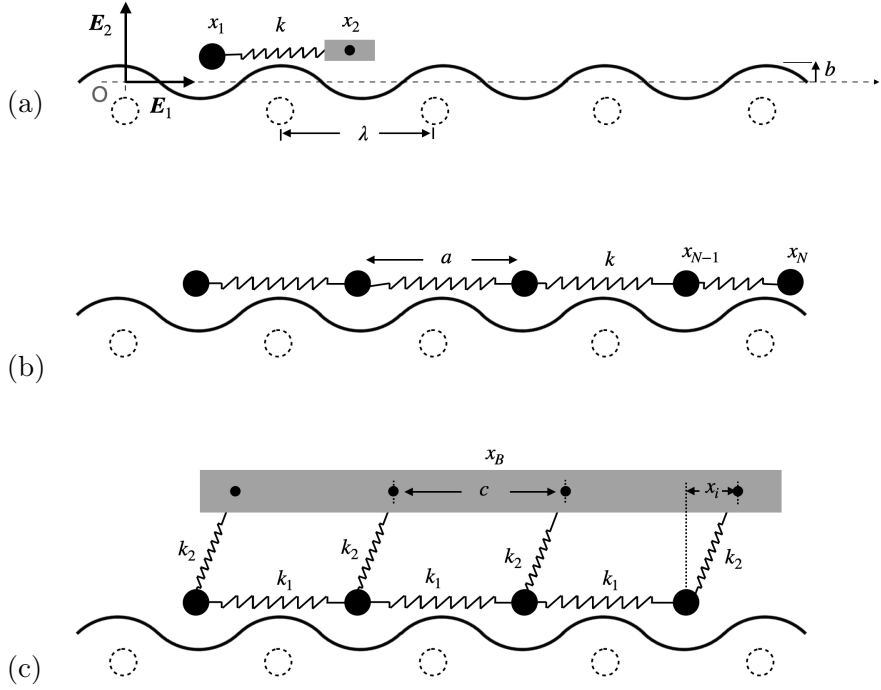

The origin of sticking and sliding friction can be explained analytically through a set of classical analytical models, illustrated in Fig.~\ref{Theoretical model of friction}. These models offer fundamental insights into the interfacial mechanics of systems governed by non-covalent interactions and are widely used to study friction at the atomic scale.

\subsubsubsection{Prandtl-Tomlinson model}

One of the earliest analytical models proposed to explain the origin of atomic-scale friction is the Prandtl-Tomlinson model \citep{prandtl1928, Tomlinson1929}. In this model, the interaction between an atom or point mass $m$ (which may be interpreted as an AFM tip) and the underlying substrate is described by a periodic potential field that mimics the atomic corrugation of the surface. The atom, connected to a spring (representing the AFM cantilever stiffness), is then dragged across this potential landscape with a constant velocity $v$ relative to the
substrate such that $x_2 = vt$. 
The total potential energy of the system, $\Pi_{\text{PT}}$, is described as the sum of the quadratic elastic energy contribution $\Pi_{\text{int}}$ from the deformed spring and the periodic interaction energy $\Pi_{\text{c}}$,
\begin{equation}
  \ds \Pi_{\text{PT}}=\frac{k}{2}(x_2-x_1)^2 -b\cos{\left(\frac{2\pi}{\lambda}x_1\right)}~,
    \label{eq35}
\end{equation}
where $b$ is the interaction potential amplitude, $\lambda$ is the substrate lattice constant, $k$ is the effective spring constant, and $x_2$ is the reference position of the loading device as shown in Fig.~\ref{Theoretical model of friction}a. Equilibrium position of the tip is given by ($-\partial _{x_1}\Pi_{\text{PT}}$ = 0)
\begin{equation}
\label{equillibrium and stability equation a}
 \ds  k(x_2-x_1) - \frac{ 2\pi b}{\lambda} \sin{\frac{2\pi}{\lambda}x_1} = 0 \, ,
\end{equation}
 and the stability criterion is 
 \begin{equation}
 \ds \frac{\partial^2 \Pi_{\text{PT}}}{\partial x_1^2} = k+ \frac{4 \pi^2 b}{\lambda^2} \cos{\frac{2\pi}{\lambda}x_1}  > 0 \, .
\label{equillibrium and stability equation b}
\end{equation}

\begin{figure}[H]
\begin{center} \unitlength1cm
\begin{picture}(0,4)
\put(-7.2,0.4){\includegraphics[height=35mm]{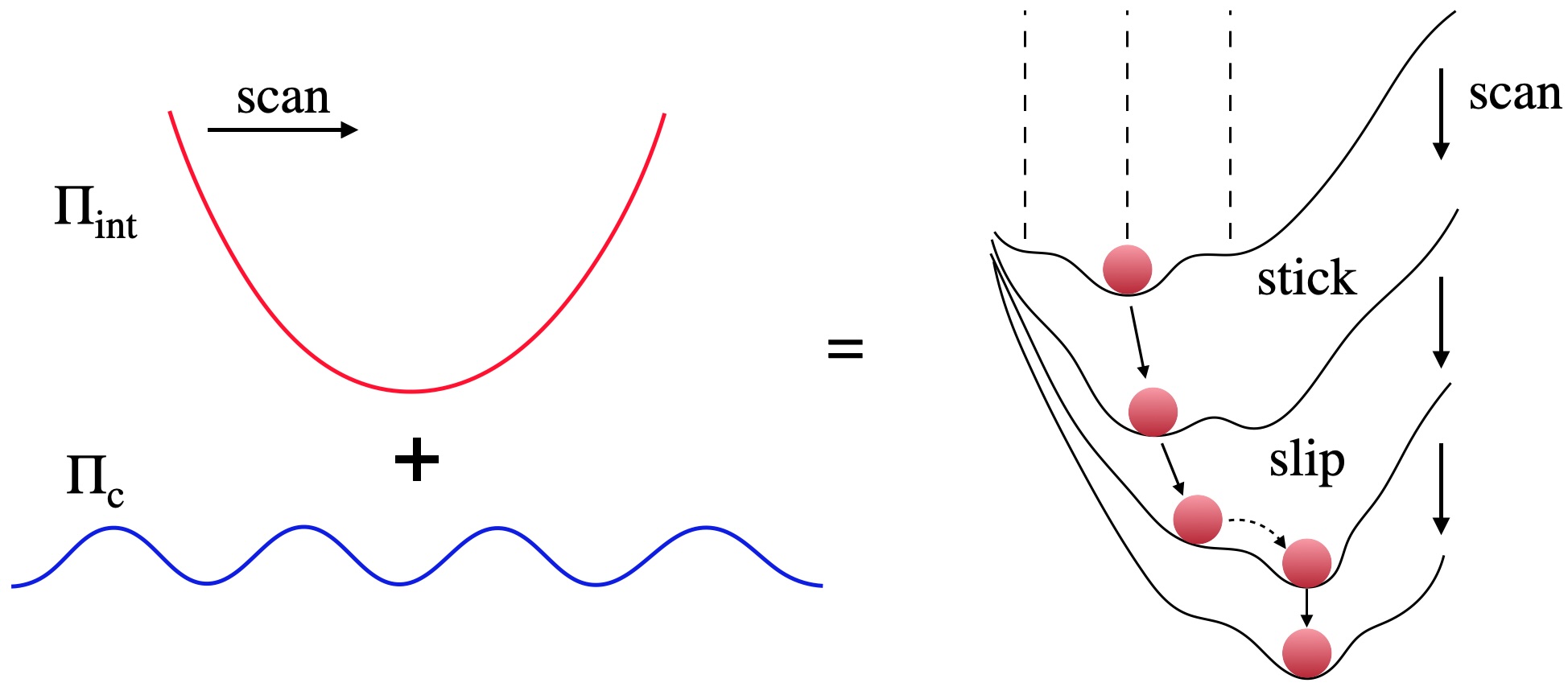}}
\put(1.5,0){\includegraphics[height=40mm]{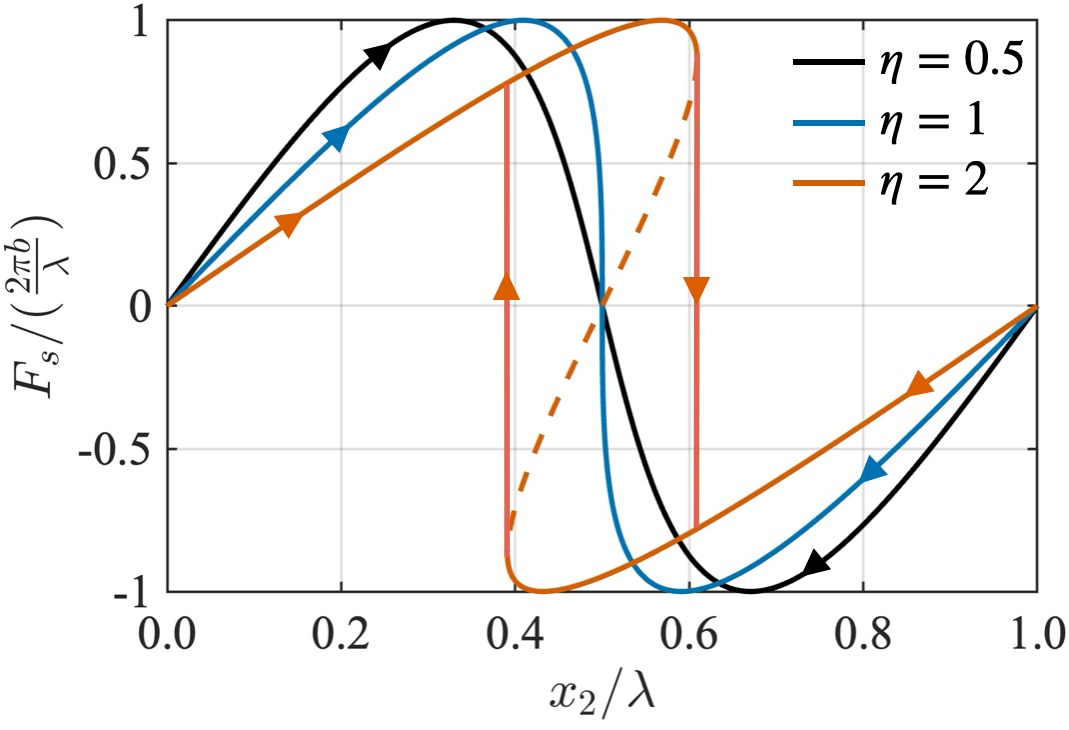}}
\put(-7.8,0){(a)}
\put(1.5,0){(b)}
\end{picture}
\caption{(a) Illustration of the evolution of $\Pi_{\text{PT}}$ during successive forward scan (sliding) displacement for $\eta > 1$. (b) Variation of normalized $F_\mrs$ for different values of $\eta$. Arrows indicate forward and backward sliding. (a) adapted from \citet{medyanik2006predictions} with permission from American Physical Society.} 
\label{stick slip and lateral force and different eta }
\end{center}
\end{figure}

The static friction force is ${F_\mrs} =  k(x_2-x_1)$, and its time average gives the kinetic friction force $\emph{F}_\mrk$.  The maximum static friction force is equal to $2\pi b/\lambda$. Depending on the corrugation amplitude ($2b$) and the elastic energy, the model exhibits different metastable states (see Fig.~\ref{stick slip and lateral force and different eta }a). The resulting solution or the sliding behavior of the system can be characterized by the dimensionless parameter
\begin{equation}
  \ds  \eta=\frac{4\pi^2 \, b}{ \lambda^2\,k}~.
  \label{PT parameter 1D}
\end{equation}

For $\eta<1$, potential $\Pi_{\text{PT}}$ exhibits only one minimum, posing a unique sliding solution and the time-dependent sliding motion is smooth (see Fig.~\ref{stick slip and lateral force and different eta }b). Therefore, the tip moves without any dissipation of energy, and there is no kinetic friction. However, the surface pins the point mass, hence, a finite amount of force is required to de-pin the system, showing that static friction is not equal to zero. On the other hand, for $\eta > 1$, potential $\Pi_{\text{PT}}$ exhibits multiple minima, and the sliding behavior is characterized by stick-slip transitions (see Fig.~\ref{stick slip and lateral force and different eta }b), resulting in both $F_\mrs$ and $F_\mrk$ to be nonzero. 
Also, a clear transition from non-dissipative to dissipative behavior is observed as $\eta>1$ (see Fig.~\ref{stick slip and lateral force and different eta }b).

Having established the athermal PT model, we now extend the framework to finite temperature. At finite temperature, thermal activation enables the tip to overcome non-vanishing energy barriers prior to reaching the athermal instability point, resulting in reduced frictional forces \cite{muser2003statistical}. In the PT model, temperature $T$ is typically incorporated via Langevin dynamics \cite{sang2001thermal},
\begin{equation}
m\ddot{x}_1 + m\gamma\dot{x}_1 + \frac{\partial \Pi_{\mathrm{PT}}}{\partial x_1} = \xi(t)~,
\end{equation}
where $\xi(t)$ is the random thermal force with $\langle \xi(t) \rangle = 0$, and satisfies the fluctuation-dissipation relation
\begin{equation}
\ds \langle \xi(t)\xi(t') \rangle = 2m\gamma k_\textrm{B} T \delta(t-t')~.
\end{equation}
Here, $k_\textrm{B}$ is the Boltzmann constant, $\gamma$ is a damping coefficient representing energy dissipation (e.g., phononic or electronic), and $\langle \cdot \rangle$ denotes ensemble averaging. This formulation captures thermally activated motion, although it typically requires numerical solution over a range of velocities. However, analytical insight can be obtained from a rate-theory perspective. In this framework, slip events correspond to thermally activated transitions between metastable states, governed by a Kramers-type rate \cite{gnecco2000velocity},
\begin{equation}
\frac{\mathrm{d}p(t)}{\mathrm{d}t} = -f_0 \exp\!\left(-\frac{\Delta \Pi_{{\text{PT}}}(t)}{k_\textrm{B} T}\right)p(t)~,
\end{equation}
here $f_0$ is the characteristic attempt frequency and $p(t)$ is the probability that the system remains in a metastable state. Near the instability threshold, the barrier can be approximated as $\Delta \Pi_{{\text{PT}}}  \propto (F_{\text{s}} - F_{\text{k}})^{3/2}$ \cite{sang2001thermal}. This leads to an implicit relation between friction force, temperature, and velocity \cite{riedo2003interaction},
\begin{equation}
\ln\!\left(\frac{v_0}{v}\right) - \frac{1}{2}\left(1 - \frac{F_{\text{k}}}{F_{\text{s}}}\right)
= \frac{1}{\beta k_\textrm{B} T}(F_{\text{s}} - F_{\text{k}})^{3/2}~,
\end{equation}
where $\beta$ is a constant that depends on the shape of the interaction potential $\Pi_{\textrm{c}} $, and the parameter $v_0$ has been defined as
\begin{equation}
v_0 := \frac{2 f_0 \beta k_\textrm{B} T}{3k \sqrt{F_{\text{s}}}}~.
\end{equation}
For low velocities ($v \ll v_0$), this simplifies to
\begin{equation}
F_{\text{k}} = F_{\text{s}} - \left(\beta \,k_\textrm{B} \,T\, \ln\!\frac{v_0}{v}\right)^{2/3}~.
\end{equation}
These results show that temperature promotes thermally activated barrier crossing, smooths the stick-slip transition, and reduces friction, which is also denoted as \textit{thermolubricity} \cite{krylov2005thermally}.


\subsubsubsection{Modified Prandtl-Tomlinson model}

\begin{figure}[H]
\begin{center} \unitlength1cm
\begin{picture}(0,4)
\put(-7.2,0){\includegraphics[height=40mm]{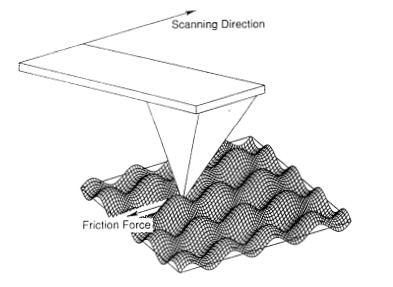}}
\put(0,0){\includegraphics[height=40mm]{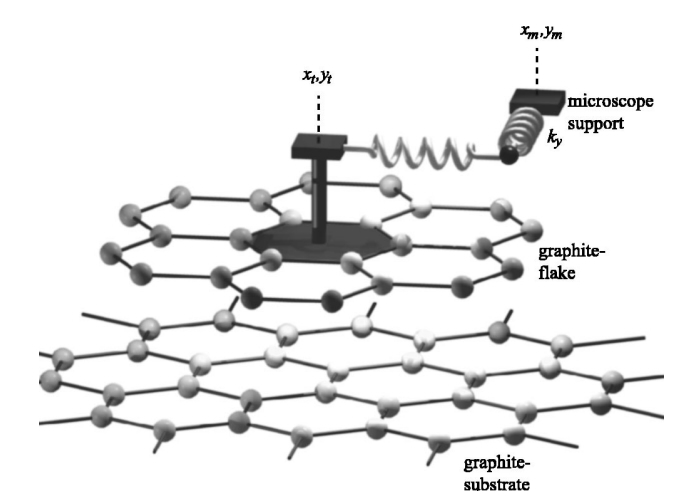}}
\put(-7.8,0){(a)}
\put(-0.6,0){(b)}
\end{picture}
\caption{(a) Schematic representation of AFM measurement of friction. (b) Modified PT model capturing multi-atom contact, i.e., for a graphene flake over a graphite substrate. Reprinted from \citet{verhoeven2004model} with permission from American Physical Society.} 
\label{PT 2D}
\end{center}
\end{figure}

 The PT model can be extended to 2D or 3D for the frictional study of flat surfaces \citep{gyalog1995mechanism, gyalog1997atomic, prioli2003influence, fusco2004power, PhysRevB.71.045413, steiner2010angular, almeida2016giant}; see Fig.~\ref{PT 2D}. 
For a 2D Bravais lattice, Eq.~\eqref{energy magnitude and variation form} describes how the interaction energy $\Psi(\br)$ varies in space, and the system's total potential thus becomes
\begin{equation}
   \ds  \Pi_{\text{M-PT}}=  \frac{1}{2}(\bR-\br) \cdot \bK(\bR-\br)+ \Psi(\br) ,
\end{equation}
where $\bK$ is the stiffness matrix capturing the elasticity of the AFM cantilever, $\bR$ and $\br$ are the given support position and position of the tip, respectively. The tangential force follows from $ \nabla_{\br}\Pi_{\text{M-PT}}$ = $\mathbf{0}$, giving
\begin{equation}
 \ds \bF= \bK(\bR-\br)\quad \text{and} \quad\bF= - \nabla_{\br} \Psi~.
 \label{deformation, potential}
\end{equation}
Equation~\eqref{deformation, potential} establishes a relation between $\bR$ and $\br$,
 \begin{equation}
 \ds \bR(\br)= \bK^{-1}\,\nabla_{\br}\Psi + \br~.
\end{equation}
The function $\bR(\br)$ is non-invertible when the contact interactions are sufficiently strong, as it consists of both a periodic and a linear component. As a result, several tip positions are possible for the given support position, and the mapping $\bR$   $ \rightarrow$ $\br$ contains folds. The tip can no longer follow the edge of a fold when the support is moved across it; instead, an irreversible jump to a new stable location occurs. These jumps cause hysteresis to develop in the quasi-static limit and generate sawtooth patterns of the friction force, similar to the 1D PT model above; see Fig.~\ref{stick slip and lateral force and different eta }b. The modified PT model can mimic the observed behavior of a graphene flake sliding over graphite \citep{verhoeven2004model} as shown in Fig.~\ref{PT 2D}b. The study shows friction anisotropy and 60° periodicity due to the 2D nature of the energy landscape.\\
Other generalizations of the PT model explain the velocity dependence of friction \citep{PhysRevLett.91.084502, PhysRevLett.93.230802, PhysRevE.71.065101, ptak2019velocity} and modulations of the normal load \citep{socoliuc2006atomic, lantz2009dynamic}. The study of \citet{andersson2020understanding} considered the PT model with one extra degree of freedom to capture the internal dynamics of the sheet. \citet{huang2022origin} proposed a PT model that accounts for the deformation caused by the appearance of Moir{\'e} patterns in heterointerfaces of Gr/h-BN. In the study by \citet{wang2024effective}, a PT-like parameter $\eta$ is also proposed for the two-dimensional sliding.

The PT only contains only a single atom or a small rigid atomic group and therefore does not account for deformation of the contacting solids. This changes in the following two models.

\subsubsubsection{Frenkel-Kontorova model}\label{FK model} 
The FK model \citep{kontorova1938theory} was originally introduced to describe the dislocation in solids and has found application in describing the sliding of crystalline interfaces, see \citet{braun2004frenkel}. This model is an extension of the PT model, in which several atoms are elastically coupled to the loading device, as shown in Fig.~\ref{Theoretical model of friction}b. This provides a more accurate representation than the PT model and can even model simple defects such as kinks (topological solitons).  
Following the works of \citet{peyrard1983critical, strunz1995sliding} and  \citet{haibin2002static}, the potential energy of the FK model with $N$ atoms, connected by springs with stiffness $k$ is 
\begin{equation}
\ds \Pi_{\text{FK}}=\frac{1}{2}\sum_{i=1}^{N}k(x_{i+1}-x_{i}-a)^2 -b\sum_{i=1}^{N}\cos{\bigg(\frac{2\pi}{\lambda}x_i\bigg)}~,
\end{equation}
with periodic boundary conditions imposed such that 
\begin{equation}
    x_{N+1} = x_1 + L \quad \text{and} \quad x_{0} = x_N -L~,
    \label{FK/FKT periodic BC}
\end{equation} 
where $L$ = $Na$ = $M\lambda$ for $N,M \in \mathbb{Z}$. The force on atom $i$ is $-\partial_{x_i}\Pi_{\text{FK}}$ and hence the equilibrium equation for all $i$ = 1, 2, ...,\,$N$ is given by
\begin{equation}
\ds k(x_{{i+1}}-2x_i+x_{i-1}) - \frac{2\pi b}{\lambda}\sin{\bigg(\frac{2\pi}{\lambda}x_i\bigg)} = 0~.
\label{ground state FK}
\end{equation}
Here, the static friction force corresponds to the net resisting force arising from the periodic interaction potential, and is expressed as
\begin{equation}
\ds F_{\text{s}}=\frac{2\pi b}{\lambda}\sum_{i=1}^{N}\sin{\bigg(\frac{2\pi}{\lambda}x_i\bigg)}~.
\end{equation}
The upper bound for the static friction $F_{\text{s}}$ is $F_{\text{s}}^{\text{max}}=2\pi b N/\lambda $. The kinetic friction $F_\mrk$ is the averaged force, i.e.
\begin{equation}
\ds F_\mrk = \lim _{x_N \rightarrow \infty} \frac{1}{x_N}\int_0^{x_N}F_\mrs \,{\text{d}}x~.
\end{equation}

The ground state obtained from Eq.~\eqref{ground state FK} is commensurate, as the ratio $a/\lambda$ is rational. The ground state is incommensurate if the ratio is irrational. In the commensurate case, the minimum energy configuration will correspond to a topological soliton (kink) formation upon relaxation. These kinks govern tribological processes in the FK model as chains locally extend and compress to achieve the minimum energy configuration. Kinks are generated at one end of the chain during the finite extension at the pulling end and then propagate to the free end. Each full kink propagation results in a shift of the whole chain by lattice constant $a$. At a critical chain stiffness $k$ = $k_\mrc$ \footnote{This stiffness, together with $k_{\text{cr}}$ in Eq.~\eqref{critical stiffness adhesion} and $k$ in Eq.~\eqref{PT parameter 1D} can be understood within a unified physical framework: In all these cases, the total potential energy comprises an internal contribution that favors smooth deformation, and an interfacial contribution that promotes localization in the minima of a corrugated energy landscape. When the internal energy penalty for deformation is small compared to the energy barriers imposed by the interaction potential, multiple metastable states exist (see Fig.~\ref{stick slip and lateral force and different eta }a), leading to discontinuous transitions between configurations. Conversely, when the internal energy dominates, metastable states do not exist, and the system evolves in a continuous manner. Thus, these critical parameters mark the transition between regimes where energy minimization leads to either continuous evolution or discontinuous jumps between metastable states. }, the ground state of the system undergoes a transition between the states \enquote{mobile} and \enquote{pinned}, known as Aubry transition \citep{peyrard1983critical}. The value of $k_\mrc$ is obtained in a manner similar to Eqs.~\eqref{critical stiffness adhesion} and \eqref{PT parameter 1D}. It has been highlighted in \citet{braun2004frenkel} that the critical stiffness value $k_\mrc$ is minimum for the case when the lattice constant ratio is ($\lambda/a=(1+\sqrt{5}/2$), i.e. the irrational golden mean. This means that for the case of $k \ge k_\mrc$, the hull function is analytic, and the phase $\phi$ is arbitrary. The ground state is, therefore, translationally invariant for any $x_N$, and kinks are mobile, resulting in a zero depinning force. Thus, any infinitesimal small force leads to sliding \citep{aubry1978}. \citet{hirano1990atomistic} termed this phenomenon superlubricity, as it corresponds to extremely low friction coefficients. On the other hand, for the case $k$ $\le$ $k_\mrc$, the hull function is non-analytic for incommensurate contact and develops a dense set of jumps. The phase is no longer arbitrary; therefore, a finite force is needed to de-pin this state. \\\citet{benassi2015breakdown} examined the robustness of superlubricity in edge-driven systems using a FK model. Their study highlights that beyond a critical system size, elastic deformations in the solid induce commensurate dislocations at the interface, leading to a transition from superlubric to high-friction behavior. The critical length at which this transition occurs was derived analytically and confirmed through simulations.  Continuum interaction models developed by \citet{popov2011commensurate, espanol2017discrete} and \citet{Xue2022} exhibit conceptual similarities to the FK model. The FK model has been extended to two dimensions by \citet{gornostyrev1999fluctuation}, enabling the study of more complex interfacial phenomena in crystalline systems. 

Thermal effects in the FK model can be summarized as follows: Temperature introduces irreversible dissipation by enabling energy transfer from the sliding motion into internal degrees of freedom, such as lattice vibrations. Thermal fluctuations promote activated barrier crossing, leading to more frequent slip events and the excitation of phonon modes, which act as an energy sink and give rise to phononic friction. Consequently, even in structurally incommensurate (superlubric) systems, thermal fluctuations induce stochastic atomic rearrangements and phonon excitation, resulting in a small but finite friction force at finite temperatures \citep{vanossi2013colloquium}. For a comprehensive treatment of FK dynamics, including thermal effects and collective excitations such as kink formation, see \citet{BRAUN19981}.

\subsubsubsection{Frenkel-Kontorova-Tomlinson model}
In the FK model the atoms are not all connected to the sliding body. This limitation is overcome in the FKT model (see Fig.~\ref{Theoretical model of friction}c). The monolayer is now described by a chain of length $L$ consisting $N$ particles with harmonic nearest-neighbor interactions defined by springs of stiffness $k_1$. The interaction of each
particle with the loading device is also harmonic, and denoted 
by springs of stiffness $k_2$. The interaction potential is defined by period $\lambda$, amplitude $b$, and chain lattice constant $c$. The potential energy of the FKT model can be written as 
\begin{equation}
  \ds  \Pi_{\text{FKT}}=\frac{1}{2}\sum_{i=1}^{N}k_1(x_{i+1}-x_{i}-c)^2 + \frac{1}{2}\sum_{i=1}^{N}k_2(x_i-x_B)^2-b\sum_{i=1}^{N}\cos{\bigg(\frac{2\pi}{\lambda}(x_i + x_B+i\, c)\bigg)}~.
\end{equation}
 For rational $c$ and $\lambda$, and under similar periodic boundary conditions to those in Eq.~\eqref{FK/FKT periodic BC}, the corresponding stationary state for all $i$ = 1, 2, ...,\,$N$ is given by 
 \begin{equation}
  \ds        k_1(x_{i+1} -2x_i+ x_{i-1})-k_2x_i - \frac{2\pi b}{\lambda }  \sin{\bigg(\frac{2\pi}{\lambda}(x_i + x_B+i\, c)\bigg)}=0~.
 \end{equation}
Following the FK model, the static friction is given by
\begin{equation}
\ds F_\mrs= \frac{2\pi b}{\lambda} \sum_{i=1}^{N}\sin{\bigg(\frac{2\pi}{\lambda}(x_i + x_B+i\, c)\bigg)}\,,
\end{equation}
with upper bound $F_{\text{s}}^{\text{max}}= 2\pi b N/\lambda$. The kinetic friction in the FKT model is then given by
\begin{equation}
\ds F_\mrk = \lim_{x_B\rightarrow \infty}\frac{1}{x_B}\int_0^{x_B}F_\mrs\,{\text{d}}x~.
\end{equation}
Similar to the FK model, the concept of Aubry transitions \citep{aubry1978, sharma1984aubry} can also be applied to the FKT model, see \citet{weiss1996dry}.  

\subsubsection{Computational modeling approaches}\label{3_2_2}
The low-dimensional analytical models discussed in Section \ref{sec_3_2_1} provide a qualitative understanding of atomic friction. However, a deeper understanding of tangential friction in crystalline materials such as graphene requires modeling techniques that can capture phenomena across multiple length and time scales. Computational approaches provide a powerful framework for this purpose, ranging from quantum mechanical methods that describe electronic interactions, to atomistic simulations that resolve individual atomic trajectories, and continuum-scale methods that address large systems efficiently. In the following subsections, these three approaches are discussed for tangential contact.

\subsubsubsection{Density functional theory
}\label{sec_3_2_2_1}

DFT is a quantum mechanical simulation method used to investigate the electronic structure of a system. The computation can be performed with software packages such as VASP \citep{PhysRevB.54.11169, KRESSE199615}, SIESTA \citep{soler2002siesta}, and Quantum ESPRESSO \citep{giannozzi2009quantum, giannozzi2017advanced}. These packages find the binding energy and the
electron density of the system. This approach has gained significant importance in nanomaterial characterization, particularly at small length scales, where it is typically applied to a unit cell or a small representative cell of only a few nanometers in size. The variation of the binding energy (corresponding to the potential energy) of such a cell is mapped by displacing the two contacting materials on a grid, and its change is defined as $\Delta \Pi_\text{c}= \Pi_\text{c}^{\text{max}}-\Pi_\text{c}^{\text{min}}$. From this, the tangential and normal component of the binding force acting on the cell are obtained as 
$F_t={\Delta \Pi_\text{c}}/{\Delta \tau}$ and $F_N={\Delta \Pi_\text{c}}/{\Delta g_n}$. At the electronic scale, the characteristics of the electronic structure strongly influence material properties and can play a central role in adhesion and friction. 
Even though, it is very expensive to model a experimental setup with electronic structure methods due to their demanding computational resources, several experimental studies have been effectively supported by DFT calculations. For example, DFT was used to evaluate the potential energy and its barriers at Gr/Gr, h-BN/h-BN and $\text{MoS}_2$/AFM tip interfaces, explaining differences in static friction observed in AFM experiments \citep{falin2017mechanical, vazirisereshk2019origin}. Other studies have explored strategies to tune the electronic configuration of layered systems in order to modify interlayer interactions and reduce sliding friction. For instance, \citet{wang2016understanding, gargiulo2017structural, yoo2019atomic} report that in twisted bilayer graphene below a critical crossover angle, the MSL undergoes a structural transformation into an array of commensurate domains separated by soliton boundaries, leading to electronic reconstruction. 

\begin{figure}[H]
\begin{center} \unitlength1cm
\begin{picture}(0,8.5)
\put(-8,0){\includegraphics[height=80mm]{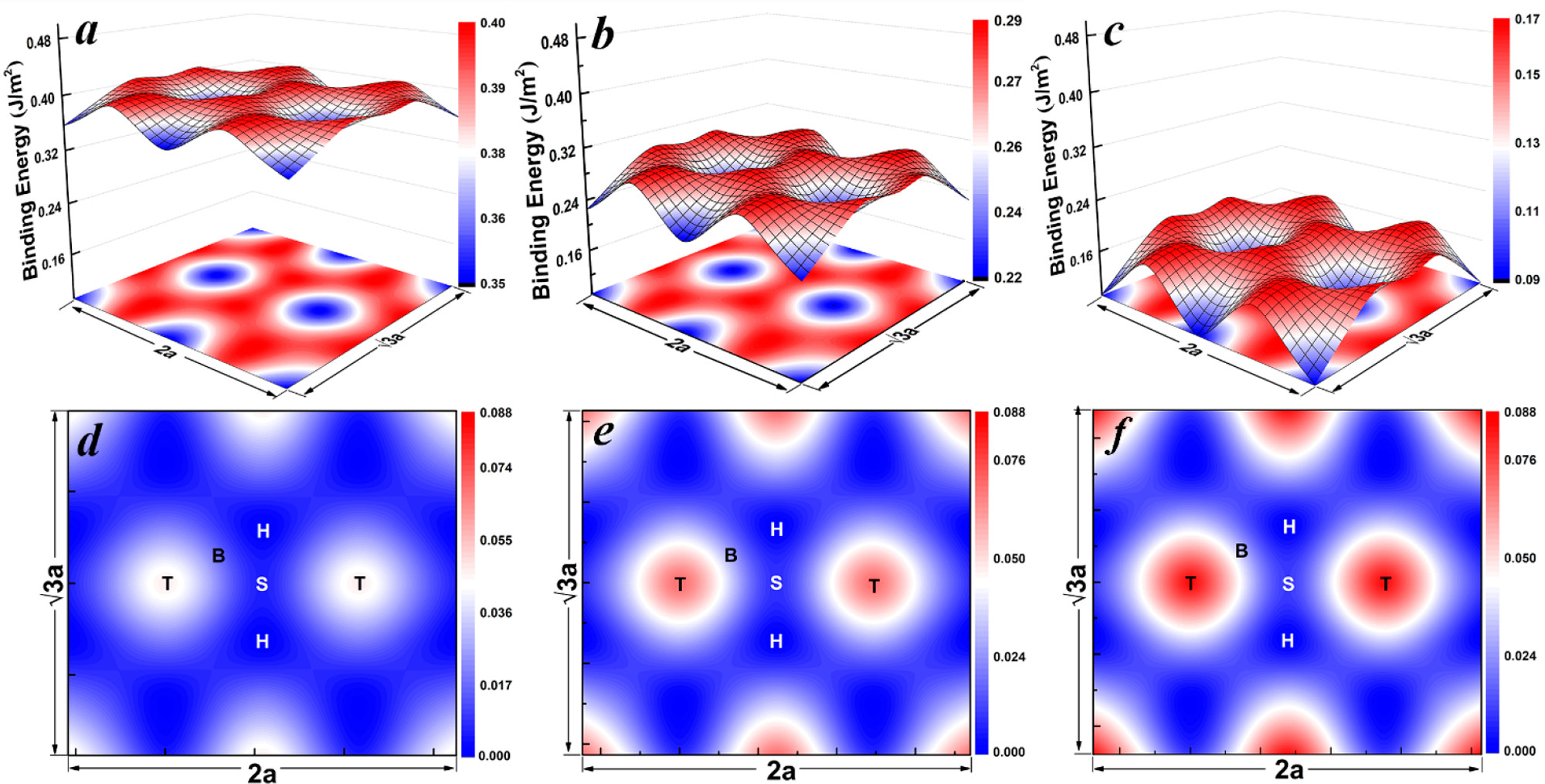}}
\end{picture}
\caption{Binding energy (a–c) and potential energy (sliding barrier) (d–f) of a Gr/Gr bilayer under biaxial strain: (a, d) 12 $\%$ compressive strain; (b, e) relaxed state; (c, f) 12 $\%$ tensile strain. Here, T, B, S, and H denote the top (AA), bridge, saddle-point (SP), and hollow (AB) stacking configurations, respectively. Reprinted  from \citet{cheng2020strain} with permission from American Chemical Society. }
\label{binding and pes}
\end{center}
\end{figure}
Similar effects of electronic reconstruction on interfacial tribology have been demonstrated in strained and lattice-mismatched systems \citep{reguzzoni2012potential, wang2017superlubricity, cheng2020strain, wang2022influences, li2023lattice}. When a dilatational strain is applied to the graphene substrate of a Gr/Gr interface, \citet{cheng2020strain} reported that the binding energy increases under biaxial compression (Fig.~\ref{binding and pes}a), whereas it decreases under biaxial tension (Fig.~\ref{binding and pes}c). In contrast, the strain-induced changes in the potential energy exhibit an opposite trend, as seen in Figs.~\ref{binding and pes}d and \ref{binding and pes}f. Consequently, tensile strain tends to promote vertical separation of the layers, while compressive strain facilitates lateral sliding in the Gr/Gr bilayer. The coefficient of friction between $\text{MoS}_2$ layers has been related to load and interlayer spacing \citep{levita2014sliding}, with similar findings reported for Gr/GaSe \citep{li2022structural}, $\text{ZrS}_2$, and Gr/Ge(111) systems \citep{xu2022first}. In general, the magnitude and distribution of the interfacial charge density across different stacking configurations critically determine the friction force, as confirmed by recent correlations between potential energy corrugation and charge-density variations \citep{sun2023charge}. 

Full DFT or quantum-region-based hybrid modeling results in high computational cost, which severely limits the system size and simulation time \cite{hao2024accurate}. As a result, it remains challenging to directly simulate large-scale tribological phenomena, motivating the use of classical MD methods, which are discussed next.

\begin{figure}[H]
\begin{center} \unitlength1cm
\begin{picture}(0,4.5)
\put(-7.2,0){\includegraphics[height=40mm]{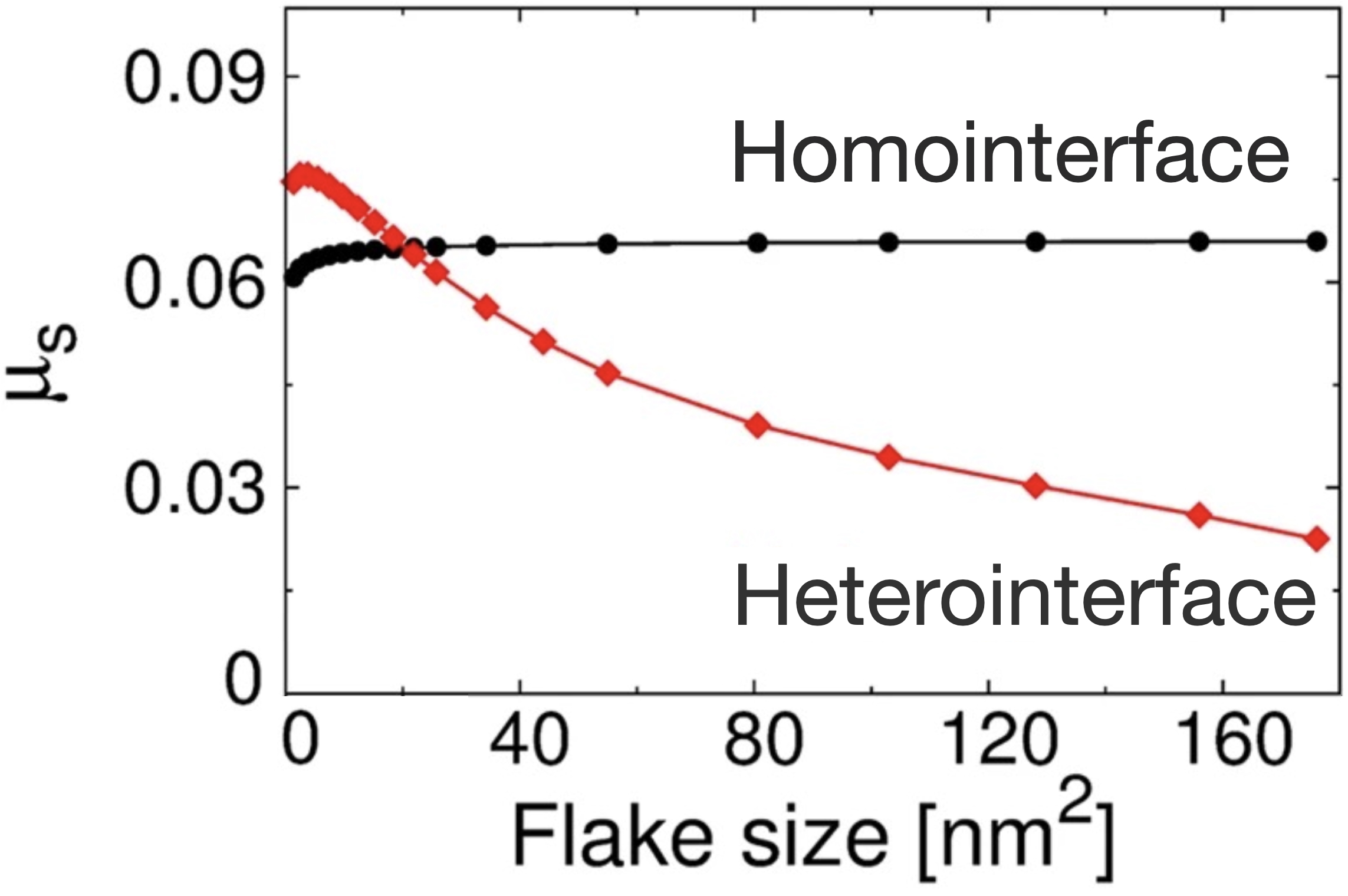}}
\put(0,0){\includegraphics[height=40mm]{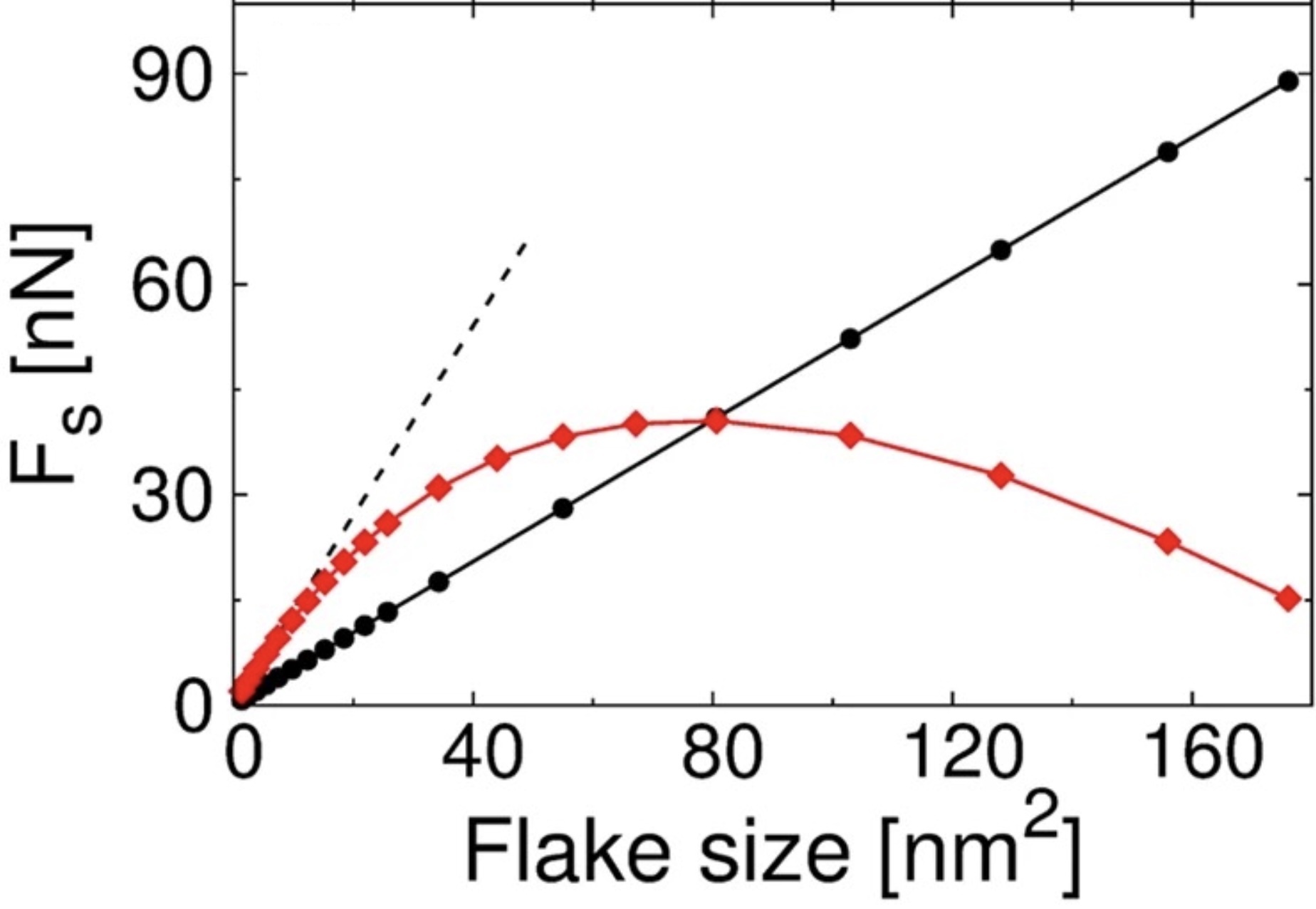}}
\put(-6.4,0){\fontsize{10pt}{20pt}\selectfont (a)  }
\put(0.0,0){\fontsize{10pt}{20pt}\selectfont (b) }
\end{picture}
\caption{Friction coefficients and friction forces for different flake sizes of hexagonal shape for both homointerface (Gr/Gr, black) and heterointerface (Gr/h-BN, red) in subfigures (a,b) respectively. The normal load is  0.1 nN/atom. The dashed lines represent the variation in friction force without MSL consideration. Reprinted  from \citet{mandelli2017sliding} with permission from Nature Portfolio.}
\label{flake size behaviour}
\end{center}
\end{figure}

\begin{figure}[H]
\centering
\includegraphics[scale=0.5]{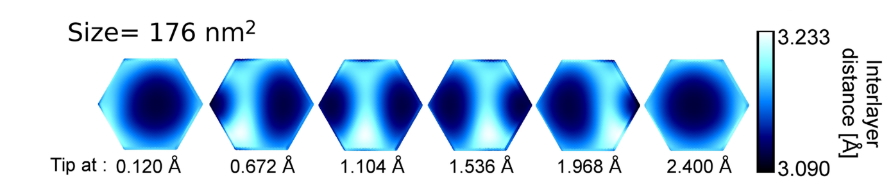} 
\caption{Snapshots capturing the soliton-like smooth sliding motion of the Moir{\'e} superstructure ridges that emerge when a Gr/h-BN interface is sheared. Reprinted from \citet{mandelli2017sliding} with permission from Nature Portfolio.}
\label{snapshot of soliton}
\end{figure}

\subsubsubsection{Molecular dynamics}\label{sec_3_2_2_2}
 In MD simulations, the atomic interactions are described through established interatomic potentials or force fields such as the LJ potential introduced earlier. Generally, two types of potentials are required for contacting surfaces: intra-body and inter-body potentials. Most widely used intra-body potentials are MM3 \citep{allinger1989molecular}, Tersoff \citep{tersoff1989modeling}, the first and second
generation REBO (reactive empirical bond order)  \citep{brenner1990empirical, brenner2002second}, AIREBO (adaptive intermolecular reactive empirical bond order) \citep{Stuart2000} and ReaxFF \citep{chenoweth2008reaxff}, while common inter-body potentials are LJ \citep{lennard1931cohesion}, KC \citep{Kolmogorov2005}, Lebedeva \citep{Lebedeva2011}, and DRIP (dihedral-angle-corrected registry-dependent) \citep{wen2018dihedral}. In all cases the average tangential tractions are determined from
\begin{equation}
 \ds   \bt_\mrt =\sum_{i=1}^N \bF_i/A\,,
\end{equation}
where $\bF_i$ is the tangential component of the vdW force acting on atom $i$ due to the bottom layer, $A$ is the contact surface area, and $N$ is the total number of particles in contact. 
Several studies have investigated the tribological response of 2D materials under different sliding directions and loading conditions \citep{leven2016interlayer, mandelli2017sliding, song2018robust, Cao2018, ouyang2018nanoserpents, mandelli2019princess}. In Fig.~\ref{flake size behaviour}, the observed reduction in both the friction coefficient and the friction force with increasing flake size for heterointerface arises from the increase of area fraction of Moir\'e regions \citep{wang2024colloquium}. The boundaries of these Moir\'e patterns form incommensurate domains, which, upon relaxation, give rise to ridges that eventually evolve into smooth, soliton-like sliding structures (see Fig.~\ref{snapshot of soliton}). In contrast, homointerface exhibit a linear increase.

Other works \citep{mandelli2017sliding, wang2019strain, wang2019generalized, ru2020interlayer, yang2021rotational, Dey2023, yan2023origin} have examined the effect of interfacial misorientation on friction reduction in 2D materials. The influence of strain engineering on lubricity has been explored in several studies \citep{zhang2017stacking, wang2019robust, zhang2019tuning, peng2020friction, li2021effect, xu2022effect}. Fig.~\ref{Strain, angle, friction dependence} shows the case of bilayer graphene, where increasing the applied strain on the graphene substrate or the stacking angle between the bilayers also increases the corresponding area fraction occupied by Moir{\'e} regions and consequently reduces friction force. This behavior follows from the strain dependence of the nominal Moir{\'e} pattern described by its period $A^\text{M}$ and lattice vectors $\bA_{i}^\text{M}$($i$\,=\,1,2) given by  \citep{yankowitz2012emergence, wang2019robust}
\begin{equation}
A^\text{M}=\|\bA_{1}^\text{M}\|=\|\bA_{2}^\text{M}\|=\frac{(1+\varepsilon_\mrb)\mra_{\text{cc}}}{\sqrt{1+(1+\varepsilon_\mrb)^2-2(1+\varepsilon_\mrb)\cos{\theta}) }}~,
\label{strain moire relation}
\end{equation}
where $\theta$ is the inter-layer twist and $\varepsilon_\mrb$ is the bi-axial strain. Further details on the influence of Moir{\'e} patterns on superlubricity can be found in the review article \cite{yan2024moire}, which discusses the influence of material size, shape, and edge-pinning on superlubric behavior.

\begin{figure}[H]
\begin{center} \unitlength1cm
\begin{picture}(0,6)
\put(-7.5,0){\includegraphics[height=55mm]{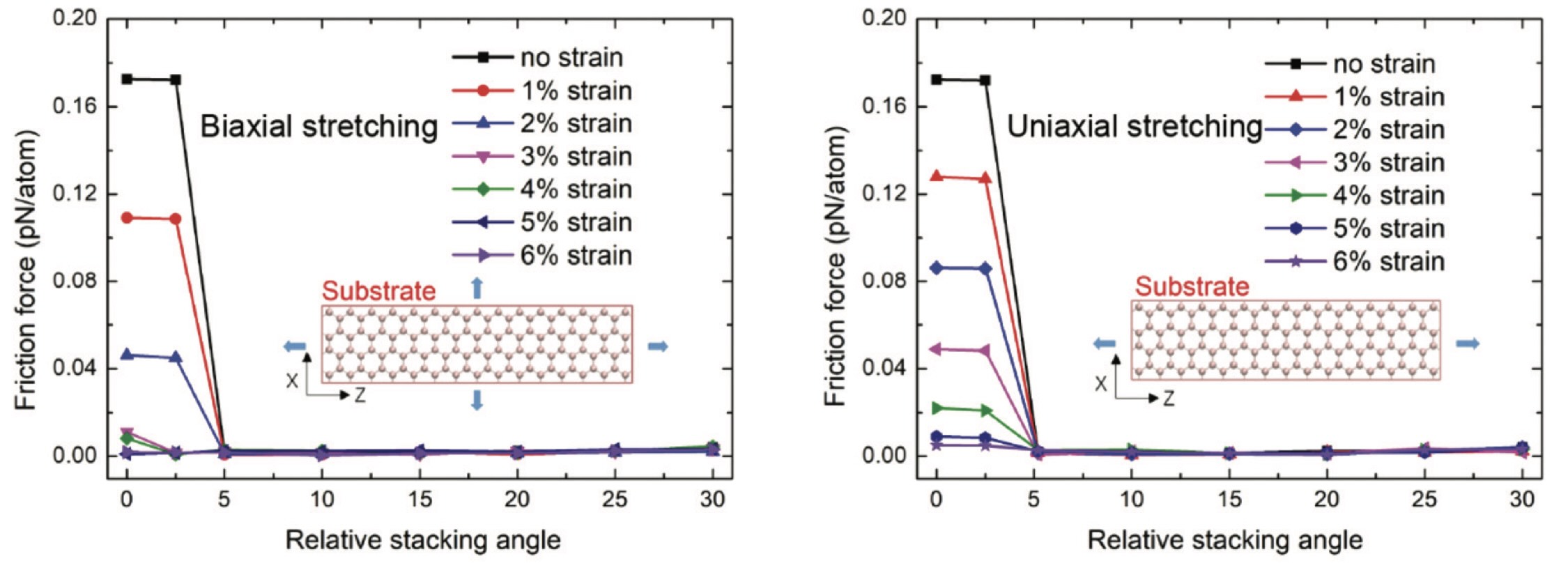}}
\put(-7.5,0){\fontsize{10pt}{20pt}\selectfont (a) }
\put(0.5,0){\fontsize{10pt}{20pt}\selectfont (b) }
\end{picture}
\caption{Stacking angle dependence of friction for different strained substrates under a normal load of 0.2 nN per atom. (a) Biaxial strain (b) uniaxial strain. Reprinted  from \citet{wang2019robust} with permission from Royal Society of Chemistry.}
\label{Strain, angle, friction dependence}
\end{center}
\end{figure}

Simulations like these show that 2D layered material heterointerfaces and homointerfaces can be used in actual applications to achieve robust superlubricity under high pressure ($\approx50$ GPa) independent of the relative interfacial orientation. 

Following the atomistic description provided by MD simulations, we now turn to continuum formulations with FEM-based models, where tangential tractions are incorporated through continuum interaction laws.

\begin{figure}[H]
\begin{center} \unitlength1cm
\begin{picture}(0,5.5)
\put(-7.8,0){\includegraphics[height=50mm]{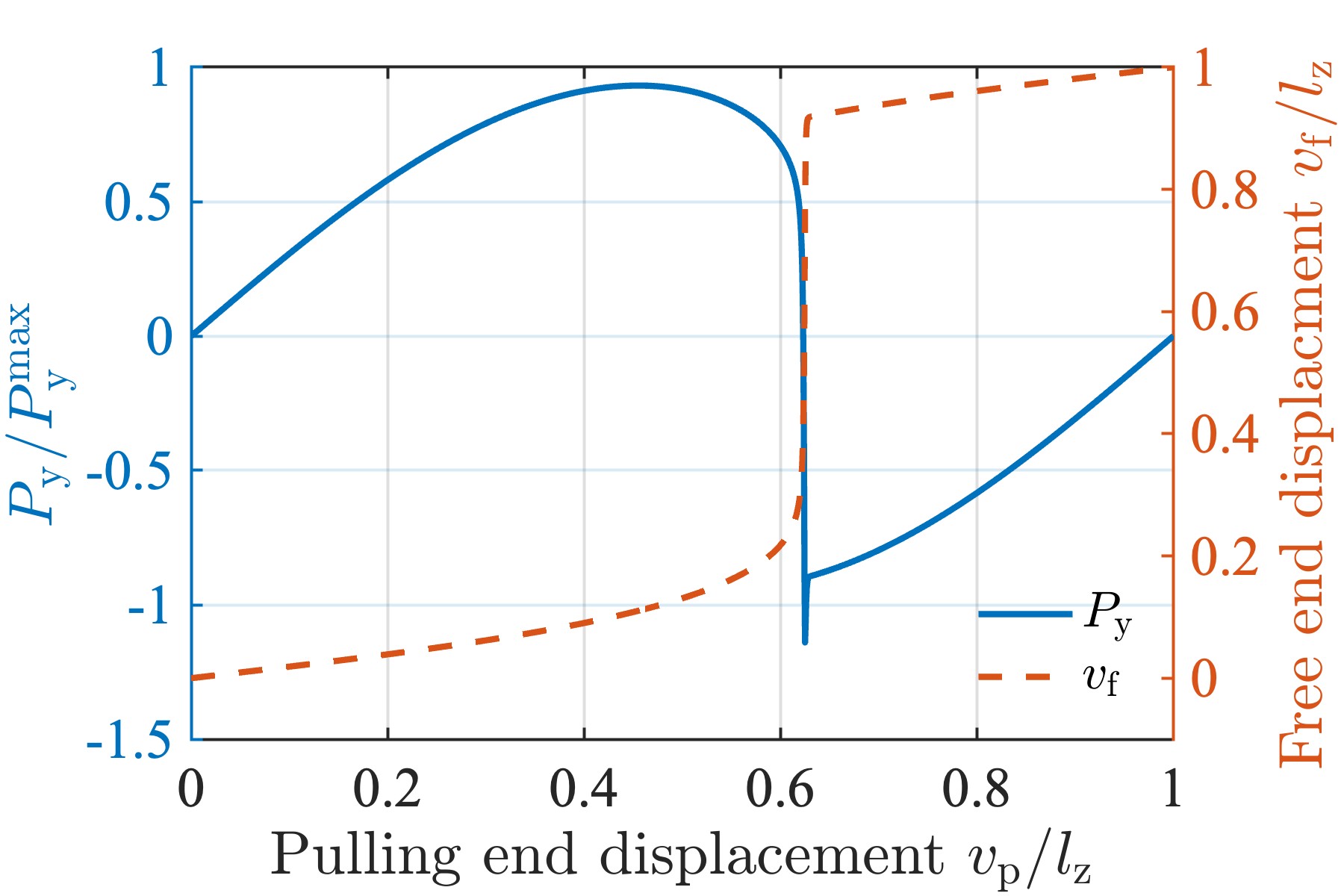}}
\put(0.2,0){\includegraphics[height=50mm]{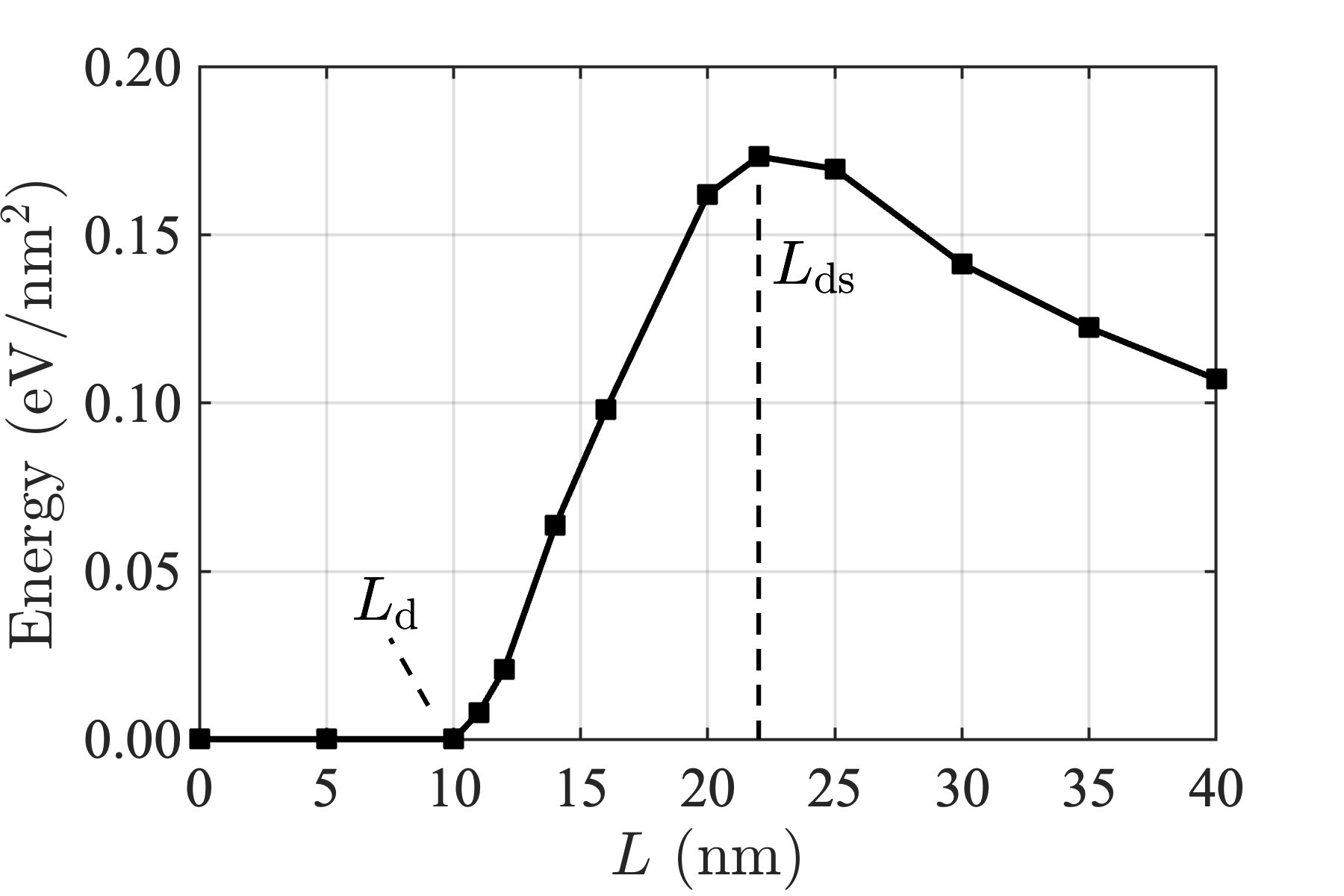}}
\put(-7.7,0){(a)}
\put(.3,0){(b)}
\end{picture}
\caption{(a) Variation of normalized pulling force with the normalized free end displacement of a 16 nm long GNR with a laterally constrained
boundary. (b) Variation of dissipated energy with ribbon length. Reprinted from \citet{yadav2025investigating} with permission from Elsevier.} 
\label{Slip and critical length for dissipative behavior}
\end{center}
\end{figure}

\begin{figure}[H]
\begin{center} \unitlength1cm
\begin{picture}(0,5.3)
\put(-7.8,0){\includegraphics[height=55mm]{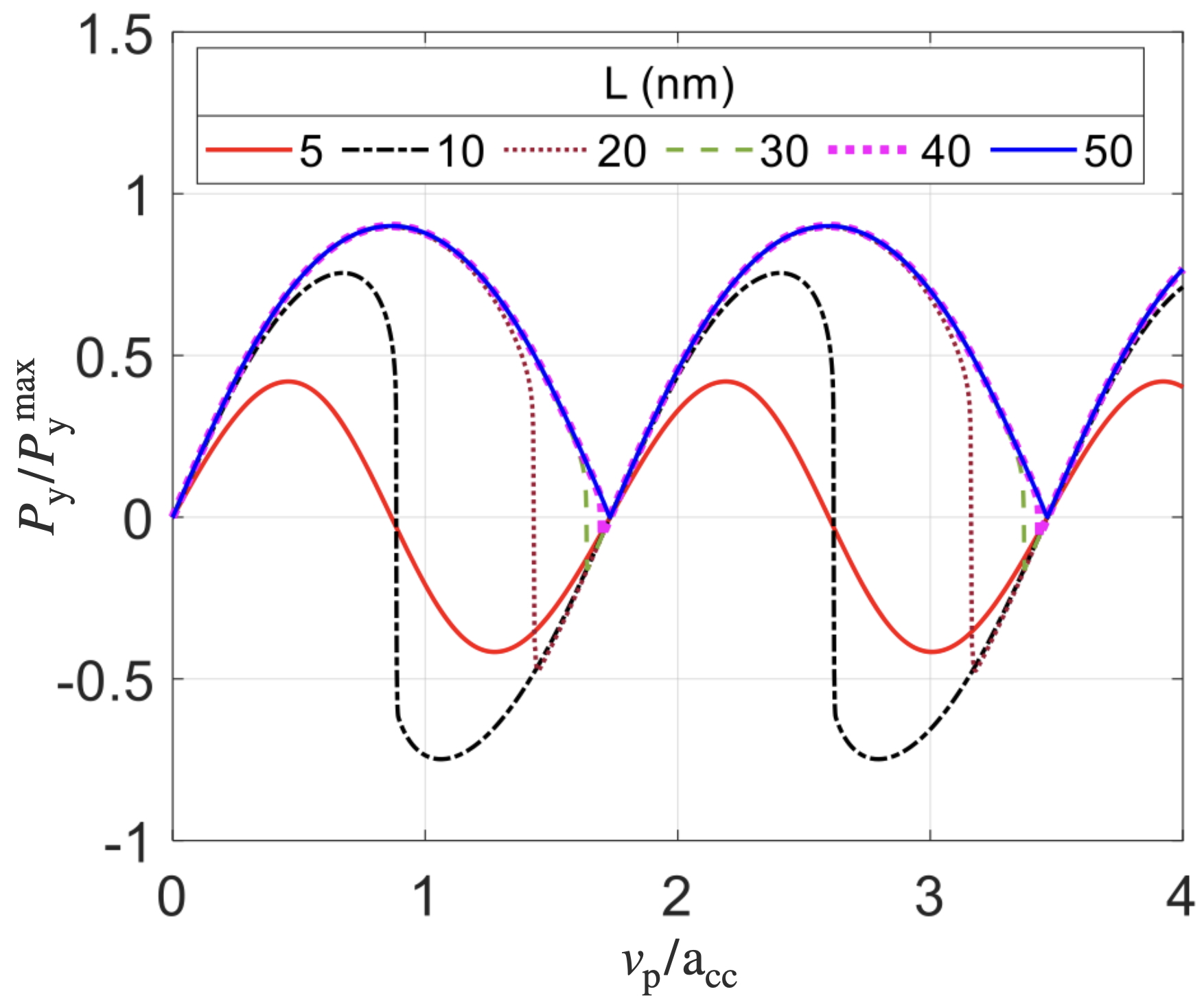}}
\put(0.2,0){\includegraphics[height=55mm]{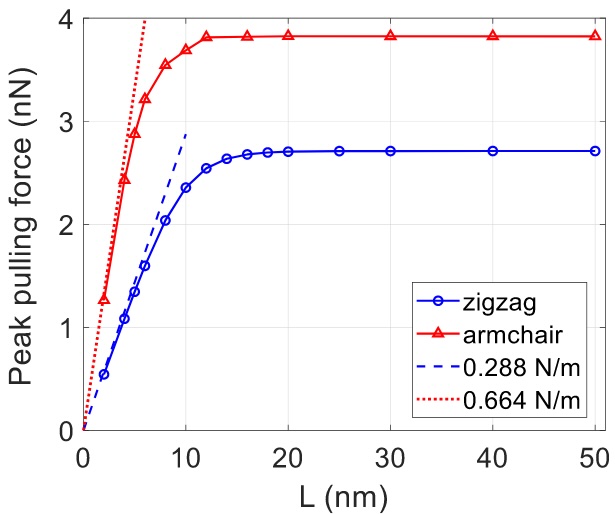}}
\put(-7.7,0){(a)}
\put(.3,0){(b)}
\end{picture}
\caption{(a) Normalized pulling force-displacement response of GNRs with different lengths, corresponding to the zigzag sliding direction. (b) The pulling peak force versus the length of the GNR for unconstrained sliding in both the zigzag and armchair directions. Reprinted from \citet{Xue2022} with permission from Elsevier.} 
\label{Pulling force and its maximum vs. length}
\end{center}
\end{figure}

\subsubsubsection{Finite element methods}\label{sec_3_2_2_3}
Finally, continuum models that describe sliding resistance in 2D crystalline materials, focusing specifically on bilayer graphene are discussed. \citet{Xue2022} investigated the variation and eventual saturation of the tangential pulling force in GNRs of different lengths, considering different boundary conditions and sliding directions. As the GNR is pulled, the material undergoes deformation, which depends on both the ribbon length and the interfacial pressure, since the strength of the adhesion energy is governed by the equilibrium interlayer gap. Notably, the entire sheet does not move uniformly; rather, only a localized region near the pulled edge initially deforms \citep{yadav2025investigating}. This can be seen in Fig.~\ref{Slip and critical length for dissipative behavior}a, where the free end displacement $v_{\mathrm{f}}$ of the GNR remains relatively small as compared to the pulling end displacement $v_{\mathrm{p}}$. This behavior is attributed to the classical shear lag effect \citep{cox1952elasticity}. This effect is more prominent in longer ribbons (longer than $\approx$ 50 nm). When the local GNR deformation exceeds a critical strain, typically in the range of 1–2$\%$ \citep{Wang2017size, ouyang2018nanoserpents, Xue2022}, the sheet buckles out of the contact interface, forming a topological soliton. The specific strain-threshold depends on the interaction potential and material properties. Once formed, the soliton propagates toward the free end of the GNR, resulting in global sliding of the sheet. When the width of the soliton resulting from deformation is comparable to the ribbon's length, it interacts with the free end of the GNR and becomes unstable before it fully develops, leading to snap through instabilities see Fig.~\ref{Slip and critical length for dissipative behavior}a and Fig.~\ref{Pulling force and its maximum vs. length}a, corresponding to stick-slip behavior. The transition from smooth sliding to stick-slip behavior with increasing GNR length is shown in Fig.~\ref{Pulling force and its maximum vs. length}a. Also, shown is the critical length $L_{\mathrm{s}}$ beyond which the peak pulling force, saturates, see Fig.~\ref{Pulling force and its maximum vs. length}b. 

\begin{figure}[H]
\begin{center} \unitlength1cm
\begin{picture}(0,5)
\put(-4,0){\includegraphics[height=50mm]{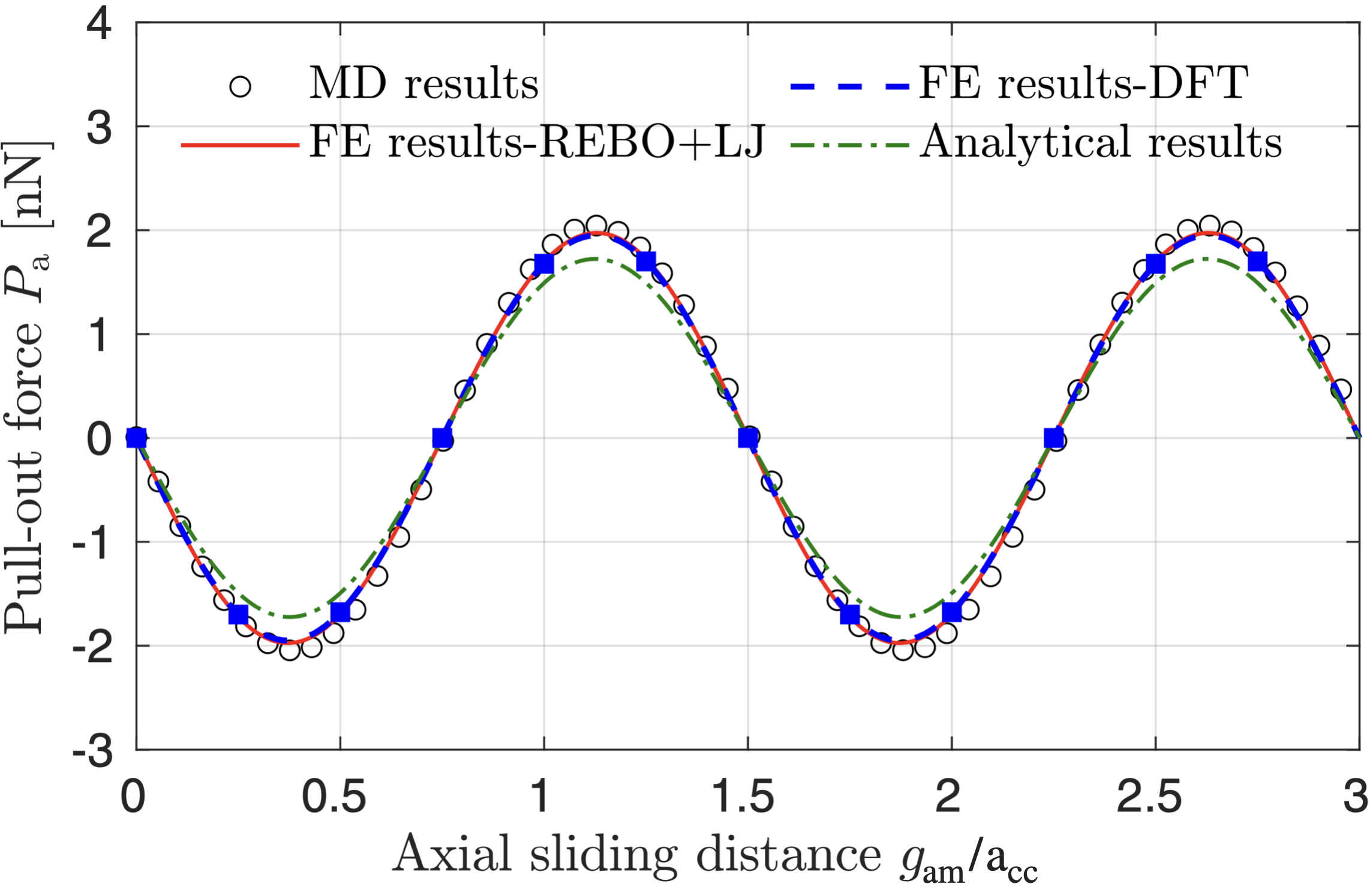}}
\end{picture}
\caption{Pull-out force for CNT(26,0) from within CNT(35,0) in dependence of the axial sliding distance. The axial contact tractions at 12 positions marked by blue squares are shown in Fig.~\ref{traction variation at 12 instance}. Reprinted from \citet{mokhalingam2024continuum} with permission from American Physical Society. } 
\label{tangential traction DWCNT comparison}
\end{center}
\end{figure}

\begin{figure}[H]
\begin{center} \unitlength1cm
\begin{picture}(0,4)
\put(-8,0){\includegraphics[height=40mm]{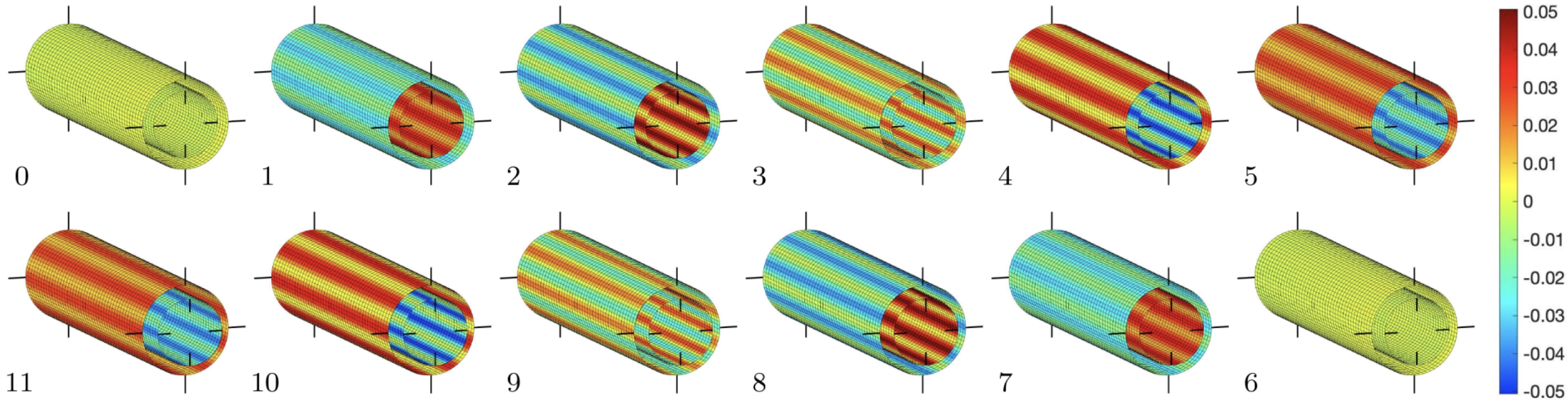}}
\end{picture}
\caption{Axial contact tractions in GPa for pull-out displacement $g_{\mra \mrm}$ $\in$ [0, 1, 2,..., 11]$\cdot \mathrm{a}_{\text{cc}}/4$.  Reprinted from \citet{mokhalingam2024continuum} with permission from American Physical Society. } 
\label{traction variation at 12 instance}
\end{center}
\end{figure}

\citet{Xue2022} carried out FEM simulations in ABAQUS, where GNR were modeled using shell elements with a user-defined subroutine to incorporate the periodic vdW interaction. \citet{mokhalingam2024continuum} developed a finite element (FE) formulation for the contact interactions of DWCNTs and compared  the CNT pull-out forces and twisting moments with analytical predictions and MD simulations. The pull-out force for a CNT(26,0) from within a CNT(35,0) is shown in Fig.~\ref{tangential traction DWCNT comparison}. Both periodicity and amplitude of the pull-out force $P_\mra$ determined from FE simulations show good agreement with the corresponding analytical and MD results. The axial contact tractions $t_\mra$ shown in Fig.~\ref{traction variation at 12 instance} vary in circumferential direction due to the lattice mismatch arising from the different diameters of the two CNTs.

In a recent study \citep{yadav2025investigating}, 1D Euler-Bernoulli beam elements were employed to investigate the effect of GNR size on sliding behavior. To address instabilities, continuation and dynamic relaxation were employed. This study also identified key parameters and characteristic length scales associated with the interfacial mechanical behavior of bilayer graphene. Two such scales -- $L_{\mathrm{d}}$ and $L_{\mathrm{ds}}$ -- are shown in Fig.~\ref{Slip and critical length for dissipative behavior}b. Here, $L_{\mathrm{d}}$ corresponds to the critical length of GNR beyond which sticking is observed, while $L_{\mathrm{ds}}$ corresponds to the saturation length beyond which solitons form.

\subsubsubsection{Comparative evaluation}

In all computational approaches, tangential tractions arise from the lateral variation of the interlayer interaction energy with relative in-plane displacements and are subsequently employed to study frictional and mechanical responses. DFT provides the most accurate description of registry-dependent interlayer energy landscapes; however, it is restricted to small system sizes and typically 0K. This accuracy is associated with the ability to capture both atomic and electronic structure, including effects such as electron exchange, charge transfer, and electronic state alignment, which significantly influence the average and variation of the binding energy \cite{wolloch2018interfacial, sun2023charge, boretti2025quantum}.

In contrast, classical MD simulations incorporate essential electronic interactions only in the average sense through phenomenological models and attribute lubrication behavior in 2D materials primarily to the corrugation of the interfacial potential energy surface due to atomic registry \cite{onodera2009computational}. MD resolves atomic discreteness and finite-temperature dynamics, and its accuracy depends on the suitability of the chosen potentials. These potentials are typically obtained from fits to DFT data or experiments conducted for specific standard scenarios; therefore, they do not capture the evolving nature of the electronic structure. This establishes a contrast between MD approaches and quantum-mechanical modeling, which is further supported by experiments revealing collective, nonlocal behavior and many-body dispersion effects in vdW interactions that cannot be explained by simple pairwise models \cite{hermann2017first, yuan2020effects}. For example, measurements of vdW forces across SiO$_2$ films show that, contrary to pairwise predictions of a $\sim$1\,nm cutoff, the interaction persists up to thicknesses of 10 to 20\,nm, indicating significantly longer-ranged, nonlocal behavior \cite{loskill2013macroscale}. Similarly, experiments demonstrate that a single layer of graphene or MoS$_2$ can almost completely screen vdW interactions from the underlying substrate. This effect is attributed to dominant in-plane electronic fluctuations of the 2D layer, which effectively decouple the interaction from the substrate \cite{tsoi2014van}. These observations highlight the limitations of classical descriptions and the role of electronic effects. From a DFT perspective, friction arises from gradients in the evolving charge density rather than solely from the corrugation of the interfacial potential energy surface associated with atomic registry \cite{sun2023charge}. This leads to differences in adhesion, energy barriers, and the resulting frictional response.

 While continuum models provide computationally efficient approximations, incorporating such quantum-mechanical effects remains important for accurate multiscale descriptions of interfacial behavior. One such recent development in multiscale approaches is the coarse-grained model for determining the impact of  MSL on its mechanical properties of tBLG \cite{yan2025coarse}. FEM-based models do not resolve atoms explicitly; instead, they incorporate tangential tractions through interaction energy functionals informed by atomistic calculations \cite{carr2018relaxation, Xue2022, morovati2022interlayer, mokhalingam2024continuum, yadav2025investigating}, allowing efficient analysis of large-scale systems and coupled normal and tangential deformations. Additionally, for thermal effects, thermomechanical constitutive assumptions are required, such as temperature-dependent rate laws or coupling to heat transfer \cite{yan2022thermodynamic, ahmed2025quantifying}. 
 
A quantitative comparison of interlayer shear strength (e.g., friction forces, friction coefficients) across different two-dimensional materials would be desirable. However, such comparisons are challenging due to the strong dependence of friction on multiple factors, including size effects and scaling laws \cite{wang2019generalized}, Moir\'e superlattice effects (size and shape) \cite{bai2022moire, yan2024moire}, elasticity and strain solitons \cite{bai2022one, morovati2022interlayer}, edge effects such as atomic pinning \cite{liao2022uitra}, and environmental conditions \cite{yang2023tunable}. As a result, direct one-to-one quantitative comparison between simulations and experiments is often not feasible, since experimental systems operate at significantly larger length and time scales than atomistic simulations, and exact matching of geometry, size, and boundary conditions is rarely possible. This limitation is also reflected in the literature, where identical scenarios are seldom available for direct comparison.

Nevertheless, overall trends -- such as the dependence of friction on size, strain, and relative orientation -- are consistently observed across both experimental and computational studies. For instance, in an experimental study on Gr/MoS$_2$, \citet{liao2022uitra} reported a monotonic decrease in interlayer shear strength with increasing flake area, which is qualitatively similar to that observed for the Gr/h-BN system shown in Fig.~\ref{flake size behaviour}b. Early experimental studies \cite{verhoeven2004model, dienwiebel2004superlubricity} further demonstrated the critical influence of relative twist on superlubricity, where slight misalignment leads to an orders-of-magnitude reduction in friction. This trend is also captured in MD simulations \cite{wang2019generalized} (see Fig.~\ref{Strain, angle, friction dependence}a). In addition, \citet{yang2023tunable} reported that the friction coefficient of heterointerfaces is approximately five times lower than that of homointerfaces under 20-30\% relative humidity and moderate edge passivation conditions. Despite the qualitative agreement observed in several studies, quantitative discrepancies between experiments and computational predictions generally persist. This can be attributed to the inherent limitations of current computational approaches, including restricted system sizes, simplified boundary conditions, and the inability to fully capture environmental effects and evolving electronic interactions at experimentally relevant scales.

All models discussed are based on the same physical principles of energy minimization and force balance and are therefore best viewed as complementary, differing primarily in their resolution, efficiency, and representation of interlayer interactions. 
The key characteristics of the modeling approaches discussed above are briefly summarized in Table~\ref{model comparison table}.
\begin{table}[H]
\centering
{\fontsize{7}{12}\selectfont
\renewcommand{\arraystretch}{1}
\begin{tabular}{
>{\RaggedRight\arraybackslash}p{1.2cm}
>{\RaggedRight\arraybackslash}p{1.6cm}
>{\RaggedRight\arraybackslash}p{2.5cm}
>{\RaggedRight\arraybackslash}p{2.5cm}
>{\RaggedRight\arraybackslash}p{2.8cm}
>{\RaggedRight\arraybackslash}p{3cm}
}
\hline
\textbf{Model} & \textbf{Length scale (m)} & \textbf{Computational scaling$^\dagger$} & \textbf{Features} & \textbf{Capability} & \textbf{Main references} \\
\hline

DFT 
& $10^{-10}$-$10^{-9}$
& $\mathcal{O}(N^3)$ for $N$ basis functions
& Electronic structure and binding energy
& Adhesion, corrugation, and registry-dependent energetics
& \cite{KRESSE199615, pan2021scaling, del2023deep, al2024review} \\

MD 
& $10^{-9}$-$10^{-7}$
& $\mathcal{O}(N)$-$\mathcal{O}(N^2)$ for $N$ atoms
& Atomistic dynamics with empirical potentials
& Sliding dynamics, friction trends, and thermal effects
& \cite{plimpton1995fast} \\

PT & Single DOF
& $\mathcal{O}(1)$
& Spring-slider model
& Stick-slip mechanism
& \cite{prandtl1928, Tomlinson1929} \\

FK & $\sim 10^{-9}$
&$\mathcal{O}(N)$ for $N$ DOFs
& Atomic chain in periodic potential
& Solitons and Aubry transition
& \cite{braun2004frenkel} \\

FKT & $\sim 10^{-9}$
&$\mathcal{O}(N)$ for $N$ DOFs
& Driven FK chain with external spring
& Collective stick-slip and sliding friction
& \cite{vanossi2013colloquium} \\

CGC/SSIP & $10^{-7}$-$10^{-5}$
& $\mathcal{O}(N^{1.5})$-$\mathcal{O}(N^2)$ for $N$ DOFs
& Reduced-order interaction models
& Efficient contact representation
& \cite{sauer2006, grill2020computational} \\

Fourier-based FEM & $10^{-3}$-$10^{0}$
& $\mathcal{O}(N^{1.5})$-$\mathcal{O}(N^2)$ for $N$ DOFs
& Lattice energy representation
& Large-scale contact analysis
& \cite{mergel2021contact, mokhalingam2024continuum, yadav2025investigating} \\

\hline
\end{tabular}
}
\caption{Comparison of interlayer interaction modeling approaches across scales, highlighting typical system sizes and representative computational characteristics reported in the literature. ($^\dagger$ These scalings correspond to standard implementations; advanced algorithms, approximations, and efficient numerical strategies can further reduce these computational costs.)}
\label{model comparison table}
\end{table}

\section{Summary and future work} \label{sec_4}

The paper surveys inter-layer interaction models for graphene and other 2D materials. Models describing vdW adhesion between elastic bodies with smooth surfaces are reviewed, with particular emphasis on the determination of normal and tangential tractions between the layers at both continuous and discrete interfaces. The importance of simplifying assumptions and the knowledge of crystal structure is highlighted in the modeling of vdW contacts. These models reformulate pairwise interatomic adhesion energies into integral or analytical expressions suitable for continuum or multiscale descriptions. The origin of adhesive friction is examined for both interface types, arising from distance-dependent adhesion in continuous interfaces and from lateral variations of the interlayer interaction energy in discrete interfaces. It is further highlighted that elasticity and system size play key roles in governing interfacial mechanics. For discrete interfaces, strain engineering strongly influences tangential tractions, leading to the formation of incommensurate domains and Moir\'e patterns that give rise to superlubricity. 

A critical aspect of modeling is its validation against experiments and advanced characterization techniques. Since current models rely on simplified configurations and often neglect important factors such as size, environmental conditions, surface roughness, and chemical complexity, a gap remains between idealized models and realistic systems. However, no single method is sufficient to address all these challenges, and future progress is expected to rely on hybrid and multiscale approaches that combine quantum accuracy with large-scale simulations. Future developments in tribochemical simulations should therefore focus on bridging this gap. A primary challenge is the limited system size accessible to quantum methods, which currently restricts simulations to hundreds of atoms and prevent the direct study of tribofilm formation. Although reactive MD enables larger systems, the associated complexity and computational cost remain significant. Thus, improvements in computational efficiency and scalability are essential. In addition, integrating DFT and MD simulations can help bridging the gap between experimental observations and real-time material-environment interactions, providing reliable guidance for the design of advanced lubricating materials \cite{hao2024accurate}. Furthermore, coupling atomistic methods with continuum techniques such as finite element methods offers a promising route to extend tribochemical insights to macroscopic scales \cite{ta2021computational}. 

Nonetheless, this study serves as a useful repository for researchers and the broader community and is expected to support and stimulate further research in vdW interfaces. Promising directions for future work include scaling ultralow friction toward macroscale contact design, exploring heterointerfaces and different materials through first-principles studies to identify superior material combinations \citep{xu2022first}, engineering functional interfaces with enhanced tribological performance \citep{luo2025engineering}, modeling surface rippling and corrugation induced by thermal effects \citep{mauri2020thermal, yan2022thermodynamic, wang2023bending}, and investigating the role of interfacial defects and grain boundaries \citep{Song2024, ying2025scaling}. From an application perspective, these developments are directly relevant to the design of next-generation low-friction coatings, NEMS/MEMS, flexible electronics, and energy-efficient mechanical components. In particular, achieving robust superlubricity at larger scales, controlling friction through interface engineering, and designing environmentally adaptive lubricating systems remain key technological goals.

\section*{CRediT authorship contribution statement} 
\textbf{Gourav Yadav}: Writing – original
draft, Writing – review and editing, Formal
analysis, Conceptualization. \textbf{Shakti S. Gupta}: Writing – review and editing, Conceptualization, Supervision. \textbf{Roger A. Sauer}:
Writing – review and editing, Conceptualization, Supervision.

\section*{Declaration of competing interest}
The authors declare no known competing financial or personal interests that could have influenced the work reviewed in this paper.


\section*{Acknowledgments} 
The authors thanks Alexandar Borković and Aningi Mokhalingam for their comments. Gourav Yadav gratefully acknowledges the financial support from Ruhr University Bochum, Germany, received during his appointment as guest researcher.

\appendix

\titleformat{\section}
 {\normalfont\Large\bfseries}
  {Appendix~\Alph{section}:}
  {0.5em}
  {}

\section{Interlayer interaction energy of graphene}\label{Appendix A}
\setcounter{equation}{0}
\setcounter{figure}{0}
\setcounter{table}{0}

\makeatletter
\renewcommand{\theequation}{\Alph{section}\arabic{equation}}
\renewcommand{\thefigure}{\Alph{section}\arabic{figure}}
\renewcommand{\thetable}{\Alph{section}\arabic{table}}
\makeatother

\begin{figure}[H]
\begin{center} \unitlength1cm
\begin{picture}(0,5.5)
\put(-6.5,-0.3){\includegraphics[height=55mm]{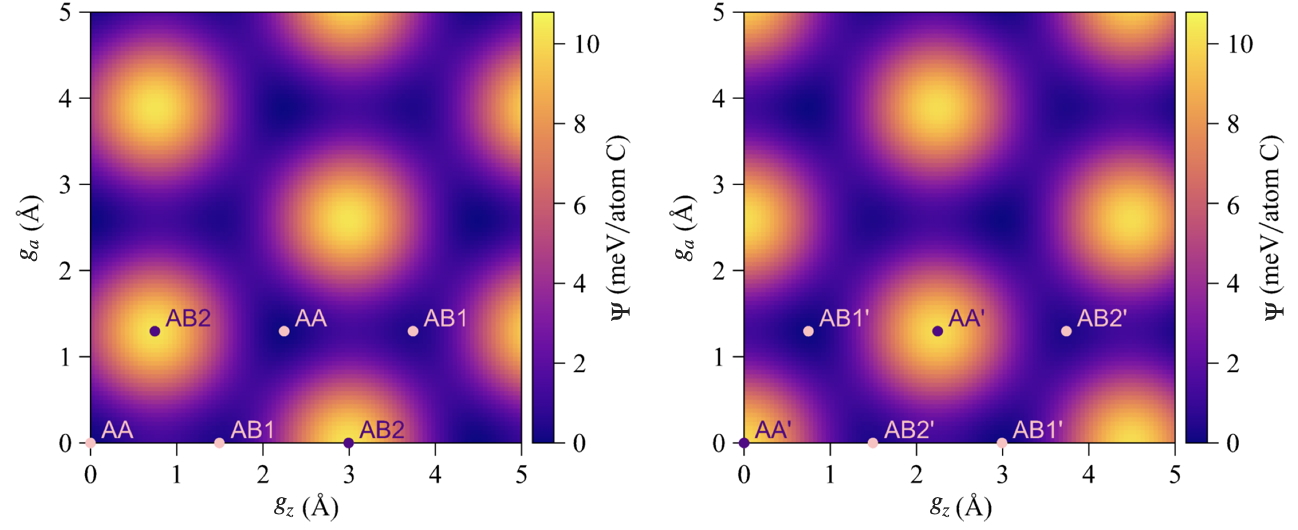}}
\put(-6.5,-0.5){(a)}
\put(0.5,-0.5){(b)}
\end{picture}
\vspace{5mm}
\caption{Adhesive interlayer interaction energy $\Psi$ (in meV per carbon atom) of hydro fluorinated bilayer graphene. Figures (a) and (b) correspond to the co-aligned and counter-aligned layers (the configuration where the hydrogen atoms of one layer are positioned just above the hydrogen atoms of the second layer in AB-type configuration), respectively. Here, the energy is given relative to the $\text{AA}$ and $\text{AB1}$ stacking, respectively. Reprinted from \citet{lebedev2020universal} with permission from American Physical Society.}
\label{energy varaition with stacking bilayer Graphene}
\end{center}
\end{figure}

\begin{figure}[H]
\begin{center} \unitlength1cm
\begin{picture}(0,4.2)
\put(-7.5,-0.3){\includegraphics[width=45mm, trim={0 0 3mm 0},clip]{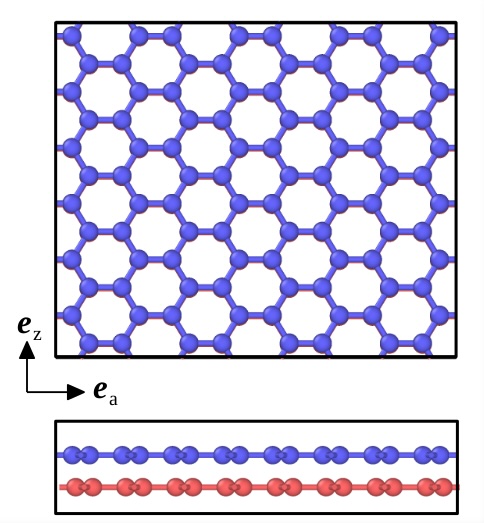}}
\put(-2,-0.3){\includegraphics[width=45mm, trim={0 0 3mm 0},clip]{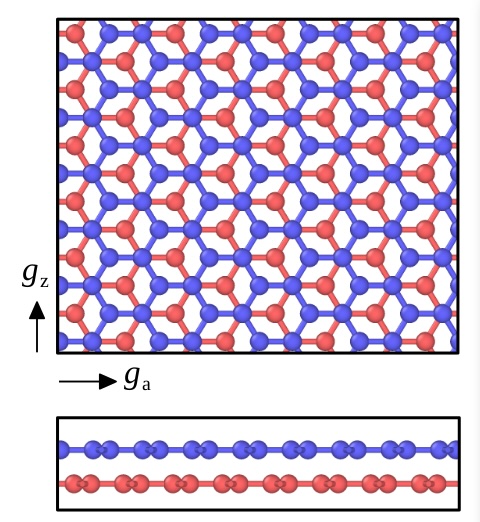}}
\put(3.5,-0.3){\includegraphics[width=45mm, trim={0 0 3mm 0},clip]{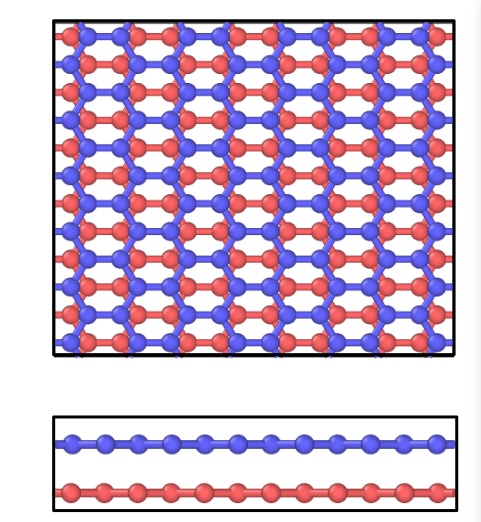}}
\put(-7.6,-0.5){(a)}
\put(-2.1,-0.5){(b)}
\put(3.5,-0.5){(c)}
\end{picture}
\vspace{5mm}
\caption{Extreme energy stackings of bi-layered graphene: (a)~AA, (b)~AB, and (c)~SP stacking. Here, $g_\mra$ and $g_\mrz$ refer to the relative displacement between the two sheets along the armchair and zigzag directions (denoted $\be_\mra$ and $\be_\mrz$), respectively.  Reprinted from \citet{mokhalingam2024continuum} with permission from American Physical Society.} 
\label{stacking}
\end{center}
\end{figure}

When crystalline 2D materials come into contact, a variety of configurations emerge based on their relative atomic arrangements. Some of these configurations correspond to global energy maxima, while others correspond to global or local minima in the adhesive (binding) energy, see Fig.~\ref{energy varaition with stacking bilayer Graphene}. For hexagonal lattice structures, typical stacking configurations are shown in Fig.~\ref{stacking}. The AA stacking configuration corresponds to the global energy maximum, while the AB and saddle-point (SP) stackings correspond to the global and local minimum, respectively see Fig.~\ref{energy varaition with stacking bilayer Graphene}.
The adhesion energy of a single atom on a crystalline substrate can be determined if the energy of the unit cell it belongs to is expressed in terms of its stacking configuration with respect to the substrate. Building on this concept, where specific stacking configurations such as $\Psi_{\text{AA}}$ and $\Psi_{\text{AB}}$ represent known energy extrema, \citet{san2014spontaneous, jung2014ab} proposed that the adhesion energy should vary periodically, following the symmetry of the lattice. As a result, the continuous variation of adhesion energy can be expressed using Fourier harmonics, requiring only the known energies at the AA and AB stacking points.

It is important to note that this formulation assumes that the contacting bodies are rigid; the actual energy in a relaxed deformed configuration may differ slightly, typically by a few percent, as discussed by \citet{lebedev2020universal}. This harmonic expansion approach can also be extended to heterointerfaces and strained systems where Moiré superlattices are formed \citep{san2014spontaneous, kumar2015elastic, lebedev2020universal}; see Appendix \ref{Appendix B}.\\
The formulation for homointerfaces proceeds by expanding Eq.~\eqref{energy as fourier sum} into
\begin{equation}
\ds \Psi(\boldsymbol{r})=2 Re\Biggl[\Psi_1\Bigl(\exp(i \, \bH_{1}\cdot\boldsymbol{\tau}) +\exp(i \, \bH_{2}\cdot\boldsymbol{\tau}) +\exp(-i(\bH_1+\bH_2)\cdot\boldsymbol{\tau})\Bigr) \Biggr] + \Psi_0 ~.
\label{expanded fourier sum energy _2 }
\end{equation}
By writing $\Psi_1$ as a complex quantity ($\Psi_1 = c + i d$) and using the centrosymmetric stacking points together with the peroidcity of the lattice, the adhesion energy corresponding to the AA stacking ($\boldsymbol{\tau}$ = $\boldsymbol{0} $) is given by 
\begin{equation}\label{AA}
\Psi_{\text{AA}} = 6 Re (\Psi_1)+\Psi_0~,
\end{equation}
and for the AB/BA stackings 
\begin{subequations} \label{AB and BA}
\begin{align} 
\Psi_{\text{AB}} &= 2Re\Biggl[\Psi_1\biggl(\exp{(i 2\pi/3)}+\exp{(i 2\pi/3)}+\exp{(-i 4\pi/3)}\biggr) \Biggr] +\Psi_0~,\\
\Psi_{\text{BA}} &= 2Re\Biggl[\Psi_1\biggl(\exp{(-i 2\pi/3)}+\exp{(-i 2\pi/3)}+\exp{(i 4\pi/3)}\biggr) \Biggr]+\Psi_0~.
\end{align}
\end{subequations}
Solving Eq.~\eqref{AA} and Eq.~\eqref{AB and BA} for $\Psi_0$ and $\Psi_1$ gives
\begin{equation}
\Psi_1= -\frac{(\Psi_{\text{AB}}-\Psi_{\text{AA}})}{9},\,\,\,\,\,\,\,\,\,\,\,\,\, \Psi_0=\frac{\Psi_{\text{AA}}+2\Psi_{\text{AB}}}{3}~.
\label{eq23}
\end{equation}
Because graphene exhibits symmetric stackings at the AB and BA configurations, the energy functional must be even about the origin (corresponding to the AA stacking). Therefore, the coefficients of the sinusoidal terms must vanish, as a result Eq.~\eqref{expanded fourier sum energy _2 } simplifies to
\begin{equation}
    \ds \Psi(\boldsymbol{\tau}, g_\mrn)= \Psi_0(g_\mrn)+2\Psi_1(g_\mrn)\Big(\cos{(\bH_1 \cdot \boldsymbol{\tau})}+ \cos{(\bH_2 \cdot \boldsymbol{\tau})}+ \cos\big({(\bH_1+\bH_2) \cdot \boldsymbol{\tau}\big)}\Big)~,
    \label{energy in reciprocal lattice vectors expanded}
\end{equation}
which is equal to Eq.~\eqref{energy magnitude and variation form}.

\section{Interlayer interaction energy of Moir{\'e} superlattice}\label{Appendix B}
\setcounter{equation}{0}
\setcounter{figure}{0}
\setcounter{table}{0}

\makeatletter
\renewcommand{\theequation}{\Alph{section}\arabic{equation}}
\renewcommand{\thefigure}{\Alph{section}\arabic{figure}}
\renewcommand{\thetable}{\Alph{section}\arabic{table}}
\makeatother
Following the work of \citet{jung2014ab, jung2015origin, jung2017moire}, the adhesion energy in terms of the MSL vectors for heterointerfaces is discussed here. Under general twisting and deformation, the primitive lattice vectors of the two layers transform according to
\begin{subequations}
    \begin{align}
\ds \ba_i = \,&\bR_{\theta} \cdot \bA_i~,  \\
 \ds  \boldsymbol{b}_i =\, &(1+\delta) (\mI+\boldsymbol{\varepsilon}) \cdot \bB_i \hspace{0.5cm}  (i=1,2)~.
\end{align}
 \label{deformed lattice vector}
\end{subequations}
Here, $\delta$ is the initial lattice mismatch, $\theta$ is the misalignment angle and $\boldsymbol{\varepsilon}$ is measuring in-plane strain between the layers. Then, by using the definition of the reciprocal lattice vector above, the reciprocal lattice vectors $\mg_i$ and $\mh_i$ of the upper and deformed lower layer, respectively, are obtained as \cite{kittel2018introduction}
\begin{subequations}
    \begin{align}
\ds \bg_i =\, & 2\pi\frac{\bR_{\frac{\pi}{2}}\cdot \ba_j}{\|\ba^{T}_1 \cdot \bR_{\frac{\pi}{2}} \cdot \ba_2 \|}~, \hspace{0.5cm}(i\neq j)~ ,  \\
 \bh_i = \,&2\pi\frac{\bR_{\frac{\pi}{2}}\cdot \boldsymbol{b}_j}{\|{\boldsymbol{b}^{T}_1} \cdot \bR_{\frac{\pi}{2}} \cdot \boldsymbol{b}_2 \|}~, \hspace{0.6cm}(i\neq j)~.  
 \end{align}
 \label{two reciprocal vector}
\end{subequations}
Using Eq.~\eqref{two reciprocal vector}, the reciprocal lattice vectors $\bG^\text{M}_i$ of a Moiré supercell is obtained as \cite{lopes2007graphene}
\begin{equation}
    \bG^\text{M}_i =  \bg_i- \bh_i~.
    \label{common reciprocal vector}
\end{equation}
Using Eq.~\eqref{energy as fourier sum} the adhesion energy for the Moiré supercell can be written as \cite{steele1973physical}
\begin{equation}
\ds     \Psi^{\text{M}} (g_a,g_z,g_n)=\sum_{i=1}^3\Psi^{\text{M}}_i(z)e^{i \bG^\text{M}_i.\boldsymbol{r}}+\Psi^{\text{M}}_0(g_n)~.
\label{moire energy expression}
\end{equation}
Note that $\Psi^{\text{M}}_0$ and $\Psi^{\text{M}}_i$ have to be determined separately for different vdWHs.


\bibliographystyle{apalike} 
\bibliography{Bib}

\end{document}